\documentclass[aip,amsmath,amssymb,reprint]{revtex4-1}

\usepackage{amsmath}
\usepackage{epstopdf}
\usepackage{graphicx}
\usepackage{dcolumn}
\usepackage{bm}
\usepackage{braket}
\usepackage{gensymb}
\usepackage{array}

\usepackage[utf8]{inputenc}
\usepackage[T1]{fontenc}
\usepackage{mathptmx}

\newcolumntype{P}[1]{>{\centering\arraybackslash}p{#1}}
\newcolumntype{M}[1]{>{\centering\arraybackslash}m{#1}}

\newcommand{\bs}{\boldsymbol}

\begin{document}

\preprint{AIP/123-QED}

\title{Exciton Dynamics in Conjugated Polymers}

\author{William Barford}
\email{william.barford@chem.ox.ac.uk}
\affiliation{Department of Chemistry, Physical and Theoretical Chemistry Laboratory, University of Oxford, Oxford, OX1 3QZ, United Kingdom}

\begin{abstract}

Exciton dynamics in $\pi$-conjugated polymers encompass multiple time and length scales. Ultrafast femtosecond processes are intrachain and involve a quantum mechanical correlation of the exciton and nuclear degrees of freedom.  In contrast, post-picosecond processes involve the incoherent F\"orster transfer of excitons between polymer chains.  Exciton dynamics is also strongly determined by the spatial and temporal disorder that is ubiquitous in conjugated polymers.
Since excitons are delocalized over hundreds of atoms, a theoretical understanding of these processes is only realistically possible by employing suitably parametrized coarse-grained exciton-phonon models. Moreover, to correctly account for ultrafast processes, the exciton and phonon modes must be treated on the same quantum mechanical basis and  the Ehrenfest approximation must be abandoned. This further implies that sophisticated numerical techniques must be employed to solve these models.
This review describes our current theoretical understanding of exciton dynamics in conjugated polymer systems. We begin by describing the energetic and spatial distribution of excitons in disordered polymer systems, and define the crucial concept of a `chromophore' in conjugated polymers. We also discuss the role of exciton-nuclear coupling, emphasizing the distinction between `fast' and `slow' nuclear degrees of freedom in determining `self-trapping' and `self-localization' of exciton-polarons.
Next, we discuss ultrafast intrachain exciton decoherence caused by exciton-phonon entanglement, which leads to fluorescence depolarization on the timescale of 10-fs. Interactions of the polymer with its environment causes the stochastic relaxation and localization of high-energy delocalized excitons onto chromophores.
The coupling of excitons with torsional modes also leads to various dynamical processes. On sub-ps timescales it causes exciton-polaron formation (i.e., exciton localization and local polymer planarization), while on post-ps timescales stochastic torsional fluctuations cause exciton-polaron diffusion along the polymer chain.
Finally, we describe a first-principles, F\"orster-type model of intrachain exciton transfer and diffusion, whose starting point is a realistic description of the donor and acceptor chromophores. We survey experimental results and explain how they can be understood in terms of our theoretical description of exciton dynamics coupled to information on polymer multiscale structures. The review also contains a brief critique of computational methods to simulate exciton dynamics.

\end{abstract}

\maketitle

\section{Introduction}\label{se:1}


The theoretical study of exciton dynamics in conjugated polymer systems is both a fascinating and complicated subject.
One reason for this is that characterizing excitonic states themselves is a challenging task: conjugated polymers exhibit strong electron-electron interactions and electron-nuclear coupling, and are  subject to spatial and temporal disorder.
Another reason is that exciton dynamics is characterised by multiple (and often overlapping) time scales; it is determined by both intrinsic processes (e.g., coupling to nuclear degrees of freedom and electrostatic interactions) and  extrinsic processes (e.g., polymer-solvent interactions); and  it is both an intrachain and interchain process.
Consequently, to make progress in both characterizing exciton states and correctly describing their dynamics, simplified, but realistic models are needed.
Moreover, as even these simplified models describe many quantized degrees of freedom, sophisticated numerical techniques are required to solve them. Luckily, fundamental theoretical progress in developing numerical techniques means that simplified  one-dimensional models of conjugated polymers are now  soluble to a high degree of accuracy.

In addition to the application of various  theoretical techniques to understand exciton dynamics, a wide range of time-resolved spectroscopic techniques have also been deployed. These include fluorescence depolarization\cite{Grage03,Ruseckas05,Wells07,Dykstra09}, three-pulse photon-echo\cite{Dykstra05,Yang05, Wells08,Sperling08} and coherent electronic two-dimensional spectroscopy\cite{Consani15}. Some of the timescales extracted from these experiments are listed in Table \ref{ta:2}; the purpose of this review is to describe their associated physical processes.
\begin{table*}
{
{\renewcommand{\arraystretch}{1.2}
\begin{tabular}{|p{2.5cm}|p{2.5cm}|p{9cm}|p{1.5cm}|}
\hline
Polymer & State &  Timescales & Citation \\
\hline
MEH-PPV & Solution & $\tau_1 = 50 $ fs, $\tau_2 = 1-2 $ ps & Ref\cite{Ruseckas05} \\
\hline
MEH-PPV & Solution & $\tau_1 = 5-10 $ fs, $\tau_2 = 100-200 $ fs  & Ref\cite{Collini09} \\
\hline
PDOPT & Film & $\tau = 0.5 - 4 $ ps & Ref\cite{Westenhoff06} \\
\hline
PDOPT & Solution & $\tau_1 < 1 $ ps, $\tau_2 = 15 -23 $ ps & Ref\cite{Westenhoff06} \\
\hline
P3HT & Film & $\tau_1 = 300 $ fs, $\tau_2 = 2.5 $ ps, $\tau_3 = 40 $ ps & Ref\cite{Westenhoff06} \\
\hline
P3HT & Solution & $\tau_1 = 700 $ fs, $\tau_2 = 6 $ ps, $\tau_3 = 41 $ ps, $\tau_4 = 530 $ ps & Ref\cite{Banerji11} \\
\hline
P3HT & Solution & $\tau_1 = 60 - 200 $ fs, $\tau_2 = 1-2 $ ps, $\tau_3 = 14-20 $ ps & Ref\cite{Busby11} \\
\hline
P3HT & Solution & $\tau_1 \lesssim 100$ fs, $\tau_2 \sim 1 - 10 $ ps & Ref\cite{Wells07} \\
\hline
\end{tabular}}
\caption{Some of the dynamical timescales observed in conjugated polymers whose associated physical processes are summarized in Table \ref{ta:1}.}
\label{ta:2}
}
\end{table*}

As well as being of intrinsic interest, the experimental and theoretical activities to understand exciton dynamics in conjugated polymer systems are also motivated by the importance of this process in determining the efficiency of polymer electronic devices.   In photovoltaic devices,  large exciton diffusion lengths are necessary so that excitons can migrate efficiently to regions where charge separation can occur. However,  precisely the opposite is required in light emitting devices, since diffusion leads to non-radiative quenching of the exciton.

Perhaps one of the reasons for the failure to fully exploit polymer electronic devices has been the difficulty in establishing the structure-function relationships which  allow the development of rational design strategies. An understanding of the principles of exciton dynamics, relating this to  multiscale polymer structures, and interpreting the associated spectroscopic signatures are all key ingredients to developing structure-function relationships. An earlier review explored the connection between structure and spectroscopy\cite{Barford17}. In this review we describe our current understanding of the important dynamical processes in conjugated polymers, beginning with photoexcitation and intrachain relaxation on ultrafast timescales ($\sim 10$ fs) to sub-ns interchain exciton transfer and diffusion. These key processes are summarized in Table \ref{ta:1}.


\begin{table*}
{
{\renewcommand{\arraystretch}{1.2}
\begin{tabular}{|p{7cm}|p{5cm}|p{2.5cm}|p{1.5cm}|}
\hline
Process & Consequences &  Timescale & Section \\
\hline
Exciton-polaron self-trapping via coupling to fast C-C bond vibrations. & Exciton-site decoherence; ultrafast fluorescence depolarization. & $\sim 10$ fs & \ref{se:3.1} \\
\hline
Energy relaxation from high-energy quasi-extended exciton states (QEESs) to low-energy local exciton ground states (LEGSs) via coupling to the environment. & Stochastic exciton density localization onto chromophores. & $\sim 100-200$ fs & \ref{se:3.2}\\
\hline
Exciton-polaron self-localization via coupling to slow bond rotations in the under-damped regime. & Exciton density localization on a chromophore; ultrafast fluorescence depolarization. & $\sim 200-600$ fs & \ref{se:3.3}\\
\hline
Exciton-polaron self-localization via coupling to slow bond rotations in the over-damped regime. & Exciton density localization on a chromophore; post-ps fluorescence depolarization. & $\sim 1- 10$ ps & \ref{se:3.3}\\
\hline
Stochastic torsional fluctuations inducing exciton `crawling' and `skipping' motion. & Intrachain exciton diffusion and energy fluctuations. & $\sim 3-30$ ps & \ref{se:4}\\
\hline
Interchromophore F\"orster resonant energy transfer. & Interchromophore exciton diffusion; post-ps spectral diffusion and  fluorescence depolarization. & $\sim 10-100$ ps & \ref{se:5}\\
\hline
Radiative decay. &  & $\sim 500$ ps & \\
\hline
\end{tabular}}
\caption{The life and times of an exciton: Some of the key exciton dynamical processes, encompassing over four-orders of magnitude, that occur in conjugated polymer systems.}
\label{ta:1}
}
\end{table*}

The plan of this review is the following. We begin by briefly describing some theoretical techniques for simulating exciton dynamics and emphasize the failures of simple methods.  As already mentioned, excitons themselves are fascinating quasiparticles, so before describing their dynamics, in Section \ref{se:3} we start by describing their stationary states. We stress the role of low-dimensionality, disorder and electron-phonon coupling, and we discuss the fundamental concept of a chromophore. Next, in Section \ref{se:4}, we describe the sub-ps processes of intrachain exciton decoherence, relaxation and  localization, which - starting from an arbitrary photoexcited state - results in an exciton forming a chromophore. We next turn to describe the exciton (and energy) transfer processes occurring on post-ps timescales. First, in Section \ref{se:5}, we describe the primarily adiabatic intrachain motion of excitons caused by stochastic torsional fluctuations, and second, in Section \ref{se:6}, we describe nonadiabatic interchain exciton transfer. We conclude and address outstanding questions in  Section \ref{se:7}.


\section{A Brief Critique of Theoretical Techniques}\label{se:2}

A theoretical description of exciton dynamics in conjugated polymers poses considerable challenges, as it requires a rigorous treatment of electronic excited states and their coupling to the nuclear degrees of freedom. Furthermore, conjugated polymers consist of thousands of atoms and tens of thousands of electrons. Thus, as the Hilbert space grows exponentially  with the number of degrees of freedom, approximate treatments of excitonic dynamics are therefore inevitable. There are two broad approaches to a theoretical treatment. One approach is to construct \emph{ab initio} Hamiltonians, with an exact as possible representation of the degrees of freedom, and then to solve these Hamiltonians with various degrees of accuracy. Another approach (albeit less common  in theoretical chemistry) is to construct effective Hamiltonians with fewer degrees of freedom, such as the Frenkel-Holstein model described in Section \ref{se:4}. These effective Hamiltonians might be parameterized via a direct mapping from \emph{ab initio} Hamiltonians (e.g., see Appendix H in ref\cite{Book}, Appendix A in ref\cite{Barford14a} and various papers by Burghardt and coworkers\cite{Binder14,Binder20b}) or else semiempirically\cite{Barford14b}. A significant advantage of effective Hamiltonians over their \emph{ab initio} counterparts is that they can be solved for larger systems over longer timescales and to  a higher level of accuracy.

As the Ehrenfest method is a widely  used approximation to study charge and exciton dynamics in conjugated polymers, we briefly explain this method and describe the important ways in which it fails. (For a fuller treatment, see\cite{Horsfield06,Nelson20}.) The Ehrenfest method makes two key  approximations. The first approximation is to treat the nuclei classically. This means that nuclear quantum tunneling and zero-point energies are neglected, and that exciton-polarons are not correctly described (see Section \ref{se:3.3}). The second assumption is that the total wavefunction is a product of the electronic and nuclear wavefunctions. This means that there is no entanglement between the  electrons and nuclei,  and so the nuclei cannot cause decoherence of the electronic degrees of freedom (see Section \ref{se:4.1}). A simple product wavefunction also implies that the nuclei  move in a mean field potential determined by the electrons. This means that a splitting of the nuclear wave packet when passing through a conical intersection or an avoided crossing does not occur
(see Section \ref{se:4.2}), and that  there is an incorrect description of energy transfer between the electronic and nuclear degrees of freedom (see Section \ref{se:5.4}). As will be discussed in the course of this review, these failures mean that in general the Ehrenfest method is not a reliable one to treat ultrafast excitonic dynamics in conjugated polymers.

Various theoretical techniques have been proposed to rectify the failures of the Ehrenfest method; for example, the surface-hopping technique\cite{Tully90,Tully12}, while still keeping the nuclei classical, partially rectifies the failures at conical interactions. More sophisticated approaches, for example the MC-TDHF  and TEBD methods, quantize the nuclear degrees of freedom and do not assume a product wavefunction.

For a given electronic potential energy surface (PES), the multiconfigurational-time dependent Hartree-Fock (MC-TDHF) method\cite{Beck00} is an (in principle) exact treatment of nuclear wavepacket propagation, although in practice exponential scaling of the Hilbert space means that a truncation is required. In addition, this method is only as reliable as the representation of the PES.

In the time-evolving block decimation (TEBD) method\cite{Vidal03,Vidal04} a quantum state, $|\Psi\rangle$, is represented by a matrix product state (MPS)\cite{Schollwock11}. Its time evolution is determined via
\begin{equation}\label{}
  |\Psi(t+\delta t\rangle = \exp(-\textrm{i}\hat{H}\delta t/\hbar)|\Psi(t)\rangle,
\end{equation}
where $\hat{H}$ is the system Hamiltonian and the action of the evolution operator is performed via a Trotter decomposition. Since the action of the evolution operator expands the Hilbert space, $|\Psi\rangle$ is subsequently compressed via a singular value decomposition (SVD)\footnote{A related method is time-dependent density matrix renormalization group (TD-DMRG). This has been successfully applied to simulate singlet fission in carotenoids, D. Manawadu, M. Marcus and W. Barford, in preparation.}. Importantly, this approach is `numerically exact' as long as the truncation parameter exceeds $2^S$, where $S$ is the entanglement entropy, defined by $S = -\sum_{\alpha} \omega_{\alpha} \textrm{ln}_2 \omega_{\alpha}$ and $\{\omega\}$ are the singular values obtained at the SVD. The TEBD method permits the electronic and nuclear degrees of freedom to be treated as quantum variables on an equal footing. It thus rectifies all of the failures of the Ehrenfest method described above and, unlike the MC-TDHF method, it is not limited by the representation of the PES. It can, however, only be applied to quantum systems described by one-dimensional lattice Hamiltonians\cite{Mannouch18}. Luckily, as described in Section \ref{se:4}, such model Hamiltonians are readily constructed to describe exciton dynamics in  conjugated polymers.


\section{Excitons in Conjugated Polymers}\label{se:3}

Before discussing the dynamics of excitons, we begin by describing exciton stationary states in static conjugated polymers.

\subsection{Two-particle model}\label{se:3.1}

An exciton is a Coulombically bound electron-hole pair formed by the linear combination of electron-hole excitations
(for further details see\cite{Book, Abe93, Barford13}). In a one-dimensional conjugated polymer an exciton is described by the two-particle wavefunction,
  $\Phi_{mj}(r,R) = \psi_m(r)\Psi_j(R)$.

$\Psi_j(R)$ is the center-of-mass wavefunction, which will be discussed shortly. Before doing that, we first discuss the relative wavefunction, $\psi_m(r)$, which  describes a particle bound to a screened Coulomb potential, where $r$ is the electron-hole separation and $m$ is the principal quantum number. The electron and hole of an exciton in a one-dimensional semiconducting polymer are more strongly bound than in a three-dimensional inorganic  semiconductor for two key reasons.\cite{Book, Barford13} First, because of the low dielectric constant and relatively large electronic effective mass in $\pi$-conjugated systems the effective Rydberg is typically $50$ times larger than for inorganic systems. Second, dimensionality plays a role: in particular, the one-dimensional Schr\"odinger equation for the relative particle\cite{Loudon16,Barford02a} predicts a strongly bound state split-off from the Rydberg series. This  state is the $m=1$ Frenkel (`$1B_u$') exciton, with a binding energy of $\sim 1$ eV and an electron-hole wavefunction  confined to a single monomer. The first exciton in the `Rydberg' series is the $m=2$ charge-transfer (`$2A_g$') exciton.

With the exception of donor-acceptor copolymers, conjugated polymers are generally non-polar, which means that each $p$-orbital has an average occupancy of one electron. This implies an approximate \emph{electron-hole} symmetry. Electron-hole symmetry has a number of consequences for the character and properties of excitons. First, it means that the relative wavefunction exhibits electron-hole parity, i.e., $\psi_m(r) = + \psi_m(-r)$  when $m$ is odd and $\psi_m(r) = - \psi_m(-r)$ when $m$ is even. Second, the transition density, $\langle \textrm{EX}|\hat{N}_i |\textrm{GS}\rangle$, vanishes for odd-parity (i.e., even $m$) excitons. This means that such excitons are not optically active, and importantly for dynamical processes, their F\"orster exciton transfer rate (defined in Section \ref{se:6.1}) vanishes.

Since Frenkel excitons are the primary photoexcited states of conjugated polymers, their dynamics is the  subject of this review. Their delocalization along the polymer chain of $N$ monomers is described by the Frenkel Hamiltonian,
\begin{eqnarray}\label{Eq:1}
  \hat{H}_{F} &=& \sum_{n=1}^{N}\epsilon_n  \hat{N}_{n} +\sum_{n=1}^{N-1}J_n \hat{T}_{n,n+1},
\end{eqnarray}
where $n = (R/d)$  labels a monomer and $d$ is the intermonomer separation.
The energy to excite a Frenkel exciton on monomer $n$ is $\epsilon_{n}$, where
 $\hat{N}_{n} = \ket{n}\bra{n}$
is the Frenkel exciton number operator.

In principle,  excitons delocalize along the chain via two mechanisms\cite{Barford09a,Barford13}. First, for even-parity (odd $m$) singlet excitons  there is a Coulomb-induced (or through space) mechanism. This is the familiar mechanism of F\"orster  energy transfer.
The exciton transfer integral for this process is
\begin{eqnarray}\label{Eq:39}
    J_{DA} =  \sum_{\substack{i\in D \\ j\in A}}V_{ij} \left[
    {_D}\langle \textrm{GS} |\hat{N}_i| \textrm{EX} \rangle_D\right]
    \left[{_A}\langle \textrm{EX} |\hat{N}_j|\textrm{GS}\rangle_A
    \right].
\end{eqnarray}
The sum is over sites $i$ in the donor monomer  and $j$ in the acceptor monomer, and $V_{ij}$ is the Coulomb interaction between these sites.
In the point-dipole approximation Eq.\ (\ref{Eq:39}) becomes
\begin{equation}\label{Eq:4}
    J_{DA} = \frac{\kappa_{mn} \mu_0^2}{4\pi\varepsilon_{r}\varepsilon_{0} R_{mn}^3},
\end{equation}
where  $\mu_0$ is the transition dipole moment of a single monomer and ${R}_{mn}$ is the distance between the monomers $m$ and $n$.
$\kappa_{mn}$ is the orientational factor,
\begin{equation}\label{Eq:40}
\kappa_{mn} = \bs{\hat{r}_m}\cdot\bs{\hat{r}_{n}} - 3(\bs{\hat{R}_{mn}}\cdot\bs{\hat{r}_{m}})(\bs{\hat{R}_{mn}}\cdot\bs{\hat{r}_{n}}),
\end{equation}
where $\bs{\hat{r}_{m}}$ is a unit vector parallel to the dipole on monomer $m$ and $\bs{\hat{R}_{mn}}$ is a unit vector parallel to the vector joining monomers $m$ and $n$. For colinear monomers, the nearest neighbor through space transfer integral is
\begin{equation}
    J_{DA} = -\frac{2 \mu_0^2}{4\pi\varepsilon_{r}\varepsilon_{0} d^3}.
\end{equation}

Second, for all excitons there is a super-exchange (or through bond) mechanism, whose origin lies in a virtual fluctuation from a Frenkel exciton on a single monomer to a charge-transfer exciton spanning two monomers back to a Frenkel exciton on a neighboring monomer. The energy scale for this process, obtained from second order perturbation theory\cite{Book}, is
\begin{equation}\label{Eq:5}
    J_{SE}(\theta) \propto -\frac{t(\theta)^2}{\Delta E},
\end{equation}
where $ t(\theta)$ (defined in Eq.\ (\ref{Eq:10c})) is proportional to the overlap of  $p$-orbitals neighboring a bridging bond, i.e., $t(\phi) \propto \cos \theta$ and  $\theta$ is the torsional (or dihedral) angle between neighboring monomers. $\Delta E$ is the difference in energy between a charge-transfer and Frenkel exciton.

The total exciton transfer integral is  thus
\begin{equation}\label{Eq:4}
J_{n} = J_{\text{DA}} + J_{\text{SE}}^0\cos^{2}{\theta_{n}}.
\end{equation}
The bond-order operator,
\begin{eqnarray}\label{Eq:7a}
\hat{T}_{n,n+1} = \left( \ket{n}\bra{n+1} +  \ket{n+1}\bra{n}\right),
\end{eqnarray}
represents the hopping of the Frenkel exciton  between monomers $n$ and $n+1$.
Evidently, $J_{SE}$ vanishes when $\theta = 0$, but $J_{DA}$ will not. Therefore, even if $J_{SE}$ vanishes because of negligible $p$-orbital overlap between neighboring monomers, singlet even-parity excitons can still retain phase coherence over the `conjugation break'\cite{Barford10a}. This observation has important implications for the definition of chromophores, as discussed in Section \ref{se:3.2}.

Eq.\ (\ref{Eq:1}) represents a `coarse-graining' of the exciton degrees of freedom. The key assumption is that we can replace the atomist detail of each monomer (or moiety) and replace it by a `coarse-grained' site, as illustrated in Fig.\ 1. All that remains is to describe how the Frenkel exciton delocalises along the chain, which is controlled by the two sets of parameters, $\{\epsilon\}$ and $\{J\}$.
Since $J$ is negative, a conjugated polymer is equivalent to a molecular J-aggregate.
\begin{figure}\label{Fi:1}
\includegraphics[width=0.5\textwidth]{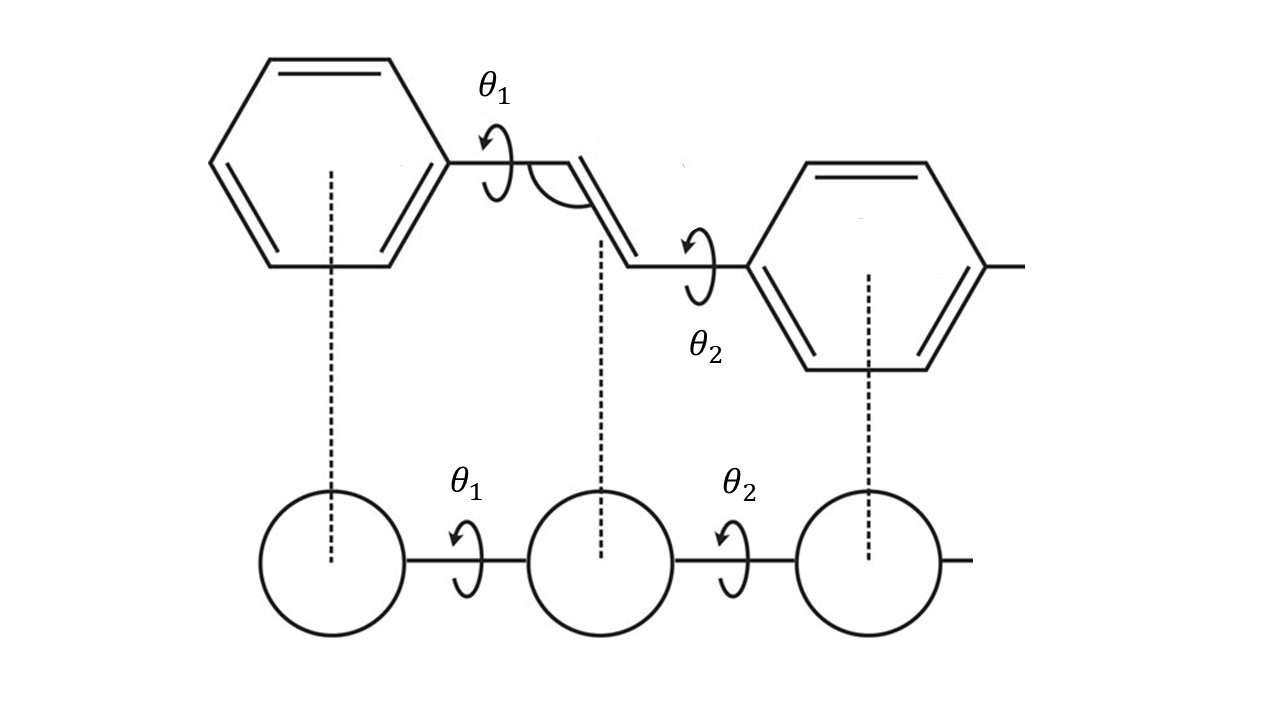}
\caption{The mapping of a polymer chain conformation to a coarse-grained linear site model. Each site corresponds to a moiety along the polymer chain, with the connection between sites characterised by the torsional (or dihedral) angle, $\theta$.}
\end{figure}

The eigenfunctions of $\hat{H}_{F}$ are the center-of-mass wavefunctions, $\Psi_j(n)$,  where  $j$ is the associated quantum number. For a linear, uniform polymer (i.e., $\epsilon_n \equiv \epsilon_0$ and $J_n \equiv J_0$)
\begin{equation}\label{Eq:251}
\Psi_j(n) = \left(\frac{2}{N+1}\right)^{1/2}\sum_{n=1}^N \sin\left(\frac{\pi j n}{N+1}\right),
\end{equation}
forming a band of states with energy
\begin{equation}\label{Eq:252}
E_j = \epsilon_0 +2J_0 \cos\left(\frac{\pi j }{N+1}\right).
\end{equation}
The family of excitons with  different $j$ values corresponds to the Frenkel exciton band with different center-of-mass momenta. In emissive polymers the  $j=1$ Frenkel exciton is generally labeled the $1^1B_u$ state.

\subsection{Role of static disorder: local exciton ground states and quasiextended exciton states}\label{se:3.2}

Polymers are rarely free from some kind of disorder and thus the form of Eq.\ (\ref{Eq:251}) is not valid for the center-of-mass wavefunction in realistic systems. Polymers in solution are necessarily conformationally disordered as a consequence of thermal fluctuations (as described in Section \ref{se:5}). Polymers in the condensed phase usually exhibit glassy, disordered conformations as consequence of  being quenched from solution. Conformational disorder implies that the dihedral angles, $\{ \theta \}$ are disordered, which
by virtue of Eq.\ (\ref{Eq:4}) implies that the exciton transfer integrals are also disordered.

As well as conformational disorder, polymers are also subject to chemical and environmental disorder (arising, for example, from density fluctuations). This type of disorder affects the energy to excite a Frenkel exciton on a monomer (or coarse-grained site).


\begin{figure}
\centering
\includegraphics[width=0.4\textwidth]{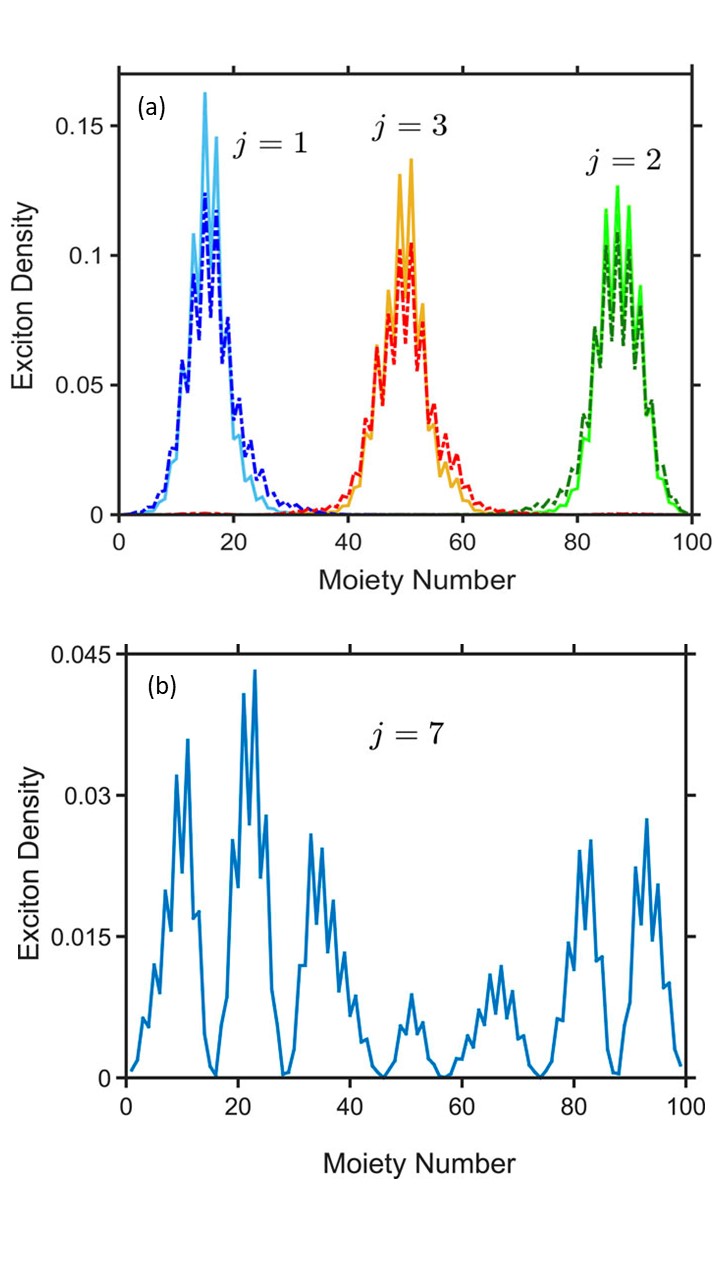}
  \caption{(a) The density of three local exciton ground states (LEGSs, dotted curves) and the three vibrationally relaxed states (VRSs, solid curves) for one particular static conformation of a PPV polymer chain made up of 50 monomers. The exciton center-of-mass quantum number, $j$, for each state is also shown.
  (b)  The exciton density of a quasiextended exciton state (QEES), with quantum number $j=7$.
  Reproduced  from J. Chem. Phys. \textbf{148}, 034901 (2018) with the permission of AIP publishing.}\label{Fi:2}
\end{figure}

As first realized by Anderson\cite{Anderson58a}, disorder localizes a quantum particle (in our case, the exciton center-of-mass particle), and determines their energetic and spatial distributions. The origin of this localization is the wave-like nature of a quantum particle and the constructive and destructive interference it experiences as it scatters off a random potential.  Malyshev and Malyshev\cite{Malyshev01a,Malyshev01b} further observed that in one-dimensional systems there are a class of states in the low energy tail of the density of states that are superlocalized,
named local exciton ground states (LEGSs\cite{Malyshev01a,Malyshev01b,Makhov10}). LEGSs are essentially nodeless, non-overlapping wavefunctions that together spatially span the entire chain.
They are \emph{local} ground states, because for the individual parts of the chain that they span there are no lower energy states. A consequence of the essentially nodeless quality of LEGSs is that the square of their transition dipole moment scales as their size\cite{Makhov10}. Thus, LEGSs define  chromophores (or spectroscopic segments), namely the irreducible parts of a polymer chain that absorb and emit light. Fig.\ \ref{Fi:2}(a) illustrates the three LEGSs for a particular conformation of PPV with 50 monomers.

Some researchers claim that `conjugation-breaks' (or more correctly, minimum thresholds in the $p_z$-orbital overlap) define the boundaries of chromophores\cite{Athanasopoulos08}. In contrast, we suggest that  it is the disorder that determines the average chromophore size, but `conjugation-breaks' can `pin' the chromophore boundaries. Thus, if the average distance between conjugation breaks is smaller than the chromophore size, chromophores will span  conjugation breaks but they may also be separated by them. Conversely, if average distance between conjugation breaks is larger than the chromophore size the chromophore boundaries are largely unaffected by the breaks. The former scenario occurs in polymers with shallow torsional potentials, e.g., polythiophene\cite{Barford10a}.

Higher energy lying states are also localized, but are nodeful and generally spatially overlap a number of low-lying LEGSs. These states are named quasiextended exciton states (QEESs) and an example is  illustrated in Fig.\ \ref{Fi:2}(b).

\begin{figure}
\centering
\includegraphics[width=0.48\textwidth]{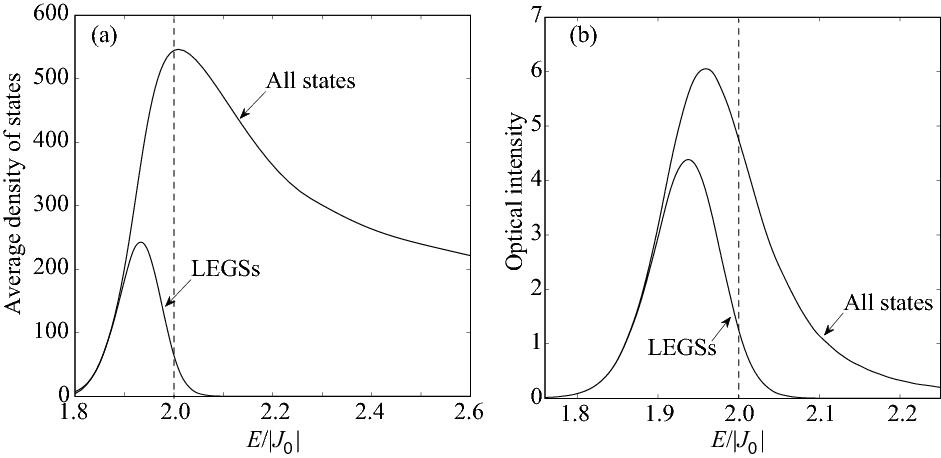}
  \caption{(a) The energy density of states and (b) the optical absorption (neglecting the vibronic progression) of the manifold of Frenkel excitons (where $|\sigma_J/J_0| = 0.1$). The width of the LEGSs density
  of states  $\sim |J_0||\sigma_J/J_0|^{4/3}$. Similarly, the width of the  optical absorption from both the LEGSs and all states $\sim
|J_0||\sigma_J/J_0|^{4/3}$. The band edge for an ordered chain is at $2|J_0|$ (indicated by the dashed lines), so LEGSs generally lie in the Lifshitz (or Urbach) tail of the density of states, i.e., $E  < 2|J_0|$.
}
\label{Fi:3}
\end{figure}

When the disorder is Gaussian distributed with a standard deviation $\sigma$, single parameter scaling theory\cite{Kramer93} provides some exact results  about the spatial and energetic distribution of the exciton center-of-mass states:
\begin{enumerate}
\item{The localization length
$L_{loc} \sim ({|J_0|}/{\sigma})^{2/3}$
at the band edge and as
$L_{loc} \sim ({|J_0|}/{\sigma})^{4/3}$
at the band center.}
\item{As a consequence of exchange narrowing, the width of the density of states occupied by LEGSs scales as $\sigma/\sqrt{L_{loc}} \sim \sigma^{4/3}$. Similarly, the optical absorption is inhomogeneously narrowed with a line width $\sim \sigma^{4/3}$.}
\item{The fraction of LEGSs scales as $1/L_{loc} \sim \sigma^{2/3}$.}
\end{enumerate}
These points are illustrated in Fig.\ \ref{Fi:3}, which shows the Frenkel exciton density of states and optical absorption for a particular value of disorder. Evidently, although LEGSs are a small fraction of the total number of states, they dominate the optical absorption.

This section has described LEGSs (or chromophores) as static objects defined by static disorder. However, as discussed in Section \ref{se:5}, dynamically torsional fluctuations also render the conformational disorder dynamic causing the LEGSs to evolve adiabatically. As a consequence, the chromophores `crawl' along the polymer chain.

\subsection{Role of electron-nuclear coupling: exciton-polarons}\label{se:3.3}

As well as disorder, another important process in determining exciton dynamics and spectroscopy is the coupling of an exciton to nuclear degrees of freedom; in a conjugated polymer these are fast C-C bond vibrations and slow monomer rotations. In this section we briefly review the origin of this coupling and then discuss exciton-polarons.

\subsubsection{Origin of electron-nuclear coupling}\label{se:3.3.1}

When a nucleus moves, either by a linear displacement or by a rotation about a fixed point, there is a change in the electronic overlap between neighboring atomic orbitals. Assuming that neighboring $p$-orbitals lie in the same plane normal to the bond  with a relative twist angle of $\theta$, the resonance integral between a pair of orbitals separated by $r$ is\cite{Mulliken49}
\begin{equation}\label{Eq:10c}
t(\theta)= t(r)\cos \theta = \beta \exp(-\alpha r) \cos \theta,
\end{equation}
where $ t(r) < 0$. The kinetic energy contribution to the Hamiltonian is
\begin{equation}\label{Eq:11}
  \hat{H}_{\textrm{ke}} = t(r)\cos \theta \times {\hat T},
\end{equation}
where the bond-order operator, ${\hat T}$, is defined in Eq.\ (\ref{Eq:7a}). Treating $r$ and $\theta$  as dynamical variables, suppose that the $\sigma$-electrons of a conjugated molecule and steric hinderances provide equilibrium values of $r=r_0$ and $\theta = \theta_0$, with corresponding elastic potentials of
\begin{equation}\label{}
  V_{\textrm{vib}} = \frac{1}{2}K_{\textrm{vib}}^{\sigma}(r - r_0)^2
\end{equation}
and
\begin{equation}\label{}
 V_{\textrm{rot}} = \frac{1}{2}K_{\textrm{rot}}^{\sigma}(\theta - \theta_0)^2.
\end{equation}
The coupling of the $\pi$-electrons to the nuclei changes these equilibrium values and the elastic constants.

To see this, we use the Hellmann-Feynman to determine the force on the bond. The linear displacement force is
\begin{eqnarray}\label{}
\nonumber
  f = -\frac{\partial E}{\partial r} && = -\big\langle\frac{\partial\hat{H}_{\textrm{ke}}}{\partial r}\big\rangle \\
                  && = \alpha  t(r)\cos \theta \langle {\hat T} \rangle - K_{\textrm{vib}}^{\sigma}(r - r_0).
\end{eqnarray}
Thus,  to first order in the change of bond length, $\delta r =  (r - r_0)$, the equilibrium distortion is
\begin{eqnarray}\label{}
  \delta r  =  \alpha  t(r_0)\cos \theta \langle {\hat T} \rangle/K_{\textrm{vib}}^{\sigma},
\end{eqnarray}
which is negative because it is favorable to shorten the bond to increase the electronic overlap.

Similarly, the torque around the bond is
\begin{eqnarray}\label{}
\nonumber
  \Gamma = -\frac{\partial E}{\partial \theta} && = -\big\langle\frac{\partial\hat{H}_{ke}}{\partial \theta}\big\rangle \\
                  && =   t(r)\sin \theta \langle {\hat T} \rangle - K_{\textrm{rot}}^{\sigma}(\theta - \theta_0)
\end{eqnarray}
and the equilibrium change of bond angle, $\delta \theta = (\theta - \theta_0)$, is
\begin{eqnarray}\label{Eq:16a}
  \delta \theta =  t(r)\sin \theta_0 \langle {\hat T} \rangle/K_{\textrm{rot}}^{\sigma},
\end{eqnarray}
which is also negative, again because it is favorable to increase the electronic overlap. Thus, the $\pi$-electron couplings act to planarize the chain.

\begin{figure}
\centering
\includegraphics[width=0.5\textwidth]{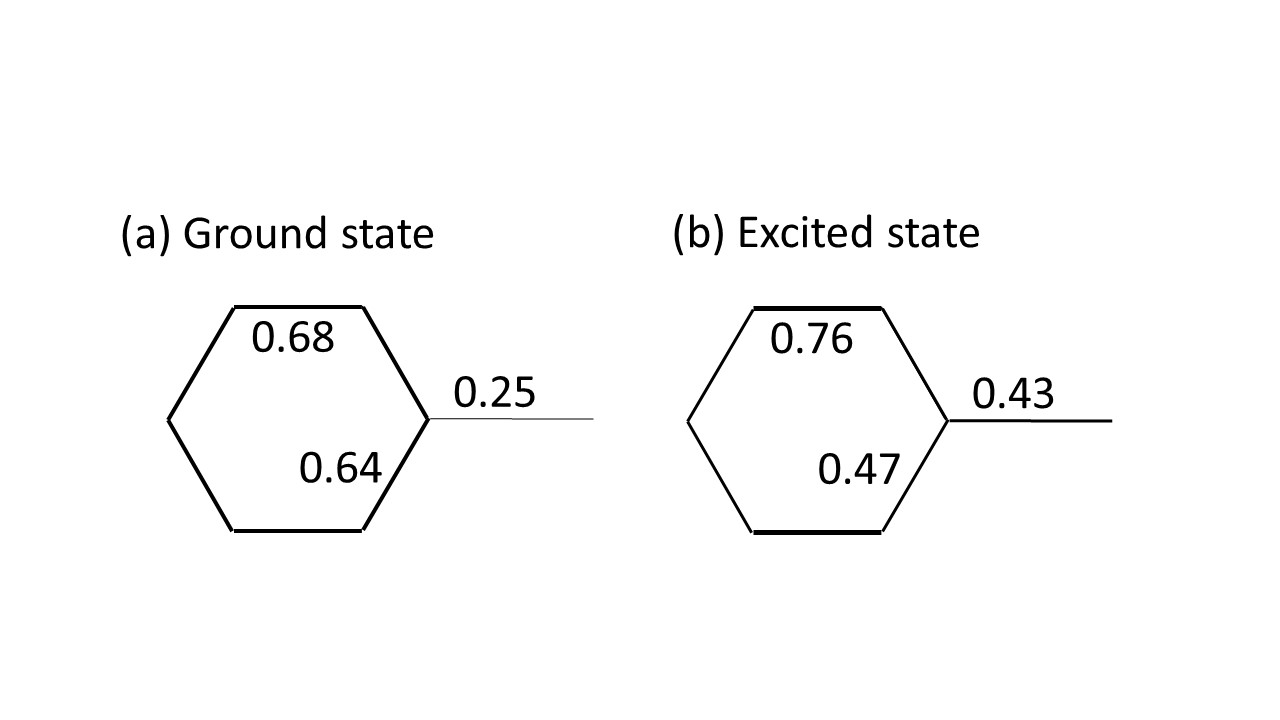}
\caption{The $\pi$-bond order expectation values, $\langle \hat{T}\rangle$, for  (a) the ground state and (b) the excited state, showing the benzenoid-quinoid transition.
As Eq.\ (\ref{Eq:16a}) and Eq.\ (\ref{Eq:17a}) indicate, the larger bond order of the bridging bond in the excited state implies a smaller dihedral angle and a stiffer torsional potential than  the ground state.}\label{Fi:40}
\end{figure}
The electron-nuclear coupling also changes the elastic constants. Assuming a harmonic potential, the new rotational spring constant is
\begin{eqnarray}\label{Eq:17a}
\nonumber
    K_{\textrm{rot}}^{\pi} = && \frac{\partial^2 E}{\partial \theta^2} \\
   = &&  - t(r_0)\cos \theta_0 \langle {\hat T} \rangle + K_{\textrm{rot}}^{\sigma}
\end{eqnarray}
and thus
$ K_{\textrm{rot}}^{\pi} > K_{\textrm{rot}}^{\sigma}$ (because $ t(r_0) < 0$).

Interestingly, as shown in Fig.\ 4, because $\langle {\hat T} \rangle_{EX} > \langle {\hat T} \rangle_{GS}$ for the bridging bond in phenyl-based systems, the torsional angle is smaller and the potential is stiffer in the excited state (as a result of the benzenoid to quinoid distortion)\cite{Beenken05}.

\subsubsection{Exciton-polarons}\label{se:3.3.2}

An  exciton  that couples to a set of harmonic oscillators, e.g., bond vibrations or torsional oscillations,  becomes `self-trapped'. Self-trapping means that the coupling between the exciton and oscillators causes a local displacement of the oscillator that is proportional to the local exciton density\cite{Rashba57a,Rashba57b,Rashba57c,Holstein59a,Holstein59b} (as illustrated in the next section). Alternatively, it is said that the exciton is dressed by a cloud of oscillators. Such a quasiparticle is named an exciton-polaron.
As there is no barrier to self-trapping in one-dimensional systems\cite{Rashba82}, there is always an associated relaxation energy.

If the exciton and oscillators are all treated quantum mechanically, then in a translationally invariant system the exciton-polaron forms a Bloch state and is \emph{not} localized. However, if the oscillators are treated classically, the non-linear feedback induced by the exciton-oscillator coupling self-localizes the exciton-polaron and `spontaneously' breaks the translational symmetry. This is a self-localized (or auto-localized) `Landau polaron'.\cite{Landau33, Campbell82} Notice that self-trapping is a necessary but not sufficient condition for self-localization.  Self-localization always occurs in the limit of vanishing oscillator frequency (i.e., the adiabatic or classical limit) and vanishing disorder.\cite{Tozer14}

Whether or not an exciton-polaron is self-localized in practice, however, depends on the strength of the disorder and the vibrational frequency of the oscillators. Qualitatively, an exciton coupling to fast oscillators (e.g., C-C bond vibrations) forms an exciton-polaron with an effective mass only slightly larger than a bare exciton\cite{Tozer14}. For realistic values of disorder, such an exciton-polaron is not self-localized. This is illustrated in Fig. \ref{Fi:2}(a), which shows the three lowest solutions of the Frenkel-Holstein model (described in Section \ref{se:4.1}), known as vibrationally relaxed states (VRSs). As we see, the density of the VRSs mirrors that of the  Anderson-localized LEGSs.
Conversely, an exciton coupling to slow oscillators (e.g.,  bridging-bond rotations) forms an exciton-polaron with a large effective mass. Such an exction-polaron is  self-localized  (as described in Section \ref{se:4.3} and shown in Fig.\ \ref{Fi:5}).

\section{Intrachain Decoherence, Relaxation and Localization}\label{se:4}

Having qualitatively described the stationary states of excitons in conjugated polymers, we now turn to a discussion of exciton dynamics.

\subsection{Role of fast C-C bond vibrations}\label{se:4.1}

After photoexcitation or charge combination after injection, the fastest process is the coupling of the exciton to C-C bond stretches. We now describe the resulting exciton-polaron formation and the loss of exciton-site coherence.

As we saw in Section \ref{se:3.3}, bond distortions couple to electrons. Using Eq.\ (\ref{Eq:11}), it can be shown\cite{Barford14b} that the
 coupling of local normal modes (e.g., vinyl-unit bond stretches or phenyl-ring symmetric breathing modes) to a Frenkel exciton is conveniently described by the Frenkel-Holstein model\cite{Holstein59a,Barford14b},
\begin{eqnarray}\label{Eq:2}
\nonumber
  \hat{H}_{FH} &=& \hat{H}_{F}
 -A\hbar\omega_{\textrm{vib}}\sum_{n=1}^{N} \tilde{Q}_n \hat{N}_{n}
    + \frac{\hbar\omega_{\textrm{vib}}}{2}\sum_{n=1}^{N} \left( \tilde{Q}_n^2 +\tilde{P}_n^2 \right).\\
\end{eqnarray}
$\hat{H}_{F}$ is the Frenkel Hamiltonian, defined in Eq.\ (\ref{Eq:1}), while
  $\tilde{Q} = (K_{\textrm{vib}}/\hbar \omega_{\textrm{vib}})^{1/2}Q$
and
  $\tilde{P} =(\omega_{\textrm{vib}}/\hbar K_{\textrm{vib}})^{1/2}P$
are the dimensionless displacement and momentum of the normal mode.
The second term on the right-hand-side of Eq.\ (\ref{Eq:2}) indicates that the normal mode couples linearly to the local exciton density\footnote{There is also a weaker and less significant coupling of the normal mode to the exciton bond-order operator\cite{Barford14b, Binder14}.}. $A$ is the dimensionless exciton-phonon coupling constant, which introduces the important polaronic parameter, namely the local Huang-Rhys factor
\begin{equation}\label{Eq:112}
  S = \frac{A^2}{2}.
\end{equation}
The final term is the sum of the elastic and kinetic energies of the harmonic oscillator, where
$\omega_{\textrm{vib}}$ and $K_{\textrm{vib}}$ are the angular frequency and force constant of the oscillator, respectively.
The Frenkel-Holstein model is another example of a coarse-grained Hamiltonian which, in addition to coarse-graining the exciton motion, assumes that the atomistic motion of the carbon nuclei can be replaced by appropriate local normal modes.

Exciton-nuclear dynamics is often modeled via the Ehrenfest approximation, which treats the nuclear coordinates as classical variables moving in a mean field determined by the exciton. However,  as described in Section \ref{se:2}, the Ehrenfest approximation fails to correctly describe ultrafast dynamical processes.
A correct description of the coupled exciton-nuclear dynamics therefore requires a full quantum mechanical treatment of the system. This is achieved by introducing the harmonic oscillator raising and lowering operators, $\hat{b}_n^{\dagger}$ and $\hat{b}_n$, for the normal modes i.e.,
$ \tilde{Q}_n \rightarrow \hat{\tilde{Q}}_n = (\hat{b}_n^{\dagger} +  \hat{b}_n)/\sqrt{2}$
and
 $ \tilde{P}_n \rightarrow \hat{\tilde{P}}_n = i(\hat{b}_n^{\dagger} -  \hat{b}_n)\sqrt{2}$.
The time evolution of the quantum system can then conveniently be simulated via the TEBD method, as briefly described in Section \ref{se:2}.

Since the photoexcited system has a different electronic bond order than the ground state, an instantaneous force is established on the nuclei.
As described in Section \ref{se:3.3}, this force creates an exciton-polaron, whose spatial size is quantified by the exciton-phonon correlation function\cite{Hoffmann02}
\begin{equation}\label{Eq:15}
C_n^{\textrm{ex-ph}}(t) \propto \sum_m \langle\hat{N}_{m} \hat{\tilde{Q}}_{m+n} \rangle.
\end{equation}
$C_n^{\textrm{ex-ph}}$ correlates the local phonon displacement, $Q$, with the instantaneous exciton density, $N$, $n$ monomers away.
$C_n^{\textrm{ex-ph}}(t)$, illustrated in Fig.\ \ref{Fi:40}, shows that the exciton-polaron is established within 10 fs (i.e., within half the period of a C-C bond vibration) of photoexcitation. The temporal oscillations, determined by the C-C bond vibrations, are damped as energy is dissipated into the vibrational degrees of freedom, which acts as a heat bath for the exciton. The exciton-phonon spatial correlations decay exponentially, extending over ca.\ 10 monomers. This short range correlation occurs because the C-C bond can respond relatively quickly to the exciton's motion. \footnote{In contrast, in the classical limit ($\omega \rightarrow 0$) the nuclei respond infinitesimally slowly to the exciton, so that the correlation length and the exciton-polaron mass diverge causing exciton-polaron self-localization.}

\begin{figure}
\includegraphics[width=9cm]{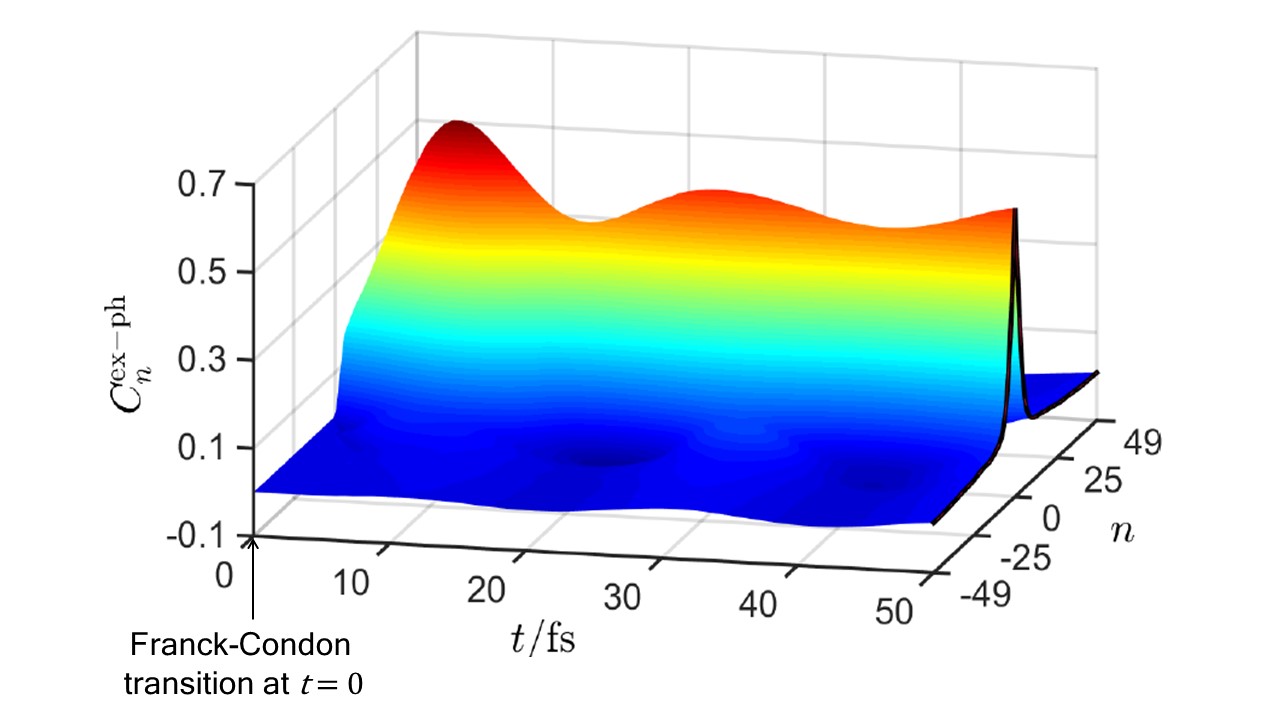}
\caption{The time-dependence of the exciton-phonon correlation function, Eq.\ (\ref{Eq:15}), after photoexcitation at time $t=0$. It fits the form $C_{n}^{\text{ex-ph}} = C_0\exp(-n/\xi)$ as $t \rightarrow \infty$, where $\xi \sim 10$. $n$ is a monomer index. The vibrational period is $20$ fs.
}
\label{Fi:40}
\end{figure}

\begin{figure}[h]
\centering
\includegraphics[width=0.45\textwidth]{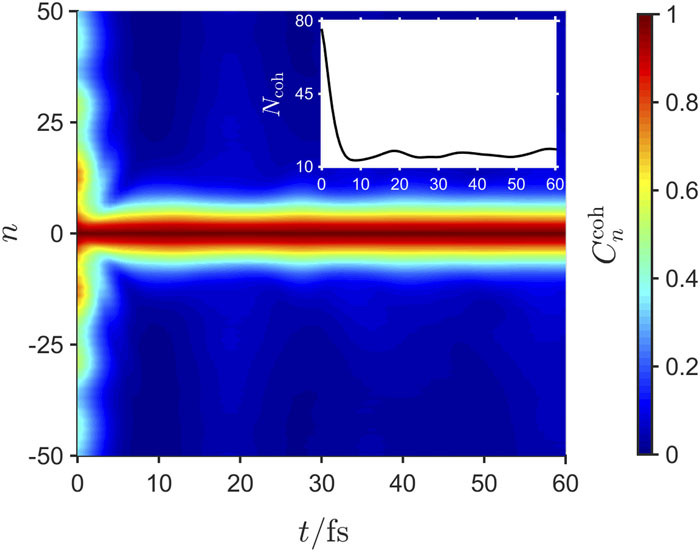}
\caption{The time dependence of the exciton coherence correlation function, $ C_n^{\textrm{coh}}$, Eq.\ (\ref{Eq:17}). The time dependence of the associated coherence number, $N^{\textrm{coh}}$ (Eq.\ (\ref{Eq:18b})), is shown in the inset.
 $N^{\textrm{coh}}$ decays within $10$ fs, i.e., within half a vibrational period. Reproduced  from J. Chem. Phys. \textbf{148}, 034901 (2018) with the permission of AIP publishing.}\label{Fi:5}
\end{figure}

The ultrafast establishment of quantum mechanically correlated exciton-phonon motion causes an ultrafast decay of off-diagonal-long-range-order (ODLRO) in the exciton site-basis density matrix. This is quantified via\cite{Kuhn97,Smyth12}
\begin{equation}\label{Eq:17}
C_n^{\textrm{coh}}(t) = \sum_m \left| \rho_{m,m+n} \right|,
\end{equation}
where $\rho_{m,m'}$ is the exciton reduced density matrix obtained by tracing over the vibrational degrees of freedom.
$C_n^{\textrm{coh}}(t)$ is displayed in Fig.\ \ref{Fi:5}, showing that ODLRO is lost within 10 fs.
The loss of ODLRO is further quantified by the coherence number, defined by
\begin{equation}\label{Eq:18b}
  N^{\textrm{coh}}  = \sum_n C_n^{\textrm{coh}},
\end{equation}
and shown in the inset of Fig.\ \ref{Fi:5}. Again, $N^{\textrm{coh}}$ decays to ca.\ 10 monomers in ca.\ 10 fs, reflecting the localization of exciton coherence resulting from the short range exciton-phonon correlations.
As discussed in Section \ref{se:4.4}, the loss of ODLRO leads to ultrafast fluorescence depolarization\cite{Mannouch18}.

We emphasise that the prediction of an electron-polaron with short range correlations is a consequence of treating the phonons quantum mechanically, while the decay of exciton-site coherences is a consequence of the exciton and phonons being quantum mechanically entangled. Neither of these predictions are possible within the Ehrenfest approximation.

\subsection{Role of system-environment interactions}\label{se:4.2}

For an exciton to dissipate energy it must  first couple to fast internal degrees of freedom (as described in the last section) and then these  degrees of freedom must couple to the environment to expell heat. For a low-energy exciton (i.e., a LEGS) this process will cause adiabatic relaxation on a single potential energy surface, forming a VRS\cite{Tretiak02,Bittner03,Sterpone08,Leener09}.
As shown in Fig.\ \ref{Fi:3}(a), however, for a kinetically hot exciton (i.e., a QEES)  this relaxation is through a dense manifold of states and is necessarily a nonadiabatic interconversion between different potential energy surfaces. As already stated in Section \ref{se:2}, the Ehrenfest approximation fails to correctly describe this process.\footnote{In fact, the Ehrenfest approximation is the cause of the unphysical bifurcation of the exciton density onto separate chromophores found in  Ehrenfest simulations of the relaxation dynamics of high energy photoexcited states\cite{Tozer12}.}

Dissipation of energy from an open quantum system arising from system-environment coupling is commonly described by a Lindblad master equation\cite{Breuer02}
\begin{equation}
\label{Eq:master}
\frac{\partial\hat{\rho}}{\partial {t}} = -\frac{i}{\hbar}\left[\hat{H},\hat{\rho}\right]-\frac{{\gamma}}{2}\sum_{n}\left(\hat{L}_{n}^{\dagger}\hat{L}_{n}\hat{\rho}+\hat{\rho}\hat{L}_{n}^{\dagger}\hat{L}_{n}-
2\hat{L}_{n}\hat{\rho}\hat{L}_{n}^{\dagger}\right),
\end{equation}
where $\hat{L}_{n}^{\dagger}$ and $\hat{L}_{n}$ are the Linblad operators, and  $\hat{\rho}$ is the system density operator.
In practice, a direct solution of the Lindblad master equation is usually prohibitively expensive, as the size of Liouville space scales as the square of the size of the associated Hilbert space.
Instead, Hilbert space scaling can be maintained by performing ensemble averages over quantum trajectories (evaluated via the TEBD method), where the action of the Linblad dissipator is modeled by quantum jumps.\cite{Daley14}

In this section we assume that the C-C bond vibrations couple directly with the environment\cite{Mannouch18,Bednarz02}, in which case  the Linblad operators are the associated raising and lowering operators (i.e., $\hat{L}_{n} \equiv \hat{b}_{n}$, introduced in the last section). In addition,
\begin{equation}
\label{Eq:ham_correc}
\hat{H}=\hat{H}_{\text{FH}}+\frac{{\gamma}\hbar}{4}\sum_{n}\left(\hat{\tilde{Q}}_{n}\hat{\tilde{P}}_{n}+\hat{\tilde{P}}_{n}\hat{\tilde{Q}}_{n}\right).
\end{equation}
(In Section \ref{se:5} we discuss coupling of the torsional modes with the environment\cite{Albu13}.)

The ultrafast localization of exciton ODLRO (or exciton-site decoherence) described in Section \ref{se:4.1} occurs via the coupling of the exciton to  internal degrees of freedom, namely the C-C bond vibrations. We showed in Section \ref{se:3.3} (see Fig.\ \ref{Fi:2}(a)) that this coupling does not cause exciton density localization. However, dissipation of energy to the environment causes an exciton in a higher energy QEES to relax onto a lower energy LEGS (i.e., onto a chromophore) and thus the exciton density becomes localized.

\begin{figure}[h]
\centering
\includegraphics[width=0.45\textwidth]{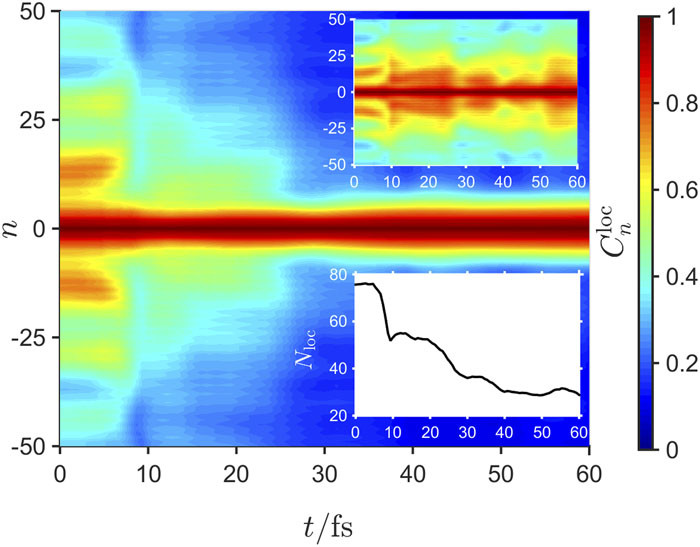}
  \caption{The time dependence of the exciton localization correlation function, $C^{\textrm{loc}}_n$ (Eq.\ (\ref{eq:17c})), for an initial high-energy QEES. The main figure corresponds to the time evolution with the dissipation time $T = \gamma^{-1} = 100$ fs. The time dependence of the  exciton density localization number, $N^{\textrm{loc}}$ (Eq.\ (\ref{eq:17d})),  is given in the lower inset.
  The upper inset  corresponds to the time evolution without external dissipation showing that in this case exciton denisty localization does not occur. Reproduced  from J. Chem. Phys. \textbf{148}, 034901 (2018) with the permission of AIP publishing.}\label{Fi:6}
\end{figure}

The spatial extent of the exciton density, averaged over an ensemble of quantum trajectories, is quantified by the correlation function\cite{Spano08a}, approximated by
\begin{equation}\label{eq:17c}
C_{n}^{\text{loc}}= \sum_m \left| \Psi_m \Psi_{m+n}^* \right|.
\end{equation}
Fig.\ \ref{Fi:6} shows the time dependence of  $C_{n}^{\text{loc}}$ with an external dissipation time $T = \gamma^{-1} = 100$ fs. The time scale for localization is seen from the time dependence of the exciton localization length\cite{Tempelaar14},
\begin{equation}\label{eq:17d}
N_{\text{loc}}={\sum_{n}\left|n\right|C_{n}^{\text{loc}}}/{\sum_{n}C_{n}^{\text{loc}}},
\end{equation}
which corresponds to the average distance between monomers for which the exciton wavefunction overlap remains non-zero, and is given in the lower inset of Fig.~\ref{Fi:6}.
Evidently, the coupling to the environment - and specifically, the damping rate - controls the timescale for energy relaxation and exciton density localization onto chromophores.
In contrast, the upper inset to Fig.\ \ref{Fi:6}  shows an absence of localization without external dissipation, indicating that exciton density localization is an extrinsic process.

Figure~\ref{Fi:6} is obtained by averaging over an ensemble of trajectories. To understand the physical process of localization onto a chromophore, Fig.\ 8 illustrates the exciton density of a single quantum trajectory for a photoexcited QEES (shown in Fig.\ 2(b)). At a time  ca.\ 20~fs  a `quantum jump' caused by the stochastic application of a Lindblad jump operator causes the exciton to  localize onto the $j=2$ LEGS, shown in Fig.\ 2(a), i.e., the high-energy extended state has randomly localized onto a chromophore because of a `measurement' by the environment.
\begin{figure}
\includegraphics[width=0.45\textwidth]{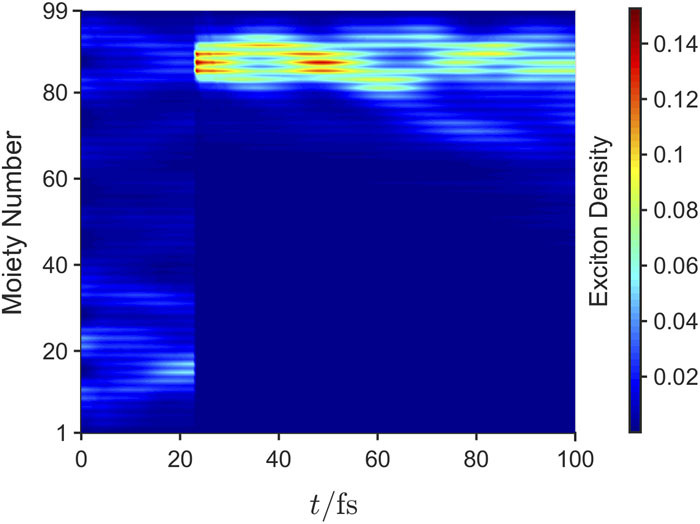}\label{Fi:17}
\caption{\label{fig:trajectory} The time dependence of the exciton density for a single trajectory of the quantum jump trajectory method.
The discontinuity in the density at ca.\ 20~fs is a `quantum jump' caused by the stochastic application of a Lindblad jump operator.
The dynamics were performed for an initial high energy QEES given in Fig.\ 2(b), showing localization onto the LEGSs (i.e., a chromophore) labeled $j=2$ in Fig.\ 2(a). Reproduced  from J. Chem. Phys. \textbf{148}, 034901 (2018) with the permission of AIP publishing.}
\end{figure}

\subsection{Role of slow bond rotations}\label{se:4.3}

By dissipating energy into the environment on sub-ps timescales, hot excitons relax into localized LEGSs, i.e., onto chromophores. The final intrachain relaxation and localization process now takes place, namely exciton-polaron formation via  coupling to the torsional degrees of freedom. For this relaxation to occur bond rotations must be allowed, which means that this process is highly dependent on the precise chemical structure of the polymer and its environment.

Assuming that bond rotations are not sterically hindered, their coupling to the excitons is conveniently modeled (via Eq.\ (\ref{Eq:5}) and Eq.\ (\ref{Eq:10c})) by supplementing the Frenkel-Holstein model (i.e., Eq.\ (\ref{Eq:2})) by\cite{Barford18}
\begin{eqnarray}\label{Eq:8}\nonumber
 \hat{H}_{\textrm{rot}} = - \sum_{n=1}^{N-1}
B(\theta_n^0)\times ({\phi}_{n+1} -  {\phi}_{n})\hat{T}_{n,n+1}
+ \frac{1}{2}  \sum_{n=1}^{N} \left( K_{\textrm{rot}}{\phi}_n^2 + {L}_n^2/I \right).\\
\end{eqnarray}
Here, ${\phi}$ is the  angular displacement of a monomer from its groundstate equilibrium value and
${L}$ is the associated angular momentum of a  monomer around its bridging bonds.

The first term on the right-hand-side of Eq.\ (\ref{Eq:8}) indicates that the change in the dihedral angle,
 $\Delta {\theta}_{n}= ( {\phi}_{n+1} -  {\phi}_{n})$,
couples linearly to the bond-order operator, $\hat{T}_{n,n+1}$,
where \begin{equation}\label{Eq:24}
B(\theta_n^0) = J_{SE}\sin 2 \theta_n^0
\end{equation}
is the  exciton-roton coupling constant and $\theta_n^0$ is the groundstate dihedral angle for the $n$th bridging bond.
The final term is the sum of the elastic and kinetic energies of the rotational harmonic oscillator.

The natural angular frequency of oscillation is $\omega_{\textrm{rot}} = (K_{\textrm{rot}}/I)^{1/2}$, where $K_{\textrm{rot}}$ is the elastic constant of the rotational oscillator and $I$ is the moment of inertia, respectively. As discussed in Section \ref{se:3.3.1}, $K_{\textrm{rot}}$ is larger for the bridging bond  in the excited state than the groundstate, because of the increase in bond order. Also notice that both the moment of inertia (and thus $\omega_{\textrm{rot}}$)  of a rotating monomer and its viscous damping from a solvent are strongly dependent on the side groups attached to it. As discussed in the next section, this observation has important implications for whether the motion is under or over damped and on its characterstic timescales.

Unlike C-C bond vibrations, being over 10 times slower torsional oscillations can be treated classically\cite{Barford18}. Furthermore, since we are now concerned with adiabatic relaxation on a single potential energy surface, we may employ the Ehrenfest approximation.
Thus, using Eq.\ (\ref{Eq:8}), the torque on each ring is
\begin{eqnarray}\label{Eq:25}
\nonumber
\Gamma_n &=& - \frac{\partial \langle \hat{H}_{rot} \rangle}{\partial {\phi}_n}\\
&=& -K_{\textrm{rot}}{\phi}_n +\lambda_n
\end{eqnarray}
where we define
\begin{equation}\label{Eq:14}
\lambda_n = B(\theta_{n-1}^0) \langle \hat{T}_{n-1,n} \rangle - B(\theta_{n}^0) \langle \hat{T}_{n,n+1} \rangle.
\end{equation}
Setting $\Gamma_n = 0$ gives the equilibrium angular displacements in the excited state as
${\phi}_n^{\textrm{eq}} = \lambda_n/K_{\textrm{rot}}$.
${\phi}_n$ is  subject to the Ehrenfest equations of motion,
\begin{equation}\label{Eq:18}
I \frac{d {\phi}_n}{d {t}} = {L}_n,
\end{equation}
and
\begin{eqnarray}\label{Eq:19}
\frac{d {L}_n}{d {t}} && =  \Gamma_n- \gamma L_n,
\end{eqnarray}
where the final term represents the damping of the rotational motion by the solvent.

\subsubsection{A single torsional oscillator}\label{se:4.3.1}

Before considering a chain of torsional oscillators, it is instructive to review the dynamics of a single, damped oscillator subject to both  restoring and  displacement forces. The equation of motion for the angular displacement is
\begin{equation}\label{}
  \frac{d^2\phi(t)}{dt^2} = -\omega_{\textrm{rot}}^2(\phi(t) - \phi_{\textrm{eq}}) -\gamma   \frac{d \phi(t)}{dt,}
\end{equation}
where $\phi_{\textrm{eq}} = \lambda/K_{rot}$ is proportional to the displacement force.

In the \emph{underdamped regime}\cite{French71}, defined by $\gamma < 2\omega_{\textrm{rot}}$,
\begin{equation}\label{}
 \phi(t) = \phi_{\textrm{eq}}\left(1-\cos(\omega t) \exp(-\gamma t/2)\right),
\end{equation}
where $\omega = (\omega_{\textrm{rot}}^2 - \gamma^2/4)^{1/2}$. In this regime, the torsional angle undergoes damped oscillations with a period $T = 2\pi/\omega$ and a decay time $\tau = 2/\gamma$.

Conversely,  in the \emph{overdamped regime}\cite{French71}, defined by $\gamma > 2\omega_{\textrm{rot}}$,
\begin{eqnarray}\label{}\nonumber
 \phi(t) = \phi_{\textrm{eq}}\left(1- \frac{1}{4\beta}\left(\gamma_1\exp(-\gamma_2 t/2) - \gamma_2\exp(-\gamma_1 t/2)\right)  \right),\\
\end{eqnarray}
where $\gamma_1 = \gamma + 2\beta$,  $\gamma_2 = \gamma - 2\beta$ and $\beta = (\gamma^2/4 - \omega_{\textrm{rot}}^2)^{1/2}$.
Now, the torsional angle undergoes damped biexponential decay with the decay times $\tau_1 = 2/\gamma_1$ and $\tau_2 = 2/\gamma_2$. In the limit of strong damping, i.e., $\gamma \gg 2\omega_{\textrm{rot}}$, there is a fast relaxation time $\tau_1 = 1/\gamma = \tau/2$ and a slow relaxation time $\tau_2 = \gamma/\omega_{\textrm{rot}}^2 \gg \tau$. In this limit, as the slow relaxation dominates at long times, the torsional angle approaches equilibrium with an effective mono-exponential decay.

For a polymer without alkyl side groups, e.g., PPP and PPV, $\omega_{\textrm{rot}} \sim  \gamma \sim 10^{13}$ s$^{-1}$ and are thus  in the underdamped regime with sub-ps relaxation. However, polymers with side groups, e.g., P3HT, MEH-PPV and PFO, have a  rotational frequency up to ten times smaller and a larger damping rate, and are thus in the overdamped regime\cite{Wells08}.

\subsubsection{A chain of torsional oscillators}\label{se:4.3.2}

An exciton delocalized along a polymer chain in a chromophore couples to multiple rotational oscillators resulting in collective oscillator dynamics.
Eq.\ (\ref{Eq:24}) and Eq.\ (\ref{Eq:14}) indicate that torsional relaxation only occurs if the monomers are in a staggered arrangement in their groundstate, i.e., $\theta_n^0 = (-1)^n\theta^0$. In this case the torque acts to planarize the chain. Furthermore, since the torsional motion is slow, the self-trapped exciton-polaron thus formed is `heavy' and in the under-damped regime becomes self-localized on a timescale of a single torsional period, i.e., $200 - 600$ fs. In this limit the relaxed staggered bond angle displacement mirrors the exciton density. Thus, the exciton is localized precisely as for a `classical' Landau polaron and is spread over $\sim 10$ monomers\cite{Barford18}.
\begin{figure*}
\centering
\includegraphics[width=0.7\textwidth]{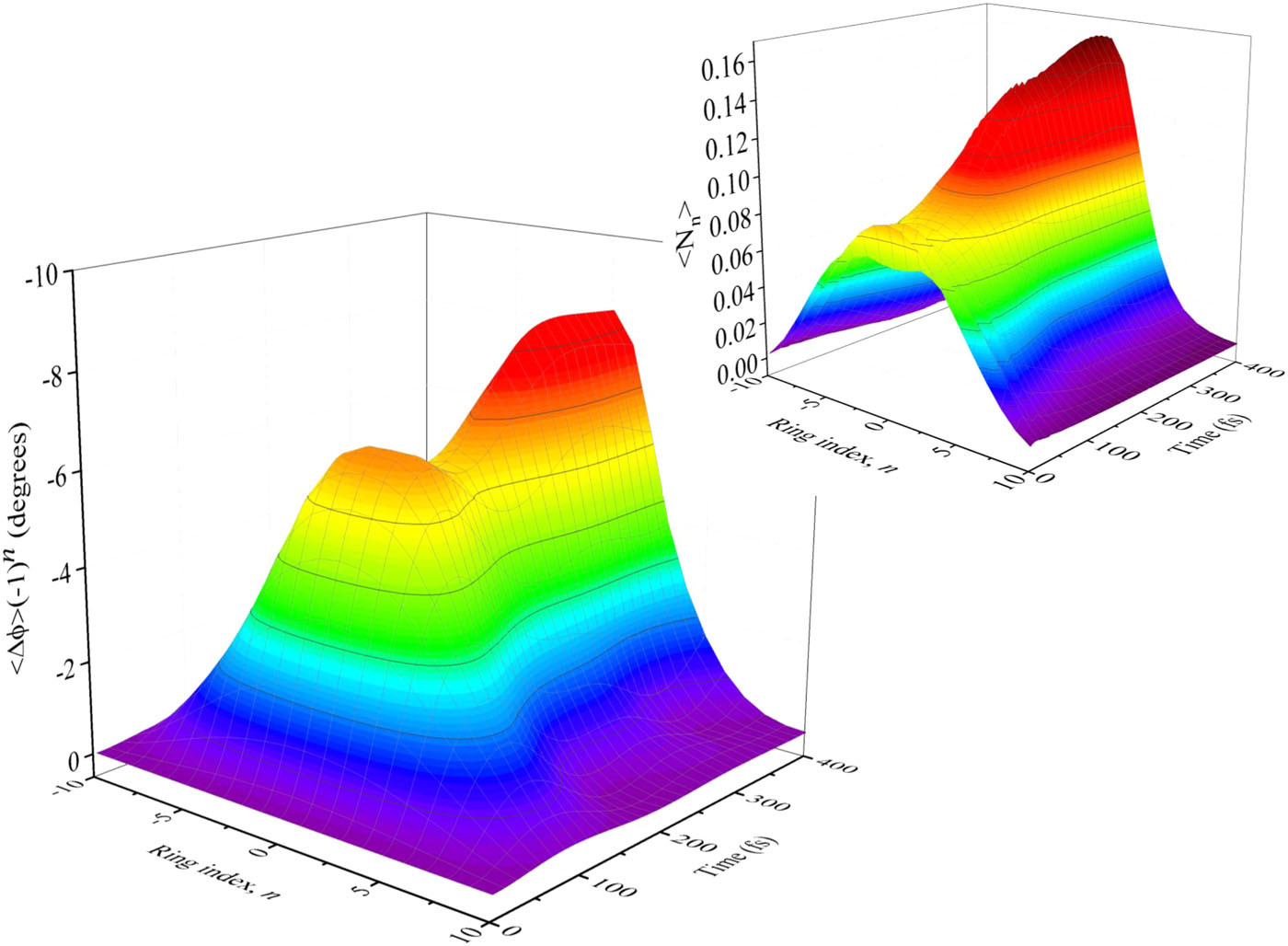}
\caption{The time-evolution of the staggered angular displacement, $\langle{\phi}_n\rangle\times(-1)^n$. The change of dihedral angle is $\Delta \theta_n = (\phi_{n+1}-\phi_n)$, showing local planarization for a PPP chain of 21 monomers.
The inset displays the time-evolution of the exciton density, $\langle N_n \rangle$,
showing exciton density localization after a single torsional period ($\sim 200$ fs). In the long-time limit (i.e., $t \gtrsim 400$ fs) $\langle{\phi}_n\rangle  \propto \langle N_n \rangle\times(-1)^n$, illustrating classical (Landau) polaron formation.
  Reproduced  from J. Chem. Phys. \textbf{149}, 214107 (2018) with the permission of AIP publishing.}\label{Fi:7}
\end{figure*}

The time-evolution of the staggered angular displacement, $\langle{\phi}_n\rangle\times(-1)^n$, is shown in Fig.\ \ref{Fi:7} illustrating that these displacements reach their equilibrated values  after two torsional periods (i.e., $t \gtrsim 400$ fs). The inset also displays the time-evolution of the exciton density, $\langle N_n \rangle$, showing exciton density localization after a single torsional period ($\sim 200$ fs).

So far we have described how exciton coupling to torsional modes causes a spatially varying planarization of the monomers that acts as a one-dimensional potential which self-localizes the exciton. The exciton `digs a hole for itself', forming an exciton-polaron\cite{Landau33}. Some researchers\cite{Westenhoff06}, however, argue that torsional relaxation causes an exciton to become more \emph{delocalized}.  A mechanism that can cause exciton delocalization occurs if the disorder-induced localization length is shorter than the intrinsic exciton-polaron size.
Then, in this case for freely rotating monomers, the stiffer elastic potential in the excited state causes a decrease both in the  variance of the dihedral angular distribution, $\sigma_{\theta}^2 = k_BT/K_{\textrm{rot}}$, and the mean dihedral angle, $\theta_0$.
This, in turn, means that the exciton band width, $|4J|$, increases and the diagonal disorder\cite{Barford14b}, $\sigma_J = J_{SE} \sigma_{\theta}\sin 2\theta_0$,
decreases. Hence, the disorder-induced localization,  $L_{\textrm{loc}} \sim (|J|/\sigma_J)^{2/3}$, increases (see Section \ref{se:3.2}).

\subsection{Summary}

The conclusions that we draw from the previous three sections are that a band edge excitation (i.e., a LEGS, which is an exciton spanning a single chromophore) undergoes ultrafast exciton site decoherence via its coupling to fast C-C bond stretches. It subsequently couples to slow torsional modes causing planarization and exciton density localization on the chromophore. A hot exciton (i.e., a QEES) also  undergoes ultrafast exciton site decoherence. However, exciton density localization within a chromophore only occurs after localization onto the chromophore via a stochastic interaction with the environment.


\subsection{Time resolved fluorescence anisotropy}\label{se:4.4}

For general polymer conformations, the loss of ODLRO (or the localization of the exciton coherence function) causes a reduction and  rotation of the transition dipole moment.
The rotation is quantified by the fluorescence anisotropy, defined by\cite{Lakowicz06}
\begin{equation}
\label{eq:r}
r = \frac{I_{\parallel}-I_{\perp}}{I_{\parallel}+2I_{\perp}},
\end{equation}
where $I_{\parallel}$ and $I_{\perp}$ are the intensities of the fluorescence radiation polarised parallel and perpendicular to the incident radiation, respectively.

For an arbitrary state of a quantum system, $|\Psi\rangle$, the integrated fluorescence intensity polarised along the $x$-axis is related to the $x$ component of the transition dipole operator, $\hat{\mu}_{x}$, by
\begin{eqnarray}\label{Eq:20}
\label{t}
I_{x} \propto  \sum_{v}\left|\langle\Psi|\hat{\mu}_{x}|\textrm{GS},v\rangle\right|^{2},
\end{eqnarray}
where $|\textrm{GS},v\rangle$ corresponds to the system in the ground electronic state, with the nuclear degrees of freedom in the state characterised by the quantum number $v$.

\begin{figure}[h]
\centering
  \includegraphics[width=0.40\textwidth]{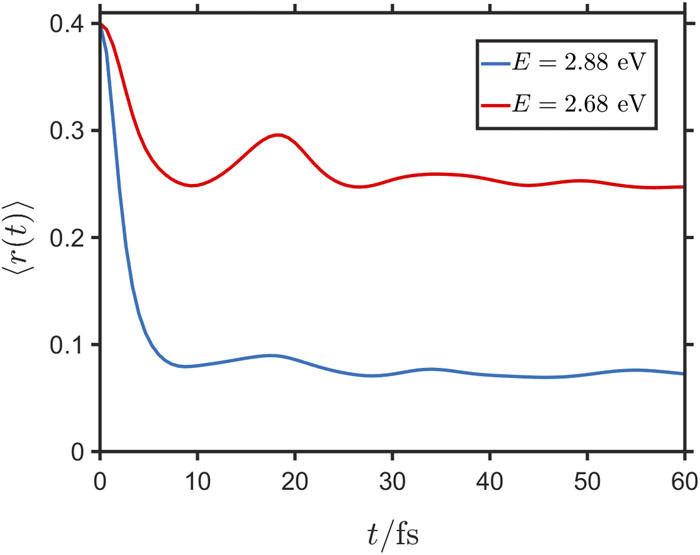}
  \caption{The time dependence of the fluorescence anisotropy, $\langle r\left(t\right)\rangle$, for two initial Frenkel excitons coupled to C-C bond stretches. The red curve corresponds to an initial LEGS, while the blue curve corresponds to a QEES. Reproduced  from J.\ Chem.\ Phys.\ \textbf{148}, 034901 (2018) with the permission of AIP publishing.}
  \label{Fi:18}
\end{figure}

The averaged fluorescence anisotropy is defined by
\begin{equation}
\label{eq:intens_av}
\langle r\left(t\right)\rangle = 0.4\times\frac{\sum_{i}I_{i}\left(t\right)r_{i}\left(t\right)}{\sum_{i}I_{i}\left(t\right)},
\end{equation}
where $I_{i}\left(t\right)$ is the total fluorescence intensity and $r_{i}\left(t\right)$ is the fluorescence anisotropy, associated with a particular conformation $i$ at time $t$. The factor of 0.4 is included on the assumption that the polymers are oriented uniformly in the bulk material.\cite{Lakowicz06} Fig.\ \ref{Fi:18} shows the simulated $\langle r\left(t\right)\rangle$ for both a high energy QEES and a low energy LEGS for an ensemble of conformationally disordered polymers.

\begin{figure}[h]
\centering
\includegraphics[width=0.55\textwidth]{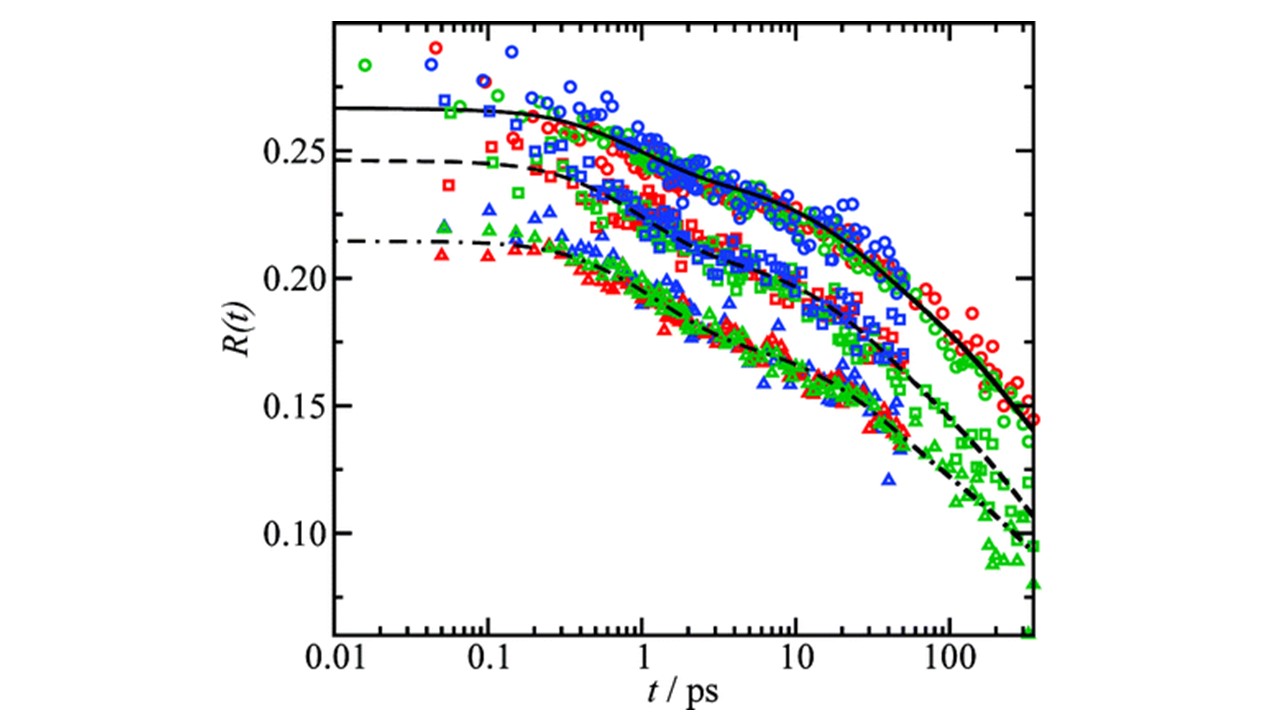}
\caption{The  experimental time dependence of the fluorescence anisotropy, $\langle R\left(t\right)\rangle$, in polythiophene in solution.
$\langle R\left(t\right)\rangle$ has decayed from $0.4$ to $\sim 0.25$ within 10 fs, consistent with the theoretical predictions shown in Fig.\ \ref{Fi:18}.
Subsequent fluorescence depolarization is caused by slower torsional relaxation on timescales of $1-10$ ps followed by possible conformational changes\cite{Wells07}.
Reproduced from J.\ Phys.\ Chem.\ C \textbf{111}, 15404 (2007) with the permission of ACS publishing.}\label{Fi:19}
\end{figure}

It is instructive to express Eq.\ (\ref{Eq:20}) as
\begin{eqnarray}\label{Eq:21}
I_{x} \propto  \sum_{m,n} s_m^x s_n^x \rho_{mn},
\end{eqnarray}
where $s^x_m$ is the $x$-component of the unit vector for the $m$th monomer and $\rho_{mn}$ is the exciton reduced density matrix.
Then, using Eq.\ (\ref{Eq:17}), Eq.\ (\ref{Eq:18b}), and Eq.\ (\ref{Eq:21}),  we observe  that the emission intensity, $I_{x}$, is related to the coherence length, $N^{\textrm{coh}}$. Thus, not surprisingly, the dynamics of $\langle r\left(t\right)\rangle$ resembles that of $N^{\textrm{coh}}(t)$ shown in Fig.\ 6. In particular, we observe a loss of fluorescence anisotropy within 10 fs, mirroring the reduction of $N^{\textrm{coh}}$ in the same time.
Furthermore, since there is greater exciton coherence localization for the QEES than for the LEGS, the former exhibits a greater loss of anisotropy.
This predicted loss of fluorescence anisotropy within 10 fs has been observed experimentally, as shown in 
Fig.\ \ref{Fi:19}.
Slower sub-ps decay of anisotropy occurs because of exciton density localization via coupling to torsional modes.\footnote{I.\  Gonzalvez Perez and W.\ Barford, in preparation.}


\section{Intrachain Exciton Motion}\label{se:5}

The last section described the relaxation and localization of higher energy excited states onto chromophores, and the subsequent torsional relaxation and localization on the chromophore. We now consider the relaxation and dynamics of these relaxed excitons caused by the stochastic torsional fluctuations experienced by a polymer in a solvent.

Environmentally-induced intrachain exciton relaxation in poly(phenylene ethynylene) was modeled by Albu and Yaron\cite{Albu13} using the Frenkel exciton model supplemented by the torsional degrees of freedom, i.e., $\hat{H} = \hat{H}_F + \hat{H}_{rot}$ (given by Eq.\ (\ref{Eq:1}) and Eq.\ (\ref{Eq:8}), respectively). Fast vibrational modes were neglected because although they cause self-trapping, they do not cause self-localization, and these modes can be assumed to respond instantaneously to the torsional modes. The polymer-solvent interactions were modeled by the Langevin equation. For chains longer than the exciton localization length the excited-state relaxation showed biexponential behavior with a shorter relaxation time of a few ps and a longer relaxation time of tens of ps.

After photoexcitation of the $n=2$ (charge-transfer) exciton in  oligofluorenes, Clark \emph{et al.}\cite{Clark12} reported torsional relaxation on sub-100 fs timescales. Since this timescale is faster than the natural rotational period of an undamped monomer, they ascribed it to the electronic energy being rapidly converted to kinetic energy via nonadiabatic transitions. They argue that this is analogous to inertial solvent reorganization.

Tozer and Barford\cite{Tozer15} using the same model as Albu and Yaron to model intrachain exciton motion in PPP where the  exciton dynamics were simulated on the assumption that at time $t+\delta t$ the new exciton target state is the eigenstate of $\hat{H}(t+\delta t)$ with the largest overlap with the previous target state at time $t$.\footnote{This latter assumption was shown by Lee and Willard\cite{Lee19} to be problematic for the non-adiabatic transport described in Section \ref{se:5.4}.}

A  more sophisticated simulation of exciton motion in poly(p-phenylene vinylene)  and oligothiophenes chains was performed by Burghardt and coworkers\cite{Binder20a, Binder20b, Binder20c, Hegger20} where high-frequency C-C bond stretches were also included, the solvent was modeled by a set of harmonic oscillators with an Ohmic spectral density, and the system was evolved via the multilayer-MCTDH  method. Their results, however, are in quantitative agreement with those of Tozer and Barford in the `low-temperature' limit (discussed in Section \ref{se:5.3}), namely activationless, linearly temperature-dependent exciton diffusion with exciton diffusion coefficients larger, but close to experimental values.

The Brownian forces excerted by the solvent on the polymer monomers have two consequences. First, as already noted in Section \ref{se:3.2}, the instantaneous spatial dihedral angle fluctuations  Anderson localize the Frenkel center-of-mass wavefunction.  Second, the temporal dihedral angle fluctuations cause the exciton to migrate via  two distinct transport processes.\footnote{This process is sometimes referred to as Environment-Assisted Quantum Transport\cite{Rebentrost09}.}

At low temperatures there is small-displacement adiabatic motion of the exciton-polaron as a whole along the polymer chain, which we will characterize as a `crawling' motion. At higher temperatures the torsional modes fluctuate enough to cause the exciton to be thermally excited out of the self-localized polaron state into a more delocalized LEGS or quasi-band QEES. While in this more delocalized state, the exciton momentarily exhibits quasi-band ballistic transport, before the wavefunction `collapses' into an exciton-polaron in a different region of the polymer chain. We  will characterize this large-scale displacement as a non-adiabatic `skipping' motion.

Before describing the details of these types of motion, we first describe a model of solvent dynamics and consider again exciton-polaron formation in a polymer subject to Brownian fluctuations.

\subsection{Solvent dynamics}\label{se:5.1}

If the solvent molecules are subject to spatially and temporally uncorrelated Brownian fluctuations, then the monomer rotational dynamics are controlled by the Langevin equation
\begin{equation} \label{Langevin}
\frac{d L_{n}(t)}{dt} =  \Gamma_{n}(t) + R_{n}(t) -\gamma L_{n}(t),
\end{equation}
where  $\Gamma_{n}(t)$ is the systematic torque given by Eq.\ (\ref{Eq:25}).
$R_{n}(t)$ is the stochastic torque on the monomer due to the random fluctuations in the solvent and $\gamma$ is the friction coefficient for the specific solvent.
From the fluctuation-dissipation theorem,  the distribution of random torques is given by
\begin{equation}\label{fluc_diss}
\langle R_{m}(t)R_{n}(0) \rangle = 2I\gamma k_{B}T\delta_{mn}\delta(t),
\end{equation}
which are typically sampled from a  Gaussian distribution with a standard deviation of
$\sigma_{R} = (2 I\gamma k_{B}T)^{\frac{1}{2}}$.
As a consequence of these Brownian fluctuations the monomer rotations are characterized by the autocorrelation function\cite{Nitzan06}
\begin{eqnarray}\nonumber
 && \langle\delta\phi(t)\delta\phi(0)\rangle =\\
&& \langle\delta\phi^{2}\rangle \left( \cos(\omega_{rot} t)+ \left(\frac{\gamma}{2\omega_{rot}}\right)\sin(\omega_{rot} t)\right) \exp(-\gamma t/2),
\end{eqnarray}
where
$\langle\delta\phi^{2}\rangle = {k_{B}T}/{K_{rot}}$,
$K_{rot}$ is the stiffness and $\omega_{rot} = \sqrt{K_{rot}/I}$ is the angular frequency of the torsional mode.

\subsection{Polaron formation}\label{se:5.2}

As we saw in Section \ref{se:4.3}, at zero temperature torsional modes couple to the exciton, forming an exciton-polaron. At finite temperatures, however,  a combination of factors  affect the localization of the exciton. First,  the exciton will still attempt to form a polaron. However, the thermally induced fluctuations in the torsional angles will affect the size of this exciton-polaron, as there is a non-negligible probability that the exciton will be excited out of its polaron potential well into a more delocalized state at high enough temperatures.
Second,  the exciton states will be Anderson localized by the instantaneous torsional disorder.

\begin{figure}
\centering
\includegraphics[width=0.4\textwidth]{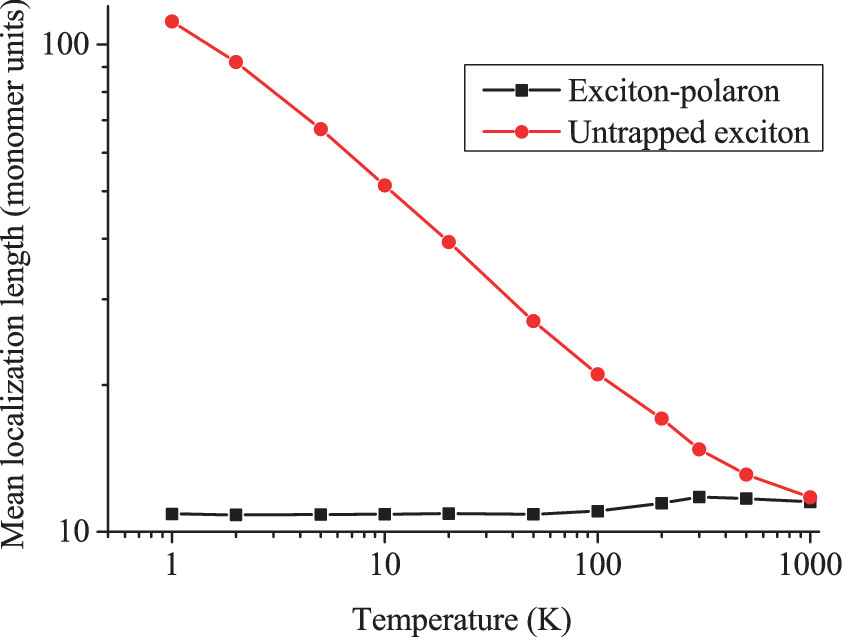}
\caption{The exciton localization length as a function of temperature for the `free' (i.e., `untrapped') exciton (red circles) and exciton-polaron (i.e., `self-trapped')  (black squares). The untrapped exciton localization length obeys $L_{\textrm{loc}}^{\textrm{free}} \propto  T^{-1/3}$. The lengths coincide when $k_BT \sim $ the exciton-polaron binding energy.
Reproduced  from J. Chem. Phys. \textbf{143}, 084102 (2015) with the permission of AIP publishing.}\label{Fi:8}
\end{figure}

Fig.\ \ref{Fi:8} shows how the average localization length varies with temperature both with and without coupling between the exciton and the torsional modes  (i.e., `self-trapped' and `free' exciton, respectively). As described in Section \ref{se:3.2}, the localization length for the `free' exciton is determined by Anderson localization. For small angular displacements from equilibrium a Gaussian distribution of dihedral angles implies a Gaussian distribution of exciton transfer integrals. Then, as confirmed by the simulation results shown in Fig.\ \ref{Fi:8}, from single-parameter scaling theory,
 $L_{\textrm{loc}}^{\textrm{free}} \propto \sigma_{\theta}^{-2/3} = \langle\delta\theta^{2}\rangle^{-1/3} \propto T^{-1/3}$.

In contrast, the localization length of the `self-trapped' exciton slowly increases with temperature because of the thermal excitation of the exciton from the self-localized polaron to a more delocalized  LEGS or QEES. The two values coincide when $k_B T$ equals the exciton-polaron binding energy (i.e., $T \sim 1500$ K in PPP).

\subsection{Adiabatic `crawling' motion}\label{se:5.3}

\begin{figure}
\centering
\includegraphics[width=0.50\textwidth]{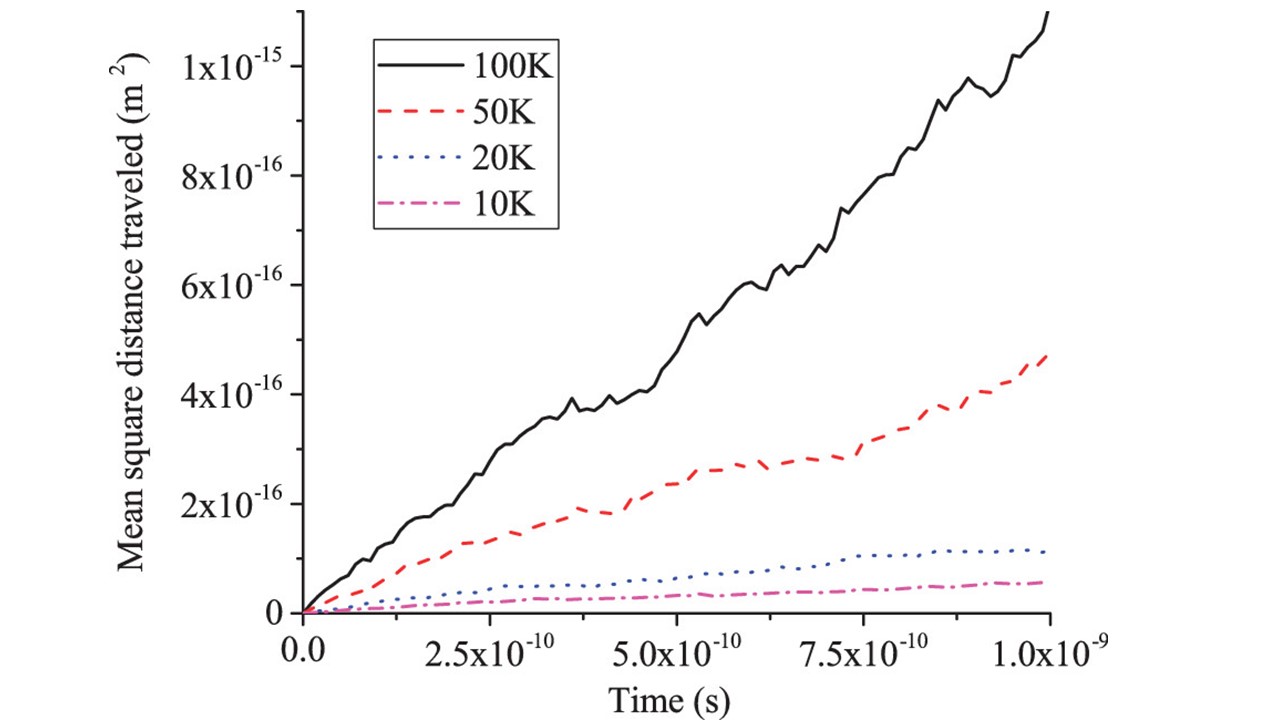}
\caption{The intrachain mean-square-distance traveled by an exciton-polaron at low temperatures caused by stochastic torsional fluctuations. The motion is diffusive, as shown by the mean-square-distance increasing linearly with time: $\langle L^{2} \rangle = 2D_A(T) t$. The gradients satisfy the  Einstein-Smoluchowski equation, $D_A(T) = \mu k_{B}T$.
Reproduced  from J. Chem. Phys. \textbf{143}, 084102 (2015) with the permission of AIP publishing.}\label{Fi:16}
\end{figure}

At low temperatures ($\lesssim 100$ K) the exciton has only a small amount of thermal energy, and not  enough to regularly break free from its polaronic torsional distortions. Thus, the exciton-polaron  migrates quasi-adiabatically and diffusively as a single unit. This is a collective motion of the exciton and the torsional degrees of freedom, as the torsional planarization accompanies the exciton. The random walk motion is illustrated in Fig.\ \ref{Fi:16}, which shows that the mean-square-distance traveled by the exciton-polaron is proportional to time, i.e.,  $\langle L^{2} \rangle = 2D_A(T) t$, where $D_A$ is the diffusion coefficient.
Since the migration of the exciton-polaron  is  an activationless process, as the gradients of Fig.\ \ref{Fi:16} indicate, at low temperatures it obeys the Einstein-Smoluchowski equation,
$D_A(T) = \mu k_{B}T$,
where $\mu$ is the mobility of the particle.

The time taken for an exciton to diffuse a distance $L$ along the chain is determined by the equation for a one-dimensional random walk, i.e.,
$\tau_D = \langle L^2 \rangle/2D$.
As shown in  Fig.\ \ref{Fi:8}, the typical exciton-polaron localization length is $\sim 12$ monomers or $\sim 6$ nm in PPP.
This characteristic length scale implies a characteristic timescale, namely the time taken for the exciton to diffuse along a chromophore length.
As shown in Table \ref{Ta:2}, these timescales are typically $3-30$ ps at room temperature depending on the solvent friction coefficient, being shorter at higher temperatures and smaller damping rates.   As we show in Section \ref{se:5}, these timescales are an order of magnitude shorter than F\"orster transfer times in the condensed phase.

As the exciton-polaron migrates along the polymer chain  it experiences a different potential energy landscape, so its energy will also fluctuate on a timescale $\sim \tau_D$.
Interestingly, these timescales are consistent with the longer timescale found experimentally in biexponential fits of relaxation processes of polymers in solution (see Table I) and correspond to the longer timescale simulated by Albu and Yaron\cite{Albu13} in longer polymers.

\begin{table}
\centering
{\renewcommand{\arraystretch}{1.2}
\begin{tabular}{|M{1.5cm}|M{1.5cm}|M{2.5cm}|M{1.5cm}|}
\hline
$\gamma$ (s$^{-1}$) & $T$ (K) & $D_A$ ($\textrm{cm}^2 \textrm{ s}^{-1}$) & $\tau_D$ (ps) \\
\hline
$10^{11}$ & 300 & $6.0\times 10^{-2}$    & 3.0  \\
$10^{11}$ & 100 & $2.0\times 10^{-2}$  & 9.0  \\
\hline
$10^{12}$ & 300 & $2.7\times10^{-2}$ & 6.7  \\
$10^{12}$ & 100 & $ 9.0\times10^{-3}$ & 20  \\
\hline
$10^{13}$ & 300 & $6.0\times10^{-3}$ & 30  \\
$10^{13}$ & 100 & $2.0\times10^{-3}$ & 90  \\
\hline
\end{tabular}}
\caption{Calculated intrachain adiabatic exciton diffusion coefficients, $D_A$, and times, $\tau_D$, in PPP from ref\cite{Tozer15}.
$\tau_D$ is the time taken for an exciton to diffuse along a chromophore of linear size $6$ nm in a solvent at temperature, $T$, with a  damping rate $\gamma$.
From simulations\cite{Tozer15}, $\tau_D  \sim \gamma^{1/2}/T$.
Hegger \emph{et al.}\cite{Hegger20} obtained $D_A \sim 10^{-2}$ $\textrm{cm}^2 \textrm{ s}^{-1}$  in oligothiophenes at 300 K  and $\gamma = 5\times10^{12}$ s$^{-1}$.}
\label{Ta:2}
\end{table}


\subsection{Nonadiabatic `skipping' motion}\label{se:5.4}

At higher temperatures,  adiabatic `crawling' migration of the exciton-polaron, as described above, still occurs. However, a second non-adiabatic mechanism for the dynamics   plays an important role. This mechanism involves the exciton-polaron being excited to a high enough energy by the thermal fluctuations to be excited out of the polaron potential well, resulting in a breakdown of the polaron and the exciton to enter an untrapped local exciton ground state (LEGS), or a higher energy quasi-extended exciton state (QEES). Once in this more delocalized state the exciton has quasi-band characteristics and travels quasi-ballistically.

As described in Section \ref{se:4}, however, on a sub-ps timescale the hot exciton will shed some of its excess kinetic energy and relax back into an exciton-polaron. As a result, the time-averaged exciton localization length calculations of Fig.\ \ref{Fi:8} show only a slight increase in localization length with increasing temperatures, as the majority of its lifetime is still spent in self-localized exciton-polarons.

The requirement that the exciton is excited out of the polaron potential well means that this process is  activated. Thus, from a simple Fermi golden rule analysis it can be shown that\cite{Barford14c}
\begin{equation}\label{High_T_D}
  D_{NA}(T) \sim T^{2/3} \exp\Bigg{(}-\frac{\Delta E}{k_{B}T}\Bigg{)},
\end{equation}
where $\Delta E$ is the exciton-polaron binding energy. At 300 K $D_{NA}$ is approximately twice as large as $D_{A}$ and thus the overall diffusion coefficient is considerably enhanced by this skipping motion.

The role of exciton transport in disordered one-dimensional systems via higher-energy quasi-band states has been discussed in ref\cite{Bednarz02}, where in that work phonons in the condensed phase environment induced non-adiabatic transitions.

\section{Interchain Exciton Motion}\label{se:6}

The stochastic, torsionally-induced intrachain exciton diffusion in polymers in solution described in the last section is not expected to be the primary cause of exciton diffusion in polymers in the condensed phase. Instead,  owing to restricted monomer rotations and the proximity of neighboring chains, exciton transfer is determined by Coulomb-induced, F\"orster-like processes. Moreover, since
dissipation rates are typically\cite{Sterpone08,Leener09} $10^{12} - 10^{13}$ s$^{-1}$, whereas exciton transfer rates are typically $10^{9} - 10^{11}$ s$^{-1}$,  exciton migration is an incoherent or diffusive process\cite{Scholes11a}.

Early models of condensed phase diffusion assumed that the donors and acceptors are point-dipoles whose energy distribution is a Gaussian random variable\cite{Movaghar86a, Movaghar86b, Meskers01}. An advantage of these models is that they allow for analytical analysis, for example predicting how the diffusion length varies with disorder and temperature\cite{Athanasopoulos19}.
They also reproduce some experimental features, such as the time-dependence of spectral diffusion. A disadvantage, however, is that there is no quantitative link between the model and actual polymer conformations and morphology.

More recent approaches have attempted to make the link between random polymer conformations and the energetic and spatial distributions of the donors and acceptors via the concept of extended chromophores\cite{Beljonne02,Beljonne05,Athanasopoulos08,Singh09} and using transition densities  to compute transfer integrals. However, the usual practice has been to arbitrarily define chromophores via a minimum threshold in the $p$-orbital overlaps, and then obtain a distribution of energies by assuming that the excitons delocalize freely on the chromophores thus defined as a `particle-in-a-box'.

As discussed in Section \ref{se:3.2}, an unambiguous link between polymer conformations and chromophores may be made by defining  chromophores via the spatial extent of local exciton ground states (LEGSs). Using this insight,  a more realistic first-principles model that accounts for polymer conformations can be developed\cite{Barford12b,Barford14c}. This is described in Section \ref{se:6.1}, while its predictions and comparisons to experimental observations are described in Section \ref{se:6.2}.

\subsection{Modified F\"orster theory}\label{se:6.1}

The F\"orster exciton transfer rate from a donor (D) to an acceptor (A) has the general Golden rule form
\begin{equation}\label{Eq:34}
    k_{DA} = \left(\frac{2\pi}{\hbar}\right)\left|J_{DA}\right|^2 \int D(E)A(E)dE,
\end{equation}
where $J_{DA}$ is the Coulomb-induced donor-acceptor transfer integral defined by Eq.\ (\ref{Eq:39}).
As we remarked in Section \ref{se:3.1}, the transition density vanishes for odd-parity singlet excitons; it also vanishes for all triplet excitons.

$D(E)$ and $A(E)$ are the donor and acceptor spectral functions, respectively,
defined by
\begin{equation}\label{}
    D(E) = \sum_{v}F_{0v}^D\delta(E+{E}^{D}_{0v})
\end{equation}
and
\begin{equation}\label{}
    A(E) = \sum_{v}F_{0v}^A\delta(E-{E}^{A}_{0v}),
\end{equation}
where $F_{0v}$ is the effective Franck-Condon factor, defined in Eq.\ (\ref{Eq:43}).
$E^A_{0v} = (E^A_{00} + v\hbar\omega_{vib})$ is the excitation energy of the acceptor,
while  $E^D_{0v} = -(E^D_{00} - v\hbar\omega_{vib})$ is the de-excitation energy of the donor.

The link between actual polymer conformations and a realistic model of exciton diffusion is made by realising that the donors and acceptors for exciton transfer are LEGSs (i.e., chromophores).
This assumption is based on the observation that  exciton transfer occurs at a much slower rate than state interconversion, so the donors are LEGSs, while the spectral overlap between LEGSs and higher energy QEESs is small, so the acceptors are also LEGSs. Moreover, the energetic and spatial distribution of LEGSs is entirely determined by the conformational and site disorder, as described in Section \ref{se:3.2}. Finally, polaronic effects are incorporated by an effective Huang-Rhys factor for each chromophore and the Condon approximation may be assumed as  C-C vibrational modes do not cause exciton self-localization.

Then, as  proved rigorously in ref\cite{Barford14c}:
\begin{enumerate}
\item{$J_{DA}$ is evaluated by invoking the Condon approximation and using the line-dipole approximation\cite{Book, Barford07}
\begin{equation}\label{Eq:42}
    J_{DA} = \left(\frac{1}{4\pi\varepsilon_{r}\varepsilon_{0}}\right)
  \sum_{\substack{n\in D \\ n'\in A}}\frac{\kappa_{nn'}}{R_{nn'}^{3}}\mu_{D}\Psi_D(n)\mu_{A}\Psi_A(n'),
\end{equation}
where $\Psi(n)$ is the LEGS center-of-mass wavefunction on monomer $n$ determined from the disordered Frenkel exciton model (Eq.\ (\ref{Eq:1})).
Since the spatial extent of $\Psi(n)$ defines a chromophore, the sum over $n$ and $n'$ is implicitly over monomers of a donor and acceptor chromophore, respectively.
$\mu_{X}$ is the transition dipole moment of a single monomer of the donor ($X=D$) or acceptor ($X=A$) chromophores (so $\mu_{X}\Psi_X(n)$ is the transition dipole moment of  monomer $n$ as part of the chromophore).
$\kappa_{nn'}$ is the orientational factor, defined in Eq.\ (\ref{Eq:40}), and $R_{nn'}$ is the separation of monomers on the  donor and acceptor chromophores. The line-dipole approximation is valid when the monomer sizes are much smaller than their separation on the donor and acceptor chromophores; it becomes the point-dipole approximation when the chromophore sizes are much smaller than their separation.}
\item{The spectral functions describe `polaronic' effects, by containing effective Franck-Condon factors which describe the  chromophores coupling to effective modes with reduced Huang-Rhys parameters:
\begin{equation}\label{Eq:43}
    F_{0v} = \frac{S_{\textrm{eff}}^v \exp(-S_{\textrm{eff}})}{v!},
\end{equation}
where
    $S_{\textrm{eff}} = S/\textrm{PN}$,
$S$ is the local Huang-Rhys parameter (defined by Eq.\ (\ref{Eq:112})) and
   $ \textrm{PN} = \left(\sum_n |\Psi_n|^4\right)^{-1}$
is the participation number (or size) of the chromophore\cite{Barford14a}.
}
\item{Similarly, the $0-0$ transition energy is defined by
   $ E_{00} = (E^{\textrm{vert}}- E^{\textrm{relax}})$,
where
$E^{\textrm{vert}}$ is determined from the Frenkel exciton model and
   $ E^{\textrm{relax}} = \hbar\omega S_{\textrm{eff}}$
is the effective reorganisation energy for the effective mode.}
\end{enumerate}

\subsection{Condensed-phase exciton diffusion}\label{se:6.2}

We might attempt to anticipate the results of the simulation of exciton diffusion  from the properties of the exciton transfer rate, $k_{DA}$. When the chromophore size, $L$, is much smaller than the donor-acceptor separation, $R$, the point-dipole approximation is valid. In this limit $k_{DA} \sim L^2/R^6$ and thus  the hopping rate \emph{increases} with increasing chromophore size. Conversely, when the chromophore size is much larger than the donor-acceptor separation, the line-dipole approximation predicts that for straight, parallel or collinear chromophores\cite{Wong04,Das10,Barford10c} $k_{DA} \sim 1/(LR)^2$ and thus  the hopping rate \emph{decreases} with increasing chromophore size.

\begin{figure}
\centering
\includegraphics[width=0.4\textwidth]{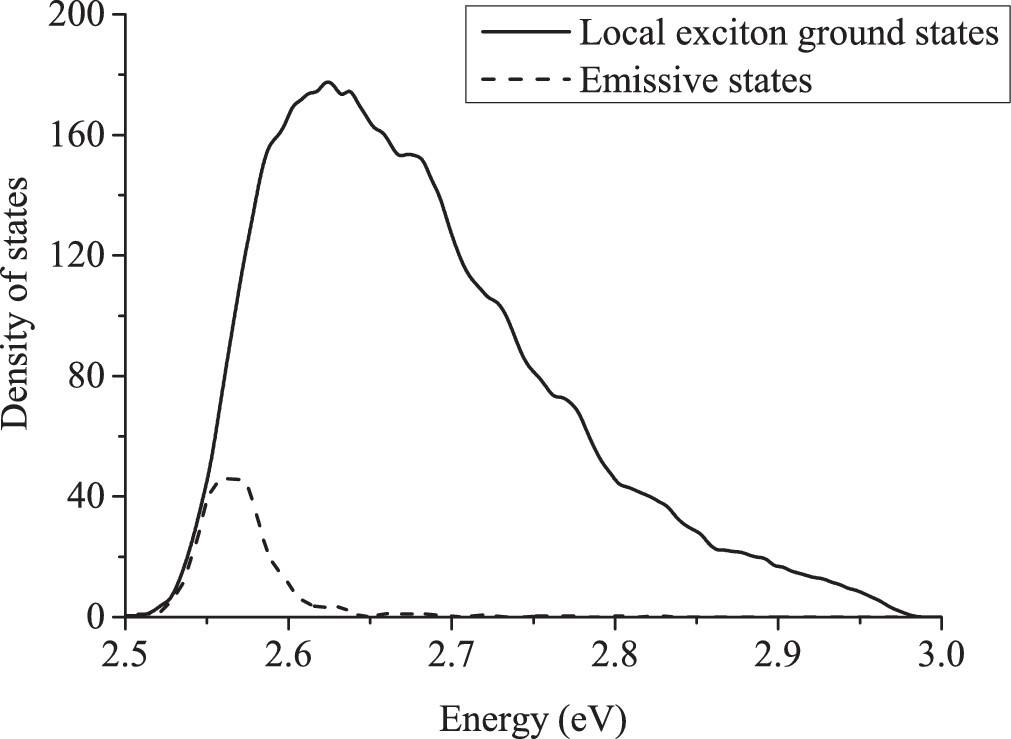}
\caption{The density of states for absorbing LEGSs (solid line) and emitting trap states (dashed line) for an ensemble of  PPV chains, with a Gaussian distribution of dihedral angles.
$\langle \theta_0 \rangle = 10^0$, $\sigma_{\theta_0} = 5^0$, and $\sigma_{\theta} = 5^0$.
Reproduced  from J. Chem. Phys. \textbf{141}, 164103  (2014) with the permission of AIP publishing.}\label{Fi:14}
\end{figure}

In practice, Monte Carlo simulations assuming a statistical model of polymer conformations find that the exciton hopping rate is essentially independent of disorder and hence of chromophore size. This is presumably because neither the assumption of straight, parallel chromophores nor point dipoles are valid. These simulations also show that the average time taken for the first exciton hop to occur after photoexcitation is $\sim 10$ ps, whereas the time intervals between hops just prior to radiative recombination is over 10 times longer, and indeed becoming so long that a radiative transition is competitive.
This increase in hopping time intervals occurs because as an exciton diffuses through the polymer system it continuously looses energy. Thus, the energetic condition for exciton transfer to occur, namely $E_A \le E_D$, becomes harder to satisfy and in general the spectral overlap between the donor and acceptor decreases. As the excitons diffuse they eventually become trapped in `emissive' chromophores, from which they radiate. As shown in Fig.\ \ref{Fi:14}, these emissive chromophores occupy the low energy tail of the LEGSs density of states. Their quasi-Gaussian distribution explains spectral diffusion\cite{Hayes95,Meskers01}: a time-dependent change in the fluorescence energy, satisfying  $E \propto - \log t$.

Typically, the average hop distance is between 4 nm (for strong disorder giving an average chromophore length of 8 nm) to 6 nm (for weak disorder giving an average chromophore length of 30 nm). On average, an exciton only makes four hops before radiating, and thus average diffusion lengths are between $\sim 8-12$ nm, being longer for more ordered systems. These theoretical predictions are consistent with experimental values obtained via various techniques\cite{Markov05,Lewis06,Scully06}
(see K\"ohler and B\"assler\cite{Kohler15} for further experimental references).
The diffusion length is remarkably insensitive to disorder, and from simulation satisfies $L_D \sim L_{\textrm{loc}}^{1/4} \sim \sigma^{-1/6}$; a result that can be explained by the spatial distribution of chromophores in randomly coiled polymers\cite{Barford12b}.

An interesting prediction of Anderson localization is that for the same mean dihedral angle lower energy chromophores are shorter than higher energy chromophores. Now, as the intensity ratio of the vibronic peaks in the emission spectrum, ${I_{00}}/{I_{01}}$, is proportional to  the chromophore size\cite{Spano11,Spano14,Barford14a,Barford14b}, i.e.,
\begin{equation}
\frac{I_{00}}{I_{01}} \propto \frac{1}{S_{\textrm{eff}}} = \frac{\langle \textrm{PN} \rangle}{S},
\end{equation}
spectral diffusion also implies that  ${I_{00}}/{I_{01}}$ reduces in time, as observed in time-resolved photoluminescence spectra in MEH-PPV (see Fig.\ 3 of ref\cite{Hayes95}).
According to simulations\cite{Barford14c}, ${I_{00}}/{I_{01}} \propto - \log t$.

\begin{figure}
\centering
\includegraphics[width=0.45\textwidth]{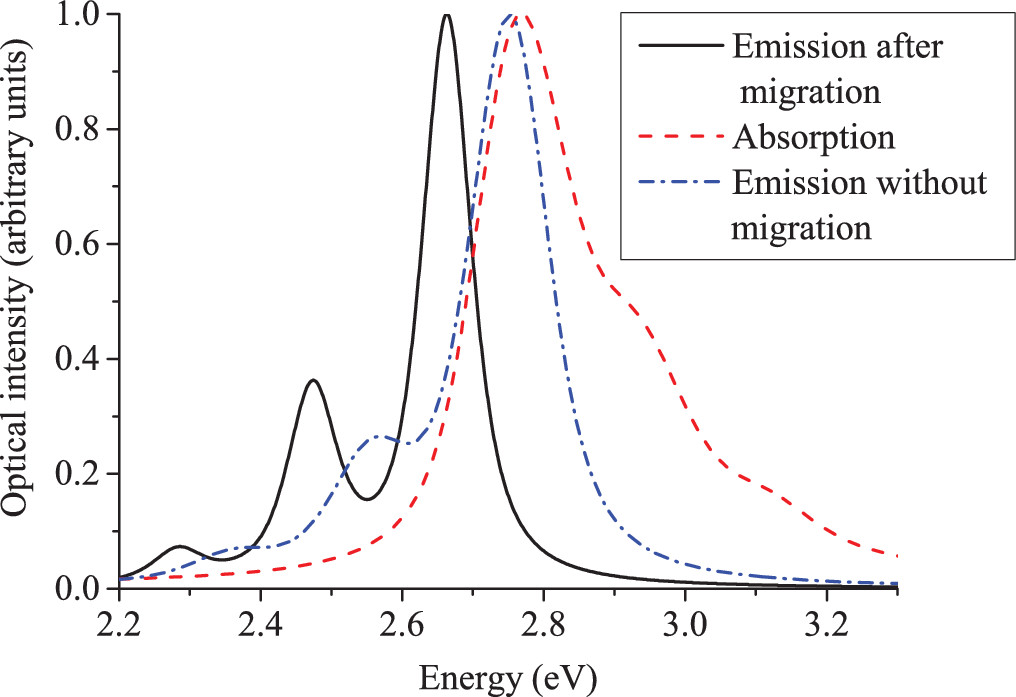}
\caption{The calculated optical spectra of PPV assuming a statistical model of random polymer conformations. Exciton migration prior to emission causes a red-shift in energy, a narrowing of the inhomogeneous broadening, and a decrease in $I_{00}/I_{01}$.
Reproduced  from J. Chem. Phys. \textbf{141}, 164103 (2014) with the permission of AIP publishing.}\label{Fi:9}
\end{figure}

Some of the key features of exciton relaxation and dynamics described in this review are  nicely encapsulated by Fig.\ \ref{Fi:9}. This figure shows the simulated  absorption to all absorbing states,  the  fluorescence via emission from all LEGSs (which occurs in the absence of exciton migration), and  the time-integrated fluorescence following exciton migration and emission from `trap' chromophores. We observe that:
\begin{itemize}
\item{The absorption spectrum and the emission spectrum assuming no exciton migration are broadly a mirror image. However, the absorption is broader and has a high energy tail as absorption  occurs to both LEGSs and  QEESs (as also shown in Fig.\ \ref{Fi:3}(b)), whereas, from Kasha's law, emission occurs only from LEGSs following interconversion from QEESs.
    }
\item{The emission following exciton migration is red-shifted, because the emissive states are in the low-energy tail of the density of states (as shown in Fig.\ \ref{Fi:14}).}
\item{The inhomogeneous broadening of the post-migration emission is narrowed, because the emissive states have a narrower density of states than LEGSs.}
\item{Similarly the intensity ratio, $I_{00}/I_{01}$, decreases, because on average emissive chromophores have shorter conjugation lengths than LEGSs.}
\end{itemize}

\section{Summary and Concluding Remarks}\label{se:7}

We have reviewed the various exciton dynamical processes in conjugated polymers. In summary, they are:

\begin{itemize}
  \item{Following photoexcitation, the initial dynamical process  is the correlation of the exciton and phonons associated with high-frequency C-C bond vibrations. This quantum mechanical entanglement causes exciton-site decoherence, which  is manifest as sub-10 fs fluorescence depolarization (see Section \ref{se:4.1}).}
  \item{Next, the energy that is transferred from the exciton to the nuclei is dissipated into the environment on a timescale determined by the strength of the system-bath interactions. For a hot exciton (i.e., a QEES) the system-bath interactions cause the entangled exciton-nuclear wavefunction to stochastically `collapse' into a particular LEGS, causing the exciton density to be localized on a `chromophore' (see Section \ref{se:4.2}).}
  \item{The fate of an exciton on a chromophore is now strongly dependent on the polymer chemical structure and the type of environment. For underdamped, freely rotating monomers, the coupling of the exciton to the low-frequency torsional modes creates an exciton-polaron, with associated  planarization and  exciton-density localization (see Section \ref{se:4.3}).}
  \item{For a polymer in solution, stochastic torsional fluctuations also causes the exciton-polaron to diffuse along the polymer chain; a process known as environment-assisted quantum transport\cite{Rebentrost09}. The diffusion coefficient is linearly proportional to temperature (see Section \ref{se:5}).}
  \item{For a polymer in the condensed phase, the dominant post-ps process is F\"orster resonant energy transfer and exciton diffusion. An exciton diffusing in the random energy landscape soon gets trapped in chromophores occupying the low-energy tail of the LEGSs density of states, exhibiting $\log t$ spectral diffusion. An exciton typically diffuses $\sim 10$ nm before radiative decay, with the diffusion length weakly increasing with decreasing disorder (see Section \ref{se:6}).}
\end{itemize}

In this review we have argued that theoretical modeling of exciton dynamics over multiple time and length scales is only realistically possible by employing suitably parametrized coarse-grained exciton-phonon models. Moreover, to correctly account for the ultrafast processes of exciton-site decoherence and the relaxation of hot excitons onto chromophores, the exciton and vibrational modes must be treated on the same quantum mechanical basis and importantly the Ehrenfest approximation must be abandoned. We have also repeatedly noted that spatial and temporal disorder play a key role in exciton spectroscopy and dynamics; and it is for this reason that exciton dynamics is conjugated polymers is essentially an incoherent process.

In a previous review\cite{Barford17} we explained how spectroscopic signatures are highly-dependent on polymer multiscale structures, and how - in principle - good theoretical modeling of excitons and spectroscopy can be used as a tool to predict these polymer structures. This review builds on that prospectus by describing how time-resolved spectroscopy can be understood via a theoretical description of exciton dynamics coupled to information on polymer multiscale structures. Again, the reverse proposition follows: time-resolved spectroscopy coupled to a theoretical description of exciton dynamics can be used to provide insights into polymer multiscale structures.


\begin{acknowledgments}
I thank Isabel Gonzalvez Perez for helping to compile Table 1.
\end{acknowledgments}

\bibliography{review}

\begin{thebibliography}{104}%
\makeatletter
\providecommand \@ifxundefined [1]{%
 \@ifx{#1\undefined}
}%
\providecommand \@ifnum [1]{%
 \ifnum #1\expandafter \@firstoftwo
 \else \expandafter \@secondoftwo
 \fi
}%
\providecommand \@ifx [1]{%
 \ifx #1\expandafter \@firstoftwo
 \else \expandafter \@secondoftwo
 \fi
}%
\providecommand \natexlab [1]{#1}%
\providecommand \enquote  [1]{``#1''}%
\providecommand \bibnamefont  [1]{#1}%
\providecommand \bibfnamefont [1]{#1}%
\providecommand \citenamefont [1]{#1}%
\providecommand \href@noop [0]{\@secondoftwo}%
\providecommand \href [0]{\begingroup \@sanitize@url \@href}%
\providecommand \@href[1]{\@@startlink{#1}\@@href}%
\providecommand \@@href[1]{\endgroup#1\@@endlink}%
\providecommand \@sanitize@url [0]{\catcode `\\12\catcode `\$12\catcode
  `\&12\catcode `\#12\catcode `\^12\catcode `\_12\catcode `\%12\relax}%
\providecommand \@@startlink[1]{}%
\providecommand \@@endlink[0]{}%
\providecommand \url  [0]{\begingroup\@sanitize@url \@url }%
\providecommand \@url [1]{\endgroup\@href {#1}{\urlprefix }}%
\providecommand \urlprefix  [0]{URL }%
\providecommand \Eprint [0]{\href }%
\providecommand \doibase [0]{http://dx.doi.org/}%
\providecommand \selectlanguage [0]{\@gobble}%
\providecommand \bibinfo  [0]{\@secondoftwo}%
\providecommand \bibfield  [0]{\@secondoftwo}%
\providecommand \translation [1]{[#1]}%
\providecommand \BibitemOpen [0]{}%
\providecommand \bibitemStop [0]{}%
\providecommand \bibitemNoStop [0]{.\EOS\space}%
\providecommand \EOS [0]{\spacefactor3000\relax}%
\providecommand \BibitemShut  [1]{\csname bibitem#1\endcsname}%
\let\auto@bib@innerbib\@empty
\bibitem [{\citenamefont {Grage}\ \emph {et~al.}(2003)\citenamefont {Grage},
  \citenamefont {Zaushitsyn}, \citenamefont {Yartsev}, \citenamefont
  {Chachisvilis}, \citenamefont {Sundstrom},\ and\ \citenamefont
  {Pullerits}}]{Grage03}%
  \BibitemOpen
  \bibfield  {author} {\bibinfo {author} {\bibfnamefont {M.~M.~L.}\
  \bibnamefont {Grage}}, \bibinfo {author} {\bibfnamefont {Y.}~\bibnamefont
  {Zaushitsyn}}, \bibinfo {author} {\bibfnamefont {A.}~\bibnamefont {Yartsev}},
  \bibinfo {author} {\bibfnamefont {M.}~\bibnamefont {Chachisvilis}}, \bibinfo
  {author} {\bibfnamefont {V.}~\bibnamefont {Sundstrom}}, \ and\ \bibinfo
  {author} {\bibfnamefont {T.}~\bibnamefont {Pullerits}},\ }\bibfield  {title}
  {\enquote {\bibinfo {title} {Ultrafast excitation transfer and trapping in a
  thin polymer film},}\ }\href {<Go to ISI>://WOS:000183483200034} {\bibfield
  {journal} {\bibinfo  {journal} {Physical Review B}\ }\textbf {\bibinfo
  {volume} {67}},\ \bibinfo {pages} {205207} (\bibinfo {year}
  {2003})}\BibitemShut {NoStop}%
\bibitem [{\citenamefont {Ruseckas}\ \emph {et~al.}(2005)\citenamefont
  {Ruseckas}, \citenamefont {Wood}, \citenamefont {Samuel}, \citenamefont
  {Webster}, \citenamefont {Mitchell}, \citenamefont {Burn},\ and\
  \citenamefont {Sundstrom}}]{Ruseckas05}%
  \BibitemOpen
  \bibfield  {author} {\bibinfo {author} {\bibfnamefont {A.}~\bibnamefont
  {Ruseckas}}, \bibinfo {author} {\bibfnamefont {P.}~\bibnamefont {Wood}},
  \bibinfo {author} {\bibfnamefont {I.~D.~W.}\ \bibnamefont {Samuel}}, \bibinfo
  {author} {\bibfnamefont {G.~R.}\ \bibnamefont {Webster}}, \bibinfo {author}
  {\bibfnamefont {W.~J.}\ \bibnamefont {Mitchell}}, \bibinfo {author}
  {\bibfnamefont {P.~L.}\ \bibnamefont {Burn}}, \ and\ \bibinfo {author}
  {\bibfnamefont {V.}~\bibnamefont {Sundstrom}},\ }\bibfield  {title} {\enquote
  {\bibinfo {title} {Ultrafast depolarization of the fluorescence in a
  conjugated polymer},}\ }\href {<Go to ISI>://WOS:000232229100081} {\bibfield
  {journal} {\bibinfo  {journal} {Physical Review B}\ }\textbf {\bibinfo
  {volume} {72}},\ \bibinfo {pages} {115214} (\bibinfo {year}
  {2005})}\BibitemShut {NoStop}%
\bibitem [{\citenamefont {Wells}\ \emph {et~al.}(2007)\citenamefont {Wells},
  \citenamefont {Boudouris}, \citenamefont {Hillmyer},\ and\ \citenamefont
  {Blank}}]{Wells07}%
  \BibitemOpen
  \bibfield  {author} {\bibinfo {author} {\bibfnamefont {N.~P.}\ \bibnamefont
  {Wells}}, \bibinfo {author} {\bibfnamefont {B.~W.}\ \bibnamefont
  {Boudouris}}, \bibinfo {author} {\bibfnamefont {M.~A.}\ \bibnamefont
  {Hillmyer}}, \ and\ \bibinfo {author} {\bibfnamefont {D.~A.}\ \bibnamefont
  {Blank}},\ }\bibfield  {title} {\enquote {\bibinfo {title} {Intramolecular
  exciton relaxation and migration dynamics in poly(3-hexylthiophene)},}\
  }\href {\doibase 10.1021/jp074657j} {\bibfield  {journal} {\bibinfo
  {journal} {Journal of Physical Chemistry C}\ }\textbf {\bibinfo {volume}
  {111}},\ \bibinfo {pages} {15404--15414} (\bibinfo {year}
  {2007})}\BibitemShut {NoStop}%
\bibitem [{\citenamefont {Dykstra}\ \emph {et~al.}(2009)\citenamefont
  {Dykstra}, \citenamefont {Hennebicq}, \citenamefont {Beljonne}, \citenamefont
  {Gierschner}, \citenamefont {Claudio}, \citenamefont {Bittner}, \citenamefont
  {Knoester},\ and\ \citenamefont {Scholes}}]{Dykstra09}%
  \BibitemOpen
  \bibfield  {author} {\bibinfo {author} {\bibfnamefont {T.~E.}\ \bibnamefont
  {Dykstra}}, \bibinfo {author} {\bibfnamefont {E.}~\bibnamefont {Hennebicq}},
  \bibinfo {author} {\bibfnamefont {D.}~\bibnamefont {Beljonne}}, \bibinfo
  {author} {\bibfnamefont {J.}~\bibnamefont {Gierschner}}, \bibinfo {author}
  {\bibfnamefont {G.}~\bibnamefont {Claudio}}, \bibinfo {author} {\bibfnamefont
  {E.~R.}\ \bibnamefont {Bittner}}, \bibinfo {author} {\bibfnamefont
  {J.}~\bibnamefont {Knoester}}, \ and\ \bibinfo {author} {\bibfnamefont
  {G.~D.}\ \bibnamefont {Scholes}},\ }\bibfield  {title} {\enquote {\bibinfo
  {title} {Conformational disorder and ultrafast exciton relaxation in
  ppv-family conjugated polymers},}\ }\href {\doibase 10.1021/jp807249b}
  {\bibfield  {journal} {\bibinfo  {journal} {Journal of Physical Chemistry B}\
  }\textbf {\bibinfo {volume} {113}},\ \bibinfo {pages} {656--667} (\bibinfo
  {year} {2009})}\BibitemShut {NoStop}%
\bibitem [{\citenamefont {Dykstra}\ \emph {et~al.}(2005)\citenamefont
  {Dykstra}, \citenamefont {Kovalevskij}, \citenamefont {Yang},\ and\
  \citenamefont {Scholes}}]{Dykstra05}%
  \BibitemOpen
  \bibfield  {author} {\bibinfo {author} {\bibfnamefont {T.~E.}\ \bibnamefont
  {Dykstra}}, \bibinfo {author} {\bibfnamefont {V.}~\bibnamefont
  {Kovalevskij}}, \bibinfo {author} {\bibfnamefont {X.~J.}\ \bibnamefont
  {Yang}}, \ and\ \bibinfo {author} {\bibfnamefont {G.~D.}\ \bibnamefont
  {Scholes}},\ }\bibfield  {title} {\enquote {\bibinfo {title} {Excited state
  dynamics of a conformationally disordered conjugated polymer: A comparison of
  solutions and film},}\ }\href {\doibase 10.1016/j.chemphys.2005.04.001}
  {\bibfield  {journal} {\bibinfo  {journal} {Chemical Physics}\ }\textbf
  {\bibinfo {volume} {318}},\ \bibinfo {pages} {21--32} (\bibinfo {year}
  {2005})}\BibitemShut {NoStop}%
\bibitem [{\citenamefont {Yang}, \citenamefont {Dykstra},\ and\ \citenamefont
  {Scholes}(2005)}]{Yang05}%
  \BibitemOpen
  \bibfield  {author} {\bibinfo {author} {\bibfnamefont {X.~J.}\ \bibnamefont
  {Yang}}, \bibinfo {author} {\bibfnamefont {T.~E.}\ \bibnamefont {Dykstra}}, \
  and\ \bibinfo {author} {\bibfnamefont {G.~D.}\ \bibnamefont {Scholes}},\
  }\bibfield  {title} {\enquote {\bibinfo {title} {Photon-echo studies of
  collective absorption and dynamic localization of excitation in conjugated
  polymers and oligomers},}\ }\href {<Go to ISI>://WOS:000226736200054}
  {\bibfield  {journal} {\bibinfo  {journal} {Physical Review B}\ }\textbf
  {\bibinfo {volume} {71}},\ \bibinfo {pages} {045203} (\bibinfo {year}
  {2005})}\BibitemShut {NoStop}%
\bibitem [{\citenamefont {Wells}\ and\ \citenamefont {Blank}(2008)}]{Wells08}%
  \BibitemOpen
  \bibfield  {author} {\bibinfo {author} {\bibfnamefont {N.~P.}\ \bibnamefont
  {Wells}}\ and\ \bibinfo {author} {\bibfnamefont {D.~A.}\ \bibnamefont
  {Blank}},\ }\bibfield  {title} {\enquote {\bibinfo {title} {Correlated
  exciton relaxation in poly(3-hexylthiophene)},}\ }\href {<Go to
  ISI>://WOS:000253764400039} {\bibfield  {journal} {\bibinfo  {journal}
  {Physical Review Letters}\ }\textbf {\bibinfo {volume} {100}},\ \bibinfo
  {pages} {086403} (\bibinfo {year} {2008})}\BibitemShut {NoStop}%
\bibitem [{\citenamefont {Sperling}\ \emph {et~al.}(2008)\citenamefont
  {Sperling}, \citenamefont {Nemeth}, \citenamefont {Baum}, \citenamefont
  {Sanda}, \citenamefont {Riedle}, \citenamefont {Kauffmann}, \citenamefont
  {Mukamel},\ and\ \citenamefont {Milota}}]{Sperling08}%
  \BibitemOpen
  \bibfield  {author} {\bibinfo {author} {\bibfnamefont {J.}~\bibnamefont
  {Sperling}}, \bibinfo {author} {\bibfnamefont {A.}~\bibnamefont {Nemeth}},
  \bibinfo {author} {\bibfnamefont {P.}~\bibnamefont {Baum}}, \bibinfo {author}
  {\bibfnamefont {F.}~\bibnamefont {Sanda}}, \bibinfo {author} {\bibfnamefont
  {E.}~\bibnamefont {Riedle}}, \bibinfo {author} {\bibfnamefont {H.~F.}\
  \bibnamefont {Kauffmann}}, \bibinfo {author} {\bibfnamefont {S.}~\bibnamefont
  {Mukamel}}, \ and\ \bibinfo {author} {\bibfnamefont {F.}~\bibnamefont
  {Milota}},\ }\bibfield  {title} {\enquote {\bibinfo {title} {Exciton dynamics
  in a disordered conjugated polymer: Three-pulse photon-echo and transient
  grating experiments},}\ }\href {\doibase 10.1016/j.chemphys.2008.02.046}
  {\bibfield  {journal} {\bibinfo  {journal} {Chemical Physics}\ }\textbf
  {\bibinfo {volume} {349}},\ \bibinfo {pages} {244--249} (\bibinfo {year}
  {2008})}\BibitemShut {NoStop}%
\bibitem [{\citenamefont {Consani}\ \emph {et~al.}(2015)\citenamefont
  {Consani}, \citenamefont {Koch}, \citenamefont {Panzer}, \citenamefont
  {Unger}, \citenamefont {Kohler},\ and\ \citenamefont {Brixner}}]{Consani15}%
  \BibitemOpen
  \bibfield  {author} {\bibinfo {author} {\bibfnamefont {C.}~\bibnamefont
  {Consani}}, \bibinfo {author} {\bibfnamefont {F.}~\bibnamefont {Koch}},
  \bibinfo {author} {\bibfnamefont {F.}~\bibnamefont {Panzer}}, \bibinfo
  {author} {\bibfnamefont {T.}~\bibnamefont {Unger}}, \bibinfo {author}
  {\bibfnamefont {A.}~\bibnamefont {Kohler}}, \ and\ \bibinfo {author}
  {\bibfnamefont {T.}~\bibnamefont {Brixner}},\ }\bibfield  {title} {\enquote
  {\bibinfo {title} {Relaxation dynamics and exciton energy transfer in the
  low-temperature phase of meh-ppv},}\ }\href {<Go to
  ISI>://WOS:000355931800033} {\bibfield  {journal} {\bibinfo  {journal}
  {Journal of Chemical Physics}\ }\textbf {\bibinfo {volume} {142}},\ \bibinfo
  {pages} {212429} (\bibinfo {year} {2015})}\BibitemShut {NoStop}%
\bibitem [{\citenamefont {Collini}\ and\ \citenamefont
  {Scholes}(2009)}]{Collini09}%
  \BibitemOpen
  \bibfield  {author} {\bibinfo {author} {\bibfnamefont {E.}~\bibnamefont
  {Collini}}\ and\ \bibinfo {author} {\bibfnamefont {G.~D.}\ \bibnamefont
  {Scholes}},\ }\bibfield  {title} {\enquote {\bibinfo {title} {Coherent
  intrachain energy migration in a conjugated polymer at room temperature},}\
  }\href {\doibase 10.1126/science.1164016} {\bibfield  {journal} {\bibinfo
  {journal} {Science}\ }\textbf {\bibinfo {volume} {323}},\ \bibinfo {pages}
  {369--373} (\bibinfo {year} {2009})}\BibitemShut {NoStop}%
\bibitem [{\citenamefont {Westenhoff}\ \emph {et~al.}(2006)\citenamefont
  {Westenhoff}, \citenamefont {Beenken}, \citenamefont {Friend}, \citenamefont
  {Greenham}, \citenamefont {Yartsev},\ and\ \citenamefont
  {Sundstrom}}]{Westenhoff06}%
  \BibitemOpen
  \bibfield  {author} {\bibinfo {author} {\bibfnamefont {S.}~\bibnamefont
  {Westenhoff}}, \bibinfo {author} {\bibfnamefont {W.~J.~D.}\ \bibnamefont
  {Beenken}}, \bibinfo {author} {\bibfnamefont {R.~H.}\ \bibnamefont {Friend}},
  \bibinfo {author} {\bibfnamefont {N.~C.}\ \bibnamefont {Greenham}}, \bibinfo
  {author} {\bibfnamefont {A.}~\bibnamefont {Yartsev}}, \ and\ \bibinfo
  {author} {\bibfnamefont {V.}~\bibnamefont {Sundstrom}},\ }\bibfield  {title}
  {\enquote {\bibinfo {title} {Anomalous energy transfer dynamics due to
  torsional relaxation in a conjugated polymer},}\ }\href {<Go to
  ISI>://WOS:000241405400051} {\bibfield  {journal} {\bibinfo  {journal}
  {Physical Review Letters}\ }\textbf {\bibinfo {volume} {97}},\ \bibinfo
  {pages} {166804} (\bibinfo {year} {2006})}\BibitemShut {NoStop}%
\bibitem [{\citenamefont {Banerji}\ \emph {et~al.}(2011)\citenamefont
  {Banerji}, \citenamefont {Cowan}, \citenamefont {Vauthey},\ and\
  \citenamefont {Heeger}}]{Banerji11}%
  \BibitemOpen
  \bibfield  {author} {\bibinfo {author} {\bibfnamefont {N.}~\bibnamefont
  {Banerji}}, \bibinfo {author} {\bibfnamefont {S.}~\bibnamefont {Cowan}},
  \bibinfo {author} {\bibfnamefont {E.}~\bibnamefont {Vauthey}}, \ and\
  \bibinfo {author} {\bibfnamefont {A.~J.}\ \bibnamefont {Heeger}},\ }\bibfield
   {title} {\enquote {\bibinfo {title} {Ultrafast relaxation of the
  poly(3-hexylthiophene) emission spectrum},}\ }\href {\doibase
  10.1021/jp1119348} {\bibfield  {journal} {\bibinfo  {journal} {Journal of
  Physical Chemistry C}\ }\textbf {\bibinfo {volume} {115}},\ \bibinfo {pages}
  {9726--9739} (\bibinfo {year} {2011})}\BibitemShut {NoStop}%
\bibitem [{\citenamefont {Busby}\ \emph {et~al.}(2011)\citenamefont {Busby},
  \citenamefont {Carroll}, \citenamefont {Chinn}, \citenamefont {Chang},
  \citenamefont {Moule},\ and\ \citenamefont {Larsen}}]{Busby11}%
  \BibitemOpen
  \bibfield  {author} {\bibinfo {author} {\bibfnamefont {E.}~\bibnamefont
  {Busby}}, \bibinfo {author} {\bibfnamefont {E.~C.}\ \bibnamefont {Carroll}},
  \bibinfo {author} {\bibfnamefont {E.~M.}\ \bibnamefont {Chinn}}, \bibinfo
  {author} {\bibfnamefont {L.~L.}\ \bibnamefont {Chang}}, \bibinfo {author}
  {\bibfnamefont {A.~J.}\ \bibnamefont {Moule}}, \ and\ \bibinfo {author}
  {\bibfnamefont {D.~S.}\ \bibnamefont {Larsen}},\ }\bibfield  {title}
  {\enquote {\bibinfo {title} {Excited-state self-trapping and ground-state
  relaxation dynamics in poly(3-hexylthiophene) resolved with broadband
  pump-dump-probe spectroscopy},}\ }\href {\doibase 10.1021/jz201168q}
  {\bibfield  {journal} {\bibinfo  {journal} {Journal of Physical Chemistry
  Letters}\ }\textbf {\bibinfo {volume} {2}},\ \bibinfo {pages} {2764--2769}
  (\bibinfo {year} {2011})}\BibitemShut {NoStop}%
\bibitem [{\citenamefont {Barford}\ and\ \citenamefont
  {Marcus}(2017)}]{Barford17}%
  \BibitemOpen
  \bibfield  {author} {\bibinfo {author} {\bibfnamefont {W.}~\bibnamefont
  {Barford}}\ and\ \bibinfo {author} {\bibfnamefont {M.}~\bibnamefont
  {Marcus}},\ }\bibfield  {title} {\enquote {\bibinfo {title} {Perspective:
  Optical spectroscopy in $\pi$-conjugated polymers and how it can be used to
  determine multiscale polymer structures},}\ }\href {<Go to
  ISI>://WOS:000399073300002} {\bibfield  {journal} {\bibinfo  {journal}
  {Journal of Chemical Physics}\ }\textbf {\bibinfo {volume} {146}},\ \bibinfo
  {pages} {130902} (\bibinfo {year} {2017})}\BibitemShut {NoStop}%
\bibitem [{\citenamefont {Barford}(2013{\natexlab{a}})}]{Book}%
  \BibitemOpen
  \bibfield  {author} {\bibinfo {author} {\bibfnamefont {W.}~\bibnamefont
  {Barford}},\ }\href@noop {} {\emph {\bibinfo {title} {Electronic and Optical
  Properties of Conjugated Polymers}}},\ \bibinfo {edition} {2nd}\ ed.\
  (\bibinfo  {publisher} {Oxford University Press},\ \bibinfo {address}
  {Oxford},\ \bibinfo {year} {2013})\BibitemShut {NoStop}%
\bibitem [{\citenamefont {Barford}\ and\ \citenamefont
  {Marcus}(2014)}]{Barford14a}%
  \BibitemOpen
  \bibfield  {author} {\bibinfo {author} {\bibfnamefont {W.}~\bibnamefont
  {Barford}}\ and\ \bibinfo {author} {\bibfnamefont {M.}~\bibnamefont
  {Marcus}},\ }\bibfield  {title} {\enquote {\bibinfo {title} {Theory of
  optical transitions in conjugated polymers. 1. ideal systems},}\ }\href {<Go
  to ISI>://WOS:000344589700005} {\bibfield  {journal} {\bibinfo  {journal}
  {Journal of Chemical Physics}\ }\textbf {\bibinfo {volume} {141}},\ \bibinfo
  {pages} {164101} (\bibinfo {year} {2014})}\BibitemShut {NoStop}%
\bibitem [{\citenamefont {Binder}\ \emph {et~al.}(2014)\citenamefont {Binder},
  \citenamefont {Romer}, \citenamefont {Wahl},\ and\ \citenamefont
  {Burghardt}}]{Binder14}%
  \BibitemOpen
  \bibfield  {author} {\bibinfo {author} {\bibfnamefont {R.}~\bibnamefont
  {Binder}}, \bibinfo {author} {\bibfnamefont {S.}~\bibnamefont {Romer}},
  \bibinfo {author} {\bibfnamefont {J.}~\bibnamefont {Wahl}}, \ and\ \bibinfo
  {author} {\bibfnamefont {I.}~\bibnamefont {Burghardt}},\ }\bibfield  {title}
  {\enquote {\bibinfo {title} {An analytic mapping of oligomer potential energy
  surfaces to an effective frenkel model},}\ }\href {<Go to
  ISI>://WOS:000339622000001} {\bibfield  {journal} {\bibinfo  {journal}
  {Journal of Chemical Physics}\ }\textbf {\bibinfo {volume} {141}},\ \bibinfo
  {pages} {014101} (\bibinfo {year} {2014})}\BibitemShut {NoStop}%
\bibitem [{\citenamefont {Binder}\ \emph {et~al.}()\citenamefont {Binder},
  \citenamefont {Bonfanti}, \citenamefont {Lauvergnat},\ and\ \citenamefont
  {Burghardt}}]{Binder20b}%
  \BibitemOpen
  \bibfield  {author} {\bibinfo {author} {\bibfnamefont {R.}~\bibnamefont
  {Binder}}, \bibinfo {author} {\bibfnamefont {M.}~\bibnamefont {Bonfanti}},
  \bibinfo {author} {\bibfnamefont {D.}~\bibnamefont {Lauvergnat}}, \ and\
  \bibinfo {author} {\bibfnamefont {I.}~\bibnamefont {Burghardt}},\ }\bibfield
  {title} {\enquote {\bibinfo {title} {First-principles description of
  intra-chain exciton migration in an oligo(para-phenylene vinylene) chain. 1.
  generalized frenkel-holstein hamiltonian},}\ }\href@noop {} {\bibfield
  {journal} {\bibinfo  {journal} {Journal of Chemical Physics}\ }\textbf
  {\bibinfo {volume} {152}}}\BibitemShut {NoStop}%
\bibitem [{\citenamefont {Marcus}, \citenamefont {Tozer},\ and\ \citenamefont
  {Barford}(2014)}]{Barford14b}%
  \BibitemOpen
  \bibfield  {author} {\bibinfo {author} {\bibfnamefont {M.}~\bibnamefont
  {Marcus}}, \bibinfo {author} {\bibfnamefont {O.~R.}\ \bibnamefont {Tozer}}, \
  and\ \bibinfo {author} {\bibfnamefont {W.}~\bibnamefont {Barford}},\
  }\bibfield  {title} {\enquote {\bibinfo {title} {Theory of optical
  transitions in conjugated polymers. 2. real systems},}\ }\href {<Go to
  ISI>://WOS:000344589700006} {\bibfield  {journal} {\bibinfo  {journal}
  {Journal of Chemical Physics}\ }\textbf {\bibinfo {volume} {141}},\ \bibinfo
  {pages} {164102} (\bibinfo {year} {2014})}\BibitemShut {NoStop}%
\bibitem [{\citenamefont {Horsfield}\ \emph {et~al.}(2006)\citenamefont
  {Horsfield}, \citenamefont {Bowler}, \citenamefont {Ness}, \citenamefont
  {Sanchez}, \citenamefont {Todorov},\ and\ \citenamefont
  {Fisher}}]{Horsfield06}%
  \BibitemOpen
  \bibfield  {author} {\bibinfo {author} {\bibfnamefont {A.~P.}\ \bibnamefont
  {Horsfield}}, \bibinfo {author} {\bibfnamefont {D.~R.}\ \bibnamefont
  {Bowler}}, \bibinfo {author} {\bibfnamefont {H.}~\bibnamefont {Ness}},
  \bibinfo {author} {\bibfnamefont {C.~G.}\ \bibnamefont {Sanchez}}, \bibinfo
  {author} {\bibfnamefont {T.~N.}\ \bibnamefont {Todorov}}, \ and\ \bibinfo
  {author} {\bibfnamefont {A.~J.}\ \bibnamefont {Fisher}},\ }\bibfield  {title}
  {\enquote {\bibinfo {title} {The transfer of energy between electrons and
  ions in solids},}\ }\href {\doibase 10.1088/0034-4885/69/4/R05} {\bibfield
  {journal} {\bibinfo  {journal} {Reports on Progress in Physics}\ }\textbf
  {\bibinfo {volume} {69}},\ \bibinfo {pages} {1195--1234} (\bibinfo {year}
  {2006})}\BibitemShut {NoStop}%
\bibitem [{\citenamefont {Nelson}\ \emph {et~al.}(2020)\citenamefont {Nelson},
  \citenamefont {White}, \citenamefont {Bjorgaard}, \citenamefont {Sifain},
  \citenamefont {Zhang}, \citenamefont {Nebgen}, \citenamefont
  {Fernandez-Alberti}, \citenamefont {Mozyrsky}, \citenamefont {Roitberg},\
  and\ \citenamefont {Tretiak}}]{Nelson20}%
  \BibitemOpen
  \bibfield  {author} {\bibinfo {author} {\bibfnamefont {T.~R.}\ \bibnamefont
  {Nelson}}, \bibinfo {author} {\bibfnamefont {A.~J.}\ \bibnamefont {White}},
  \bibinfo {author} {\bibfnamefont {J.~A.}\ \bibnamefont {Bjorgaard}}, \bibinfo
  {author} {\bibfnamefont {A.~E.}\ \bibnamefont {Sifain}}, \bibinfo {author}
  {\bibfnamefont {Y.}~\bibnamefont {Zhang}}, \bibinfo {author} {\bibfnamefont
  {B.}~\bibnamefont {Nebgen}}, \bibinfo {author} {\bibfnamefont
  {S.}~\bibnamefont {Fernandez-Alberti}}, \bibinfo {author} {\bibfnamefont
  {D.}~\bibnamefont {Mozyrsky}}, \bibinfo {author} {\bibfnamefont {A.~E.}\
  \bibnamefont {Roitberg}}, \ and\ \bibinfo {author} {\bibfnamefont
  {S.}~\bibnamefont {Tretiak}},\ }\bibfield  {title} {\enquote {\bibinfo
  {title} {Non-adiabatic excited-state molecular dynamics: Theory and
  applications for modeling photophysics in extended molecular materials},}\
  }\href {\doibase 10.1021/acs.chemrev.9b00447} {\bibfield  {journal} {\bibinfo
   {journal} {Chemical Reviews}\ }\textbf {\bibinfo {volume} {120}},\ \bibinfo
  {pages} {2215--2287} (\bibinfo {year} {2020})}\BibitemShut {NoStop}%
\bibitem [{\citenamefont {Tully}(1990)}]{Tully90}%
  \BibitemOpen
  \bibfield  {author} {\bibinfo {author} {\bibfnamefont {J.~C.}\ \bibnamefont
  {Tully}},\ }\bibfield  {title} {\enquote {\bibinfo {title}
  {Molecular-dynamics with electronic-transitions},}\ }\href {\doibase Doi
  10.1063/1.459170} {\bibfield  {journal} {\bibinfo  {journal} {Journal of
  Chemical Physics}\ }\textbf {\bibinfo {volume} {93}},\ \bibinfo {pages}
  {1061--1071} (\bibinfo {year} {1990})}\BibitemShut {NoStop}%
\bibitem [{\citenamefont {Tully}(2012)}]{Tully12}%
  \BibitemOpen
  \bibfield  {author} {\bibinfo {author} {\bibfnamefont {J.~C.}\ \bibnamefont
  {Tully}},\ }\bibfield  {title} {\enquote {\bibinfo {title} {Perspective:
  Nonadiabatic dynamics theory},}\ }\href {<Go to ISI>://WOS:000312491400001}
  {\bibfield  {journal} {\bibinfo  {journal} {Journal of Chemical Physics}\
  }\textbf {\bibinfo {volume} {137}},\ \bibinfo {pages} {22A301} (\bibinfo
  {year} {2012})}\BibitemShut {NoStop}%
\bibitem [{\citenamefont {Beck}\ \emph {et~al.}(2000)\citenamefont {Beck},
  \citenamefont {Jackle}, \citenamefont {Worth},\ and\ \citenamefont
  {Meyer}}]{Beck00}%
  \BibitemOpen
  \bibfield  {author} {\bibinfo {author} {\bibfnamefont {M.~H.}\ \bibnamefont
  {Beck}}, \bibinfo {author} {\bibfnamefont {A.}~\bibnamefont {Jackle}},
  \bibinfo {author} {\bibfnamefont {G.~A.}\ \bibnamefont {Worth}}, \ and\
  \bibinfo {author} {\bibfnamefont {H.~D.}\ \bibnamefont {Meyer}},\ }\bibfield
  {title} {\enquote {\bibinfo {title} {The multiconfiguration time-dependent
  hartree (mctdh) method: a highly efficient algorithm for propagating
  wavepackets},}\ }\href {\doibase Doi 10.1016/S0370-1573(99)00047-2}
  {\bibfield  {journal} {\bibinfo  {journal} {Physics Reports}\ }\textbf
  {\bibinfo {volume} {324}},\ \bibinfo {pages} {1--105} (\bibinfo {year}
  {2000})}\BibitemShut {NoStop}%
\bibitem [{\citenamefont {Vidal}(2003)}]{Vidal03}%
  \BibitemOpen
  \bibfield  {author} {\bibinfo {author} {\bibfnamefont {G.}~\bibnamefont
  {Vidal}},\ }\bibfield  {title} {\enquote {\bibinfo {title} {Efficient
  classical simulation of slightly entangled quantum computations},}\ }\href
  {<Go to ISI>://WOS:000185719500046} {\bibfield  {journal} {\bibinfo
  {journal} {Physical Review Letters}\ }\textbf {\bibinfo {volume} {91}},\
  \bibinfo {pages} {147902} (\bibinfo {year} {2003})}\BibitemShut {NoStop}%
\bibitem [{\citenamefont {Vidal}(2004)}]{Vidal04}%
  \BibitemOpen
  \bibfield  {author} {\bibinfo {author} {\bibfnamefont {G.}~\bibnamefont
  {Vidal}},\ }\bibfield  {title} {\enquote {\bibinfo {title} {Efficient
  simulation of one-dimensional quantum many-body systems},}\ }\href {<Go to
  ISI>://WOS:000222856400005} {\bibfield  {journal} {\bibinfo  {journal}
  {Physical Review Letters}\ }\textbf {\bibinfo {volume} {93}},\ \bibinfo
  {pages} {040502} (\bibinfo {year} {2004})}\BibitemShut {NoStop}%
\bibitem [{\citenamefont {Sch\"ollwock}(2011)}]{Schollwock11}%
  \BibitemOpen
  \bibfield  {author} {\bibinfo {author} {\bibfnamefont {U.}~\bibnamefont
  {Sch\"ollwock}},\ }\bibfield  {title} {\enquote {\bibinfo {title} {The
  density-matrix renormalization group in the age of matrix product states},}\
  }\href {\doibase 10.1016/j.aop.2010.09.012} {\bibfield  {journal} {\bibinfo
  {journal} {Annals of Physics}\ }\textbf {\bibinfo {volume} {326}},\ \bibinfo
  {pages} {96--192} (\bibinfo {year} {2011})}\BibitemShut {NoStop}%
\bibitem [{Note1()}]{Note1}%
  \BibitemOpen
  \bibinfo {note} {A related method is time-dependent density matrix
  renormalization group (TD-DMRG). This has been successfully applied to
  simulate singlet fission in carotenoids, D. Manawadu, M. Marcus and W.
  Barford, in preparation.}\BibitemShut {Stop}%
\bibitem [{\citenamefont {Mannouch}, \citenamefont {Barford},\ and\
  \citenamefont {Al-Assam}(2018)}]{Mannouch18}%
  \BibitemOpen
  \bibfield  {author} {\bibinfo {author} {\bibfnamefont {J.~R.}\ \bibnamefont
  {Mannouch}}, \bibinfo {author} {\bibfnamefont {W.}~\bibnamefont {Barford}}, \
  and\ \bibinfo {author} {\bibfnamefont {S.}~\bibnamefont {Al-Assam}},\
  }\bibfield  {title} {\enquote {\bibinfo {title} {Ultra-fast relaxation,
  decoherence, and localization of photoexcited states in $\pi$-conjugated
  polymers},}\ }\href {\doibase 10.1063/1.5009393} {\bibfield  {journal}
  {\bibinfo  {journal} {J Chem Phys}\ }\textbf {\bibinfo {volume} {148}},\
  \bibinfo {pages} {034901} (\bibinfo {year} {2018})}\BibitemShut {NoStop}%
\bibitem [{\citenamefont {Kobayashi}(1993)}]{Abe93}%
  \BibitemOpen
  \bibfield  {author} {\bibinfo {author} {\bibfnamefont {T.}~\bibnamefont
  {Kobayashi}},\ }\href@noop {} {\emph {\bibinfo {title} {Relaxation in
  Polymers}}}\ (\bibinfo  {publisher} {World Scientific},\ \bibinfo {address}
  {Singapore},\ \bibinfo {year} {1993})\BibitemShut {NoStop}%
\bibitem [{\citenamefont {Barford}(2013{\natexlab{b}})}]{Barford13}%
  \BibitemOpen
  \bibfield  {author} {\bibinfo {author} {\bibfnamefont {W.}~\bibnamefont
  {Barford}},\ }\bibfield  {title} {\enquote {\bibinfo {title} {Excitons in
  conjugated polymers: A tale of two particles},}\ }\href {\doibase
  10.1021/jp310110r} {\bibfield  {journal} {\bibinfo  {journal} {Journal of
  Physical Chemistry A}\ }\textbf {\bibinfo {volume} {117}},\ \bibinfo {pages}
  {2665--2671} (\bibinfo {year} {2013}{\natexlab{b}})}\BibitemShut {NoStop}%
\bibitem [{\citenamefont {Loudon}(2016)}]{Loudon16}%
  \BibitemOpen
  \bibfield  {author} {\bibinfo {author} {\bibfnamefont {R.}~\bibnamefont
  {Loudon}},\ }\bibfield  {title} {\enquote {\bibinfo {title} {One-dimensional
  hydrogen atom},}\ }\href {<Go to ISI>://WOS:000368479000010} {\bibfield
  {journal} {\bibinfo  {journal} {Proceedings of the Royal Society A}\ }\textbf
  {\bibinfo {volume} {472}},\ \bibinfo {pages} {20150534} (\bibinfo {year}
  {2016})}\BibitemShut {NoStop}%
\bibitem [{\citenamefont {Barford}, \citenamefont {Bursill},\ and\
  \citenamefont {Smith}(2002)}]{Barford02a}%
  \BibitemOpen
  \bibfield  {author} {\bibinfo {author} {\bibfnamefont {W.}~\bibnamefont
  {Barford}}, \bibinfo {author} {\bibfnamefont {R.~J.}\ \bibnamefont
  {Bursill}}, \ and\ \bibinfo {author} {\bibfnamefont {R.~W.}\ \bibnamefont
  {Smith}},\ }\bibfield  {title} {\enquote {\bibinfo {title} {Theoretical and
  computational studies of excitons in conjugated polymers},}\ }\href {<Go to
  ISI>://WOS:000178461000043} {\bibfield  {journal} {\bibinfo  {journal}
  {Physical Review B}\ }\textbf {\bibinfo {volume} {66}},\ \bibinfo {pages}
  {115205} (\bibinfo {year} {2002})}\BibitemShut {NoStop}%
\bibitem [{\citenamefont {Barford}\ and\ \citenamefont
  {Trembath}(2009)}]{Barford09a}%
  \BibitemOpen
  \bibfield  {author} {\bibinfo {author} {\bibfnamefont {W.}~\bibnamefont
  {Barford}}\ and\ \bibinfo {author} {\bibfnamefont {D.}~\bibnamefont
  {Trembath}},\ }\bibfield  {title} {\enquote {\bibinfo {title} {Exciton
  localization in polymers with static disorder},}\ }\href {<Go to
  ISI>://WOS:000271352100127} {\bibfield  {journal} {\bibinfo  {journal}
  {Physical Review B}\ }\textbf {\bibinfo {volume} {80}},\ \bibinfo {pages}
  {165418} (\bibinfo {year} {2009})}\BibitemShut {NoStop}%
\bibitem [{\citenamefont {Barford}\ \emph {et~al.}(2010)\citenamefont
  {Barford}, \citenamefont {Lidzey}, \citenamefont {Makhov},\ and\
  \citenamefont {Meijer}}]{Barford10a}%
  \BibitemOpen
  \bibfield  {author} {\bibinfo {author} {\bibfnamefont {W.}~\bibnamefont
  {Barford}}, \bibinfo {author} {\bibfnamefont {D.~G.}\ \bibnamefont {Lidzey}},
  \bibinfo {author} {\bibfnamefont {D.~V.}\ \bibnamefont {Makhov}}, \ and\
  \bibinfo {author} {\bibfnamefont {A.~J.~H.}\ \bibnamefont {Meijer}},\
  }\bibfield  {title} {\enquote {\bibinfo {title} {Exciton localization in
  disordered poly(3-hexylthiophene)},}\ }\href {<Go to
  ISI>://WOS:000280854600036} {\bibfield  {journal} {\bibinfo  {journal}
  {Journal of Chemical Physics}\ }\textbf {\bibinfo {volume} {133}},\ \bibinfo
  {pages} {044504} (\bibinfo {year} {2010})}\BibitemShut {NoStop}%
\bibitem [{\citenamefont {Anderson}(1958)}]{Anderson58a}%
  \BibitemOpen
  \bibfield  {author} {\bibinfo {author} {\bibfnamefont {P.~W.}\ \bibnamefont
  {Anderson}},\ }\bibfield  {title} {\enquote {\bibinfo {title} {Absence of
  diffusion in certain random lattices},}\ }\href {\doibase DOI
  10.1103/PhysRev.109.1492} {\bibfield  {journal} {\bibinfo  {journal}
  {Physical Review}\ }\textbf {\bibinfo {volume} {109}},\ \bibinfo {pages}
  {1492--1505} (\bibinfo {year} {1958})}\BibitemShut {NoStop}%
\bibitem [{\citenamefont {Malyshev}\ and\ \citenamefont
  {Malyshev}(2001{\natexlab{a}})}]{Malyshev01a}%
  \BibitemOpen
  \bibfield  {author} {\bibinfo {author} {\bibfnamefont {A.~V.}\ \bibnamefont
  {Malyshev}}\ and\ \bibinfo {author} {\bibfnamefont {V.~A.}\ \bibnamefont
  {Malyshev}},\ }\bibfield  {title} {\enquote {\bibinfo {title} {Statistics of
  low energy levels of a one-dimensional weakly localized frenkel exciton: A
  numerical study},}\ }\href {<Go to ISI>://WOS:000168814200035} {\bibfield
  {journal} {\bibinfo  {journal} {Physical Review B}\ }\textbf {\bibinfo
  {volume} {63}},\ \bibinfo {pages} {195111} (\bibinfo {year}
  {2001}{\natexlab{a}})}\BibitemShut {NoStop}%
\bibitem [{\citenamefont {Malyshev}\ and\ \citenamefont
  {Malyshev}(2001{\natexlab{b}})}]{Malyshev01b}%
  \BibitemOpen
  \bibfield  {author} {\bibinfo {author} {\bibfnamefont {A.~V.}\ \bibnamefont
  {Malyshev}}\ and\ \bibinfo {author} {\bibfnamefont {V.~A.}\ \bibnamefont
  {Malyshev}},\ }\bibfield  {title} {\enquote {\bibinfo {title} {Level and wave
  function statistics of a localized 1d frenkel exciton at the bottom of the
  band},}\ }\href {\doibase Doi 10.1016/S0022-2313(01)00303-9} {\bibfield
  {journal} {\bibinfo  {journal} {Journal of Luminescence}\ }\textbf {\bibinfo
  {volume} {94}},\ \bibinfo {pages} {369--372} (\bibinfo {year}
  {2001}{\natexlab{b}})}\BibitemShut {NoStop}%
\bibitem [{\citenamefont {Makhov}\ and\ \citenamefont
  {Barford}(2010)}]{Makhov10}%
  \BibitemOpen
  \bibfield  {author} {\bibinfo {author} {\bibfnamefont {D.~V.}\ \bibnamefont
  {Makhov}}\ and\ \bibinfo {author} {\bibfnamefont {W.}~\bibnamefont
  {Barford}},\ }\bibfield  {title} {\enquote {\bibinfo {title} {Local exciton
  ground states in disordered polymers},}\ }\href {<Go to
  ISI>://WOS:000277217200048} {\bibfield  {journal} {\bibinfo  {journal}
  {Physical Review B}\ }\textbf {\bibinfo {volume} {81}},\ \bibinfo {pages}
  {165201} (\bibinfo {year} {2010})}\BibitemShut {NoStop}%
\bibitem [{\citenamefont {Athanasopoulos}\ \emph {et~al.}(2008)\citenamefont
  {Athanasopoulos}, \citenamefont {Hennebicq}, \citenamefont {Beljonne},\ and\
  \citenamefont {Walker}}]{Athanasopoulos08}%
  \BibitemOpen
  \bibfield  {author} {\bibinfo {author} {\bibfnamefont {S.}~\bibnamefont
  {Athanasopoulos}}, \bibinfo {author} {\bibfnamefont {E.}~\bibnamefont
  {Hennebicq}}, \bibinfo {author} {\bibfnamefont {D.}~\bibnamefont {Beljonne}},
  \ and\ \bibinfo {author} {\bibfnamefont {A.~B.}\ \bibnamefont {Walker}},\
  }\bibfield  {title} {\enquote {\bibinfo {title} {Trap limited exciton
  transport in conjugated polymers},}\ }\href {\doibase 10.1021/jp802704z}
  {\bibfield  {journal} {\bibinfo  {journal} {Journal of Physical Chemistry C}\
  }\textbf {\bibinfo {volume} {112}},\ \bibinfo {pages} {11532--11538}
  (\bibinfo {year} {2008})}\BibitemShut {NoStop}%
\bibitem [{\citenamefont {Kramer}\ and\ \citenamefont
  {Mackinnon}(1993)}]{Kramer93}%
  \BibitemOpen
  \bibfield  {author} {\bibinfo {author} {\bibfnamefont {B.}~\bibnamefont
  {Kramer}}\ and\ \bibinfo {author} {\bibfnamefont {A.}~\bibnamefont
  {Mackinnon}},\ }\bibfield  {title} {\enquote {\bibinfo {title} {Localization
  - theory and experiment},}\ }\href {\doibase Doi 10.1088/0034-4885/56/12/001}
  {\bibfield  {journal} {\bibinfo  {journal} {Reports on Progress in Physics}\
  }\textbf {\bibinfo {volume} {56}},\ \bibinfo {pages} {1469--1564} (\bibinfo
  {year} {1993})}\BibitemShut {NoStop}%
\bibitem [{\citenamefont {Mulliken}\ \emph {et~al.}(1949)\citenamefont
  {Mulliken}, \citenamefont {Rieke}, \citenamefont {Orloff},\ and\
  \citenamefont {Orloff}}]{Mulliken49}%
  \BibitemOpen
  \bibfield  {author} {\bibinfo {author} {\bibfnamefont {R.~S.}\ \bibnamefont
  {Mulliken}}, \bibinfo {author} {\bibfnamefont {C.~A.}\ \bibnamefont {Rieke}},
  \bibinfo {author} {\bibfnamefont {D.}~\bibnamefont {Orloff}}, \ and\ \bibinfo
  {author} {\bibfnamefont {H.}~\bibnamefont {Orloff}},\ }\bibfield  {title}
  {\enquote {\bibinfo {title} {Formulas and numerical tables for overlap
  integrals},}\ }\href {\doibase Doi 10.1063/1.1747150} {\bibfield  {journal}
  {\bibinfo  {journal} {Journal of Chemical Physics}\ }\textbf {\bibinfo
  {volume} {17}},\ \bibinfo {pages} {1248--1267} (\bibinfo {year}
  {1949})}\BibitemShut {NoStop}%
\bibitem [{\citenamefont {Beenken}\ and\ \citenamefont
  {Lischka}(2005)}]{Beenken05}%
  \BibitemOpen
  \bibfield  {author} {\bibinfo {author} {\bibfnamefont {W.~J.~D.}\
  \bibnamefont {Beenken}}\ and\ \bibinfo {author} {\bibfnamefont
  {H.}~\bibnamefont {Lischka}},\ }\bibfield  {title} {\enquote {\bibinfo
  {title} {Spectral broadening and diffusion by torsional motion in
  biphenyl},}\ }\href {<Go to ISI>://WOS:000232532000031} {\bibfield  {journal}
  {\bibinfo  {journal} {Journal of Chemical Physics}\ }\textbf {\bibinfo
  {volume} {123}},\ \bibinfo {pages} {144311} (\bibinfo {year}
  {2005})}\BibitemShut {NoStop}%
\bibitem [{\citenamefont {Rashba}(1957{\natexlab{a}})}]{Rashba57a}%
  \BibitemOpen
  \bibfield  {author} {\bibinfo {author} {\bibfnamefont {E.~I.}\ \bibnamefont
  {Rashba}},\ }\bibfield  {title} {\enquote {\bibinfo {title} {A theory of
  impurity absorption of light in molecular crystals},}\ }\href {<Go to
  ISI>://WOS:A1957WP06000004} {\bibfield  {journal} {\bibinfo  {journal}
  {Optika I Spektroskopiya}\ }\textbf {\bibinfo {volume} {2}},\ \bibinfo
  {pages} {568--577} (\bibinfo {year} {1957}{\natexlab{a}})}\BibitemShut
  {NoStop}%
\bibitem [{\citenamefont {Rashba}(1957{\natexlab{b}})}]{Rashba57b}%
  \BibitemOpen
  \bibfield  {author} {\bibinfo {author} {\bibfnamefont {E.~I.}\ \bibnamefont
  {Rashba}},\ }\bibfield  {title} {\enquote {\bibinfo {title} {Theory of strong
  interactions of electron excitations with lattice vibrations in molecular
  crystals. 1.}}\ }\href {<Go to ISI>://WOS:A1957WP05600009} {\bibfield
  {journal} {\bibinfo  {journal} {Optika I Spektroskopiya}\ }\textbf {\bibinfo
  {volume} {2}},\ \bibinfo {pages} {75--87} (\bibinfo {year}
  {1957}{\natexlab{b}})}\BibitemShut {NoStop}%
\bibitem [{\citenamefont {Rashba}(1957{\natexlab{c}})}]{Rashba57c}%
  \BibitemOpen
  \bibfield  {author} {\bibinfo {author} {\bibfnamefont {E.~I.}\ \bibnamefont
  {Rashba}},\ }\bibfield  {title} {\enquote {\bibinfo {title} {Theory of strong
  interactions of electron excitations with lattice vibrations in molecular
  crystals. 2.}}\ }\href {<Go to ISI>://WOS:A1957WP05600010} {\bibfield
  {journal} {\bibinfo  {journal} {Optika I Spektroskopiya}\ }\textbf {\bibinfo
  {volume} {2}},\ \bibinfo {pages} {88--98} (\bibinfo {year}
  {1957}{\natexlab{c}})}\BibitemShut {NoStop}%
\bibitem [{\citenamefont {Holstein}(1959{\natexlab{a}})}]{Holstein59a}%
  \BibitemOpen
  \bibfield  {author} {\bibinfo {author} {\bibfnamefont {T.}~\bibnamefont
  {Holstein}},\ }\bibfield  {title} {\enquote {\bibinfo {title} {Studies of
  polaron motion. 1. the molecular-crystal model},}\ }\href {\doibase Doi
  10.1016/0003-4916(59)90002-8} {\bibfield  {journal} {\bibinfo  {journal}
  {Annals of Physics}\ }\textbf {\bibinfo {volume} {8}},\ \bibinfo {pages}
  {325--342} (\bibinfo {year} {1959}{\natexlab{a}})}\BibitemShut {NoStop}%
\bibitem [{\citenamefont {Holstein}(1959{\natexlab{b}})}]{Holstein59b}%
  \BibitemOpen
  \bibfield  {author} {\bibinfo {author} {\bibfnamefont {T.}~\bibnamefont
  {Holstein}},\ }\bibfield  {title} {\enquote {\bibinfo {title} {Studies of
  polaron motion. 2. the small polaron},}\ }\href {\doibase Doi
  10.1016/0003-4916(59)90003-X} {\bibfield  {journal} {\bibinfo  {journal}
  {Annals of Physics}\ }\textbf {\bibinfo {volume} {8}},\ \bibinfo {pages}
  {343--389} (\bibinfo {year} {1959}{\natexlab{b}})}\BibitemShut {NoStop}%
\bibitem [{\citenamefont {Rashba}\ and\ \citenamefont
  {Sturge}(1982)}]{Rashba82}%
  \BibitemOpen
  \bibfield  {author} {\bibinfo {author} {\bibfnamefont {E.~I.}\ \bibnamefont
  {Rashba}}\ and\ \bibinfo {author} {\bibfnamefont {M.~D.}\ \bibnamefont
  {Sturge}},\ }\href@noop {} {\emph {\bibinfo {title} {Excitons}}}\ (\bibinfo
  {publisher} {North-Holland},\ \bibinfo {address} {Amsterdam},\ \bibinfo
  {year} {1982})\BibitemShut {NoStop}%
\bibitem [{\citenamefont {Landau}(1933)}]{Landau33}%
  \BibitemOpen
  \bibfield  {author} {\bibinfo {author} {\bibfnamefont {L.~D.}\ \bibnamefont
  {Landau}},\ }\href@noop {} {\bibfield  {journal} {\bibinfo  {journal} {Z.
  Phys.}\ }\textbf {\bibinfo {volume} {3}},\ \bibinfo {pages} {664} (\bibinfo
  {year} {1933})}\BibitemShut {NoStop}%
\bibitem [{\citenamefont {Campbell}, \citenamefont {Bishop},\ and\
  \citenamefont {Fesser}(1982)}]{Campbell82}%
  \BibitemOpen
  \bibfield  {author} {\bibinfo {author} {\bibfnamefont {D.~K.}\ \bibnamefont
  {Campbell}}, \bibinfo {author} {\bibfnamefont {A.~R.}\ \bibnamefont
  {Bishop}}, \ and\ \bibinfo {author} {\bibfnamefont {K.}~\bibnamefont
  {Fesser}},\ }\bibfield  {title} {\enquote {\bibinfo {title} {Polarons in
  quasi-one-dimensional systems},}\ }\href {\doibase DOI
  10.1103/PhysRevB.26.6862} {\bibfield  {journal} {\bibinfo  {journal}
  {Physical Review B}\ }\textbf {\bibinfo {volume} {26}},\ \bibinfo {pages}
  {6862--6874} (\bibinfo {year} {1982})}\BibitemShut {NoStop}%
\bibitem [{\citenamefont {Tozer}\ and\ \citenamefont
  {Barford}(2014)}]{Tozer14}%
  \BibitemOpen
  \bibfield  {author} {\bibinfo {author} {\bibfnamefont {O.~R.}\ \bibnamefont
  {Tozer}}\ and\ \bibinfo {author} {\bibfnamefont {W.}~\bibnamefont
  {Barford}},\ }\bibfield  {title} {\enquote {\bibinfo {title} {Localization of
  large polarons in the disordered holstein model},}\ }\href {<Go to
  ISI>://WOS:000337301200010} {\bibfield  {journal} {\bibinfo  {journal}
  {Physical Review B}\ }\textbf {\bibinfo {volume} {89}},\ \bibinfo {pages}
  {155434} (\bibinfo {year} {2014})}\BibitemShut {NoStop}%
\bibitem [{Note2()}]{Note2}%
  \BibitemOpen
  \bibinfo {note} {There is also a weaker and less significant coupling of the
  normal mode to the exciton bond-order operator\cite {Barford14b,
  Binder14}.}\BibitemShut {Stop}%
\bibitem [{\citenamefont {Hoffmann}\ and\ \citenamefont
  {Soos}(2002)}]{Hoffmann02}%
  \BibitemOpen
  \bibfield  {author} {\bibinfo {author} {\bibfnamefont {M.}~\bibnamefont
  {Hoffmann}}\ and\ \bibinfo {author} {\bibfnamefont {Z.~G.}\ \bibnamefont
  {Soos}},\ }\bibfield  {title} {\enquote {\bibinfo {title} {Optical absorption
  spectra of the holstein molecular crystal for weak and intermediate
  electronic coupling},}\ }\href@noop {} {\bibfield  {journal} {\bibinfo
  {journal} {Physical Review B}\ }\textbf {\bibinfo {volume} {66}},\ \bibinfo
  {pages} {024305} (\bibinfo {year} {2002})}\BibitemShut {NoStop}%
\bibitem [{Note3()}]{Note3}%
  \BibitemOpen
  \bibinfo {note} {In contrast, in the classical limit ($\omega \rightarrow 0$)
  the nuclei respond infinitesimally slowly to the exciton, so that the
  correlation length and the exciton-polaron mass diverge causing
  exciton-polaron self-localization.}\BibitemShut {Stop}%
\bibitem [{\citenamefont {Kuhn}\ and\ \citenamefont
  {Sundstrom}(1997)}]{Kuhn97}%
  \BibitemOpen
  \bibfield  {author} {\bibinfo {author} {\bibfnamefont {O.}~\bibnamefont
  {Kuhn}}\ and\ \bibinfo {author} {\bibfnamefont {V.}~\bibnamefont
  {Sundstrom}},\ }\bibfield  {title} {\enquote {\bibinfo {title} {Pump-probe
  spectroscopy of dissipative energy transfer dynamics in photosynthetic
  antenna complexes: A density matrix approach},}\ }\href {<Go to
  ISI>://WOS:A1997XW11900006} {\bibfield  {journal} {\bibinfo  {journal}
  {Journal of Chemical Physics}\ }\textbf {\bibinfo {volume} {107}},\ \bibinfo
  {pages} {4154--4164} (\bibinfo {year} {1997})}\BibitemShut {NoStop}%
\bibitem [{\citenamefont {Smyth}, \citenamefont {Fassioli},\ and\ \citenamefont
  {Scholes}(2012)}]{Smyth12}%
  \BibitemOpen
  \bibfield  {author} {\bibinfo {author} {\bibfnamefont {C.}~\bibnamefont
  {Smyth}}, \bibinfo {author} {\bibfnamefont {F.}~\bibnamefont {Fassioli}}, \
  and\ \bibinfo {author} {\bibfnamefont {G.~D.}\ \bibnamefont {Scholes}},\
  }\bibfield  {title} {\enquote {\bibinfo {title} {Measures and implications of
  electronic coherence in photosynthetic light-harvesting},}\ }\href {\doibase
  10.1098/rsta.2011.0420} {\bibfield  {journal} {\bibinfo  {journal}
  {Philosophical Transactions of the Royal Society A}\ }\textbf {\bibinfo
  {volume} {370}},\ \bibinfo {pages} {3728--3749} (\bibinfo {year}
  {2012})}\BibitemShut {NoStop}%
\bibitem [{\citenamefont {Tretiak}\ \emph {et~al.}(2002)\citenamefont
  {Tretiak}, \citenamefont {Saxena}, \citenamefont {Martin},\ and\
  \citenamefont {Bishop}}]{Tretiak02}%
  \BibitemOpen
  \bibfield  {author} {\bibinfo {author} {\bibfnamefont {S.}~\bibnamefont
  {Tretiak}}, \bibinfo {author} {\bibfnamefont {A.}~\bibnamefont {Saxena}},
  \bibinfo {author} {\bibfnamefont {R.~L.}\ \bibnamefont {Martin}}, \ and\
  \bibinfo {author} {\bibfnamefont {A.~R.}\ \bibnamefont {Bishop}},\ }\bibfield
   {title} {\enquote {\bibinfo {title} {Conformational dynamics of photoexcited
  conjugated molecules},}\ }\href {<Go to ISI>://WOS:000177529000057}
  {\bibfield  {journal} {\bibinfo  {journal} {Physical Review Letters}\
  }\textbf {\bibinfo {volume} {89}},\ \bibinfo {pages} {097402} (\bibinfo
  {year} {2002})}\BibitemShut {NoStop}%
\bibitem [{\citenamefont {Karabunarliev}\ and\ \citenamefont
  {Bittner}(2003)}]{Bittner03}%
  \BibitemOpen
  \bibfield  {author} {\bibinfo {author} {\bibfnamefont {S.}~\bibnamefont
  {Karabunarliev}}\ and\ \bibinfo {author} {\bibfnamefont {E.~R.}\ \bibnamefont
  {Bittner}},\ }\bibfield  {title} {\enquote {\bibinfo {title}
  {Polaron-excitons and electron-vibrational band shapes in conjugated
  polymers},}\ }\href {\doibase 10.1063/1.1543938} {\bibfield  {journal}
  {\bibinfo  {journal} {Journal of Chemical Physics}\ }\textbf {\bibinfo
  {volume} {118}},\ \bibinfo {pages} {4291--4296} (\bibinfo {year}
  {2003})}\BibitemShut {NoStop}%
\bibitem [{\citenamefont {Sterpone}\ and\ \citenamefont
  {Rossky}(2008)}]{Sterpone08}%
  \BibitemOpen
  \bibfield  {author} {\bibinfo {author} {\bibfnamefont {F.}~\bibnamefont
  {Sterpone}}\ and\ \bibinfo {author} {\bibfnamefont {P.~J.}\ \bibnamefont
  {Rossky}},\ }\bibfield  {title} {\enquote {\bibinfo {title} {Molecular
  modeling and simulation of conjugated polymer oligomers: Ground and excited
  state chain dynamics of ppv in the gas phase},}\ }\href {\doibase
  10.1021/jp711848q} {\bibfield  {journal} {\bibinfo  {journal} {Journal of
  Physical Chemistry B}\ }\textbf {\bibinfo {volume} {112}},\ \bibinfo {pages}
  {4983--4993} (\bibinfo {year} {2008})}\BibitemShut {NoStop}%
\bibitem [{\citenamefont {De~Leener}\ \emph {et~al.}(2009)\citenamefont
  {De~Leener}, \citenamefont {Hennebicq}, \citenamefont {Sancho-Garcia},\ and\
  \citenamefont {Beljonne}}]{Leener09}%
  \BibitemOpen
  \bibfield  {author} {\bibinfo {author} {\bibfnamefont {C.}~\bibnamefont
  {De~Leener}}, \bibinfo {author} {\bibfnamefont {E.}~\bibnamefont
  {Hennebicq}}, \bibinfo {author} {\bibfnamefont {J.~C.}\ \bibnamefont
  {Sancho-Garcia}}, \ and\ \bibinfo {author} {\bibfnamefont {D.}~\bibnamefont
  {Beljonne}},\ }\bibfield  {title} {\enquote {\bibinfo {title} {Modeling the
  dynamics of chromophores in conjugated polymers: The case of meh-ppv},}\
  }\href {\doibase 10.1021/jp8029902} {\bibfield  {journal} {\bibinfo
  {journal} {Journal of Physical Chemistry B}\ }\textbf {\bibinfo {volume}
  {113}},\ \bibinfo {pages} {1311--1322} (\bibinfo {year} {2009})}\BibitemShut
  {NoStop}%
\bibitem [{Note4()}]{Note4}%
  \BibitemOpen
  \bibinfo {note} {In fact, the Ehrenfest approximation is the cause of the
  unphysical bifurcation of the exciton density onto separate chromophores
  found in Ehrenfest simulations of the relaxation dynamics of high energy
  photoexcited states\cite {Tozer12}.}\BibitemShut {Stop}%
\bibitem [{\citenamefont {Breuer}\ and\ \citenamefont
  {Petruccione}(2002)}]{Breuer02}%
  \BibitemOpen
  \bibfield  {author} {\bibinfo {author} {\bibfnamefont {H.-P.}\ \bibnamefont
  {Breuer}}\ and\ \bibinfo {author} {\bibfnamefont {F.}~\bibnamefont
  {Petruccione}},\ }\href@noop {} {\emph {\bibinfo {title} {The Theory of Open
  Quantum Systems}}}\ (\bibinfo  {publisher} {Oxford University Press},\
  \bibinfo {address} {Oxford},\ \bibinfo {year} {2002})\BibitemShut {NoStop}%
\bibitem [{\citenamefont {Daley}(2014)}]{Daley14}%
  \BibitemOpen
  \bibfield  {author} {\bibinfo {author} {\bibfnamefont {A.~J.}\ \bibnamefont
  {Daley}},\ }\bibfield  {title} {\enquote {\bibinfo {title} {Quantum
  trajectories and open many-body quantum systems},}\ }\href {\doibase
  10.1080/00018732.2014.933502} {\bibfield  {journal} {\bibinfo  {journal}
  {Advances in Physics}\ }\textbf {\bibinfo {volume} {63}},\ \bibinfo {pages}
  {77--149} (\bibinfo {year} {2014})}\BibitemShut {NoStop}%
\bibitem [{\citenamefont {Bednarz}, \citenamefont {Malyshev},\ and\
  \citenamefont {Knoester}(2002)}]{Bednarz02}%
  \BibitemOpen
  \bibfield  {author} {\bibinfo {author} {\bibfnamefont {M.}~\bibnamefont
  {Bednarz}}, \bibinfo {author} {\bibfnamefont {V.~A.}\ \bibnamefont
  {Malyshev}}, \ and\ \bibinfo {author} {\bibfnamefont {J.}~\bibnamefont
  {Knoester}},\ }\bibfield  {title} {\enquote {\bibinfo {title} {Intraband
  relaxation and temperature dependence of the fluorescence decay time of
  one-dimensional frenkel excitons: The pauli master equation approach},}\
  }\href {\doibase 10.1063/1.1499483} {\bibfield  {journal} {\bibinfo
  {journal} {Journal of Chemical Physics}\ }\textbf {\bibinfo {volume} {117}},\
  \bibinfo {pages} {6200--6213} (\bibinfo {year} {2002})}\BibitemShut {NoStop}%
\bibitem [{\citenamefont {Albu}\ and\ \citenamefont {Yaron}(2013)}]{Albu13}%
  \BibitemOpen
  \bibfield  {author} {\bibinfo {author} {\bibfnamefont {N.~M.}\ \bibnamefont
  {Albu}}\ and\ \bibinfo {author} {\bibfnamefont {D.~J.}\ \bibnamefont
  {Yaron}},\ }\bibfield  {title} {\enquote {\bibinfo {title} {Brownian dynamics
  model of excited-state relaxation in solutions of conjugated oligomers},}\
  }\href {\doibase 10.1021/jp400538g} {\bibfield  {journal} {\bibinfo
  {journal} {Journal of Physical Chemistry C}\ }\textbf {\bibinfo {volume}
  {117}},\ \bibinfo {pages} {12299--12306} (\bibinfo {year}
  {2013})}\BibitemShut {NoStop}%
\bibitem [{\citenamefont {Spano}\ \emph {et~al.}(2008)\citenamefont {Spano},
  \citenamefont {Meskers}, \citenamefont {Hennebicq},\ and\ \citenamefont
  {Beljonne}}]{Spano08a}%
  \BibitemOpen
  \bibfield  {author} {\bibinfo {author} {\bibfnamefont {F.~C.}\ \bibnamefont
  {Spano}}, \bibinfo {author} {\bibfnamefont {S.~C.~J.}\ \bibnamefont
  {Meskers}}, \bibinfo {author} {\bibfnamefont {E.}~\bibnamefont {Hennebicq}},
  \ and\ \bibinfo {author} {\bibfnamefont {D.}~\bibnamefont {Beljonne}},\
  }\bibfield  {title} {\enquote {\bibinfo {title} {Using circularly polarized
  luminescence to probe exciton coherence in disordered helical aggregates},}\
  }\href {<Go to ISI>://WOS:000257629100041} {\bibfield  {journal} {\bibinfo
  {journal} {Journal of Chemical Physics}\ }\textbf {\bibinfo {volume} {129}},\
  \bibinfo {pages} {024704} (\bibinfo {year} {2008})}\BibitemShut {NoStop}%
\bibitem [{\citenamefont {Tempelaar}\ \emph {et~al.}(2014)\citenamefont
  {Tempelaar}, \citenamefont {Spano}, \citenamefont {Knoester},\ and\
  \citenamefont {Jansen}}]{Tempelaar14}%
  \BibitemOpen
  \bibfield  {author} {\bibinfo {author} {\bibfnamefont {R.}~\bibnamefont
  {Tempelaar}}, \bibinfo {author} {\bibfnamefont {F.~C.}\ \bibnamefont
  {Spano}}, \bibinfo {author} {\bibfnamefont {J.}~\bibnamefont {Knoester}}, \
  and\ \bibinfo {author} {\bibfnamefont {T.~L.~C.}\ \bibnamefont {Jansen}},\
  }\bibfield  {title} {\enquote {\bibinfo {title} {Mapping the evolution of
  spatial exciton coherence through time-resolved fluorescence},}\ }\href
  {\doibase 10.1021/jz500488u} {\bibfield  {journal} {\bibinfo  {journal}
  {Journal of Physical Chemistry Letters}\ }\textbf {\bibinfo {volume} {5}},\
  \bibinfo {pages} {1505--1510} (\bibinfo {year} {2014})}\BibitemShut {NoStop}%
\bibitem [{\citenamefont {Barford}\ and\ \citenamefont
  {Mannouch}(2018)}]{Barford18}%
  \BibitemOpen
  \bibfield  {author} {\bibinfo {author} {\bibfnamefont {W.}~\bibnamefont
  {Barford}}\ and\ \bibinfo {author} {\bibfnamefont {J.~R.}\ \bibnamefont
  {Mannouch}},\ }\bibfield  {title} {\enquote {\bibinfo {title} {Torsionally
  induced exciton localization and decoherence in $\pi$-conjugated polymers},}\
  }\href {<Go to ISI>://WOS:000452539400008} {\bibfield  {journal} {\bibinfo
  {journal} {Journal of Chemical Physics}\ }\textbf {\bibinfo {volume} {149}},\
  \bibinfo {pages} {214107} (\bibinfo {year} {2018})}\BibitemShut {NoStop}%
\bibitem [{\citenamefont {French}(1971)}]{French71}%
  \BibitemOpen
  \bibfield  {author} {\bibinfo {author} {\bibfnamefont {A.~P.}\ \bibnamefont
  {French}},\ }\href@noop {} {\emph {\bibinfo {title} {Vibrations and Waves}}}\
  (\bibinfo  {publisher} {Nelson},\ \bibinfo {address} {London},\ \bibinfo
  {year} {1971})\BibitemShut {NoStop}%
\bibitem [{\citenamefont {Lakowicz}(2006)}]{Lakowicz06}%
  \BibitemOpen
  \bibfield  {author} {\bibinfo {author} {\bibfnamefont {J.~R.}\ \bibnamefont
  {Lakowicz}},\ }\href@noop {} {\emph {\bibinfo {title} {Principles of
  Fluorescence Spectroscopy}}},\ \bibinfo {edition} {3rd}\ ed.\ (\bibinfo
  {publisher} {Springer},\ \bibinfo {address} {New York},\ \bibinfo {year}
  {2006})\BibitemShut {NoStop}%
\bibitem [{Note5()}]{Note5}%
  \BibitemOpen
  \bibinfo {note} {I.\ Gonzalvez Perez and W.\ Barford, in
  preparation.}\BibitemShut {Stop}%
\bibitem [{\citenamefont {Clark}\ \emph {et~al.}(2012)\citenamefont {Clark},
  \citenamefont {Nelson}, \citenamefont {Tretiak}, \citenamefont {Cirmi},\ and\
  \citenamefont {Lanzani}}]{Clark12}%
  \BibitemOpen
  \bibfield  {author} {\bibinfo {author} {\bibfnamefont {J.}~\bibnamefont
  {Clark}}, \bibinfo {author} {\bibfnamefont {T.}~\bibnamefont {Nelson}},
  \bibinfo {author} {\bibfnamefont {S.}~\bibnamefont {Tretiak}}, \bibinfo
  {author} {\bibfnamefont {G.}~\bibnamefont {Cirmi}}, \ and\ \bibinfo {author}
  {\bibfnamefont {G.}~\bibnamefont {Lanzani}},\ }\bibfield  {title} {\enquote
  {\bibinfo {title} {Femtosecond torsional relaxation},}\ }\href {\doibase
  10.1038/Nphys2210} {\bibfield  {journal} {\bibinfo  {journal} {Nature
  Physics}\ }\textbf {\bibinfo {volume} {8}},\ \bibinfo {pages} {225--231}
  (\bibinfo {year} {2012})}\BibitemShut {NoStop}%
\bibitem [{\citenamefont {Tozer}\ and\ \citenamefont
  {Barford}(2015)}]{Tozer15}%
  \BibitemOpen
  \bibfield  {author} {\bibinfo {author} {\bibfnamefont {O.~R.}\ \bibnamefont
  {Tozer}}\ and\ \bibinfo {author} {\bibfnamefont {W.}~\bibnamefont
  {Barford}},\ }\bibfield  {title} {\enquote {\bibinfo {title} {Intrachain
  exciton dynamics in conjugated polymer chains in solution},}\ }\href {<Go to
  ISI>://WOS:000360653900005} {\bibfield  {journal} {\bibinfo  {journal}
  {Journal of Chemical Physics}\ }\textbf {\bibinfo {volume} {143}},\ \bibinfo
  {pages} {084102} (\bibinfo {year} {2015})}\BibitemShut {NoStop}%
\bibitem [{Note6()}]{Note6}%
  \BibitemOpen
  \bibinfo {note} {This latter assumption was shown by Lee and Willard\cite
  {Lee19} to be problematic for the non-adiabatic transport described in
  Section \ref {se:5.4}.}\BibitemShut {Stop}%
\bibitem [{\citenamefont {Binder}\ and\ \citenamefont
  {Burghardt}()}]{Binder20a}%
  \BibitemOpen
  \bibfield  {author} {\bibinfo {author} {\bibfnamefont {R.}~\bibnamefont
  {Binder}}\ and\ \bibinfo {author} {\bibfnamefont {I.}~\bibnamefont
  {Burghardt}},\ }\bibfield  {title} {\enquote {\bibinfo {title}
  {First-principles description of intra-chain exciton migration in an
  oligo(para-phenylene vinylene) chain. 2. ml-mctdh simulations of exciton
  dynamics at a torsional defect},}\ }\href@noop {} {\bibfield  {journal}
  {\bibinfo  {journal} {Journal of Chemical Physics}\ }\textbf {\bibinfo
  {volume} {152}}}\BibitemShut {NoStop}%
\bibitem [{\citenamefont {Binder}\ and\ \citenamefont
  {Burghardt}(2020)}]{Binder20c}%
  \BibitemOpen
  \bibfield  {author} {\bibinfo {author} {\bibfnamefont {R.}~\bibnamefont
  {Binder}}\ and\ \bibinfo {author} {\bibfnamefont {I.}~\bibnamefont
  {Burghardt}},\ }\bibfield  {title} {\enquote {\bibinfo {title}
  {First-principles quantum simulations of exciton diffusion on a minimal
  oligothiophene chain at finite temperature},}\ }\href {\doibase
  10.1039/c9fd00066f} {\bibfield  {journal} {\bibinfo  {journal} {Faraday
  Discussions}\ }\textbf {\bibinfo {volume} {221}},\ \bibinfo {pages}
  {406--427} (\bibinfo {year} {2020})}\BibitemShut {NoStop}%
\bibitem [{\citenamefont {Hegger}, \citenamefont {Binder},\ and\ \citenamefont
  {Burghardt}(2020)}]{Hegger20}%
  \BibitemOpen
  \bibfield  {author} {\bibinfo {author} {\bibfnamefont {R.}~\bibnamefont
  {Hegger}}, \bibinfo {author} {\bibfnamefont {R.}~\bibnamefont {Binder}}, \
  and\ \bibinfo {author} {\bibfnamefont {I.}~\bibnamefont {Burghardt}},\
  }\bibfield  {title} {\enquote {\bibinfo {title} {First-principles quantum and
  quantum-classical simulations of exciton diffusion in semiconducting polymer
  chains at finite temperature},}\ }\href {\doibase 10.1021/acs.jctc.0c00351}
  {\bibfield  {journal} {\bibinfo  {journal} {Journal of Chemical Theory and
  Computation}\ }\textbf {\bibinfo {volume} {16}},\ \bibinfo {pages}
  {5441--5455} (\bibinfo {year} {2020})}\BibitemShut {NoStop}%
\bibitem [{Note7()}]{Note7}%
  \BibitemOpen
  \bibinfo {note} {This process is sometimes referred to as
  Environment-Assisted Quantum Transport\cite {Rebentrost09}.}\BibitemShut
  {Stop}%
\bibitem [{\citenamefont {Nitzan}(2006)}]{Nitzan06}%
  \BibitemOpen
  \bibfield  {author} {\bibinfo {author} {\bibfnamefont {A.}~\bibnamefont
  {Nitzan}},\ }\href {Table of contents only
  http://www.loc.gov/catdir/toc/ecip062/2005030160.html Publisher description
  http://www.loc.gov/catdir/enhancements/fy0907/2005030160-d.html} {\emph
  {\bibinfo {title} {Chemical Dynamics in Condensed phases: Relaxation,
  Transfer and Reactions in Condensed Molecular Systems}}},\ Oxford graduate
  texts\ (\bibinfo  {publisher} {Oxford University Press},\ \bibinfo {address}
  {Oxford},\ \bibinfo {year} {2006})\BibitemShut {NoStop}%
\bibitem [{\citenamefont {Barford}\ and\ \citenamefont
  {Tozer}(2014)}]{Barford14c}%
  \BibitemOpen
  \bibfield  {author} {\bibinfo {author} {\bibfnamefont {W.}~\bibnamefont
  {Barford}}\ and\ \bibinfo {author} {\bibfnamefont {O.~R.}\ \bibnamefont
  {Tozer}},\ }\bibfield  {title} {\enquote {\bibinfo {title} {Theory of exciton
  transfer and diffusion in conjugated polymers},}\ }\href {<Go to
  ISI>://WOS:000344589700007} {\bibfield  {journal} {\bibinfo  {journal}
  {Journal of Chemical Physics}\ }\textbf {\bibinfo {volume} {141}},\ \bibinfo
  {pages} {164103} (\bibinfo {year} {2014})}\BibitemShut {NoStop}%
\bibitem [{\citenamefont {Hwang}\ and\ \citenamefont
  {Scholes}(2011)}]{Scholes11a}%
  \BibitemOpen
  \bibfield  {author} {\bibinfo {author} {\bibfnamefont {I.}~\bibnamefont
  {Hwang}}\ and\ \bibinfo {author} {\bibfnamefont {G.~D.}\ \bibnamefont
  {Scholes}},\ }\bibfield  {title} {\enquote {\bibinfo {title} {Electronic
  energy transfer and quantum-coherence in $\pi$-conjugated polymers},}\ }\href
  {\doibase 10.1021/cm102360x} {\bibfield  {journal} {\bibinfo  {journal}
  {Chemistry of Materials}\ }\textbf {\bibinfo {volume} {23}},\ \bibinfo
  {pages} {610--620} (\bibinfo {year} {2011})}\BibitemShut {NoStop}%
\bibitem [{\citenamefont {Movaghar}\ \emph {et~al.}(1986)\citenamefont
  {Movaghar}, \citenamefont {Grunewald}, \citenamefont {Ries}, \citenamefont
  {B\"assler},\ and\ \citenamefont {Wurtz}}]{Movaghar86a}%
  \BibitemOpen
  \bibfield  {author} {\bibinfo {author} {\bibfnamefont {B.}~\bibnamefont
  {Movaghar}}, \bibinfo {author} {\bibfnamefont {M.}~\bibnamefont {Grunewald}},
  \bibinfo {author} {\bibfnamefont {B.}~\bibnamefont {Ries}}, \bibinfo {author}
  {\bibfnamefont {H.}~\bibnamefont {B\"assler}}, \ and\ \bibinfo {author}
  {\bibfnamefont {D.}~\bibnamefont {Wurtz}},\ }\bibfield  {title} {\enquote
  {\bibinfo {title} {Diffusion and relaxation of energy in disordered organic
  and inorganic materials},}\ }\href {\doibase DOI 10.1103/PhysRevB.33.5545}
  {\bibfield  {journal} {\bibinfo  {journal} {Physical Review B}\ }\textbf
  {\bibinfo {volume} {33}},\ \bibinfo {pages} {5545--5554} (\bibinfo {year}
  {1986})}\BibitemShut {NoStop}%
\bibitem [{\citenamefont {Movaghar}, \citenamefont {Ries},\ and\ \citenamefont
  {Grunewald}(1986)}]{Movaghar86b}%
  \BibitemOpen
  \bibfield  {author} {\bibinfo {author} {\bibfnamefont {B.}~\bibnamefont
  {Movaghar}}, \bibinfo {author} {\bibfnamefont {B.}~\bibnamefont {Ries}}, \
  and\ \bibinfo {author} {\bibfnamefont {M.}~\bibnamefont {Grunewald}},\
  }\bibfield  {title} {\enquote {\bibinfo {title} {Diffusion and relaxation of
  energy in disordered-systems - departure from mean-field theories},}\
  }\href@noop {} {\bibfield  {journal} {\bibinfo  {journal} {Physical Review
  B}\ }\textbf {\bibinfo {volume} {34}},\ \bibinfo {pages} {5724--5582}
  (\bibinfo {year} {1986})}\BibitemShut {NoStop}%
\bibitem [{\citenamefont {Meskers}\ \emph {et~al.}(2001)\citenamefont
  {Meskers}, \citenamefont {Hubner}, \citenamefont {Oestreich},\ and\
  \citenamefont {Bassler}}]{Meskers01}%
  \BibitemOpen
  \bibfield  {author} {\bibinfo {author} {\bibfnamefont {S.~C.~J.}\
  \bibnamefont {Meskers}}, \bibinfo {author} {\bibfnamefont {J.}~\bibnamefont
  {Hubner}}, \bibinfo {author} {\bibfnamefont {M.}~\bibnamefont {Oestreich}}, \
  and\ \bibinfo {author} {\bibfnamefont {H.}~\bibnamefont {Bassler}},\
  }\bibfield  {title} {\enquote {\bibinfo {title} {Dispersive relaxation
  dynamics of photoexcitations in a polyfluorene film involving energy
  transfer: Experiment and monte carlo simulations},}\ }\href {\doibase DOI
  10.1021/jp0113331} {\bibfield  {journal} {\bibinfo  {journal} {Journal of
  Physical Chemistry B}\ }\textbf {\bibinfo {volume} {105}},\ \bibinfo {pages}
  {9139--9149} (\bibinfo {year} {2001})}\BibitemShut {NoStop}%
\bibitem [{\citenamefont {Athanasopoulos}, \citenamefont {B\"assler},\ and\
  \citenamefont {K\"ohler}(2019)}]{Athanasopoulos19}%
  \BibitemOpen
  \bibfield  {author} {\bibinfo {author} {\bibfnamefont {S.}~\bibnamefont
  {Athanasopoulos}}, \bibinfo {author} {\bibfnamefont {H.}~\bibnamefont
  {B\"assler}}, \ and\ \bibinfo {author} {\bibfnamefont {A.}~\bibnamefont
  {K\"ohler}},\ }\bibfield  {title} {\enquote {\bibinfo {title} {Disorder vs
  delocalization: Which is more advantageous for high-efficiency organic solar
  cells?}}\ }\href@noop {} {\bibfield  {journal} {\bibinfo  {journal} {Journal
  of Physical Chemistry Letters}\ }\textbf {\bibinfo {volume} {10}},\ \bibinfo
  {pages} {7107--7112} (\bibinfo {year} {2019})}\BibitemShut {NoStop}%
\bibitem [{\citenamefont {Beljonne}\ \emph {et~al.}(2002)\citenamefont
  {Beljonne}, \citenamefont {Pourtois}, \citenamefont {Silva}, \citenamefont
  {Hennebicq}, \citenamefont {Herz}, \citenamefont {Friend}, \citenamefont
  {Scholes}, \citenamefont {Setayesh}, \citenamefont {Mullen},\ and\
  \citenamefont {Bredas}}]{Beljonne02}%
  \BibitemOpen
  \bibfield  {author} {\bibinfo {author} {\bibfnamefont {D.}~\bibnamefont
  {Beljonne}}, \bibinfo {author} {\bibfnamefont {G.}~\bibnamefont {Pourtois}},
  \bibinfo {author} {\bibfnamefont {C.}~\bibnamefont {Silva}}, \bibinfo
  {author} {\bibfnamefont {E.}~\bibnamefont {Hennebicq}}, \bibinfo {author}
  {\bibfnamefont {L.~M.}\ \bibnamefont {Herz}}, \bibinfo {author}
  {\bibfnamefont {R.~H.}\ \bibnamefont {Friend}}, \bibinfo {author}
  {\bibfnamefont {G.~D.}\ \bibnamefont {Scholes}}, \bibinfo {author}
  {\bibfnamefont {S.}~\bibnamefont {Setayesh}}, \bibinfo {author}
  {\bibfnamefont {K.}~\bibnamefont {Mullen}}, \ and\ \bibinfo {author}
  {\bibfnamefont {J.~L.}\ \bibnamefont {Bredas}},\ }\bibfield  {title}
  {\enquote {\bibinfo {title} {Interchain vs. intrachain energy transfer in
  acceptor-capped conjugated polymers},}\ }\href {\doibase
  10.1073/pnas.172390999} {\bibfield  {journal} {\bibinfo  {journal}
  {Proceedings of the National Academy of Sciences of the United States of
  America}\ }\textbf {\bibinfo {volume} {99}},\ \bibinfo {pages} {10982--10987}
  (\bibinfo {year} {2002})}\BibitemShut {NoStop}%
\bibitem [{\citenamefont {Hennebicq}\ \emph {et~al.}(2005)\citenamefont
  {Hennebicq}, \citenamefont {Pourtois}, \citenamefont {Scholes}, \citenamefont
  {Herz}, \citenamefont {Russell}, \citenamefont {Silva}, \citenamefont
  {Setayesh}, \citenamefont {Grimsdale}, \citenamefont {Mullen}, \citenamefont
  {Bredas},\ and\ \citenamefont {Beljonne}}]{Beljonne05}%
  \BibitemOpen
  \bibfield  {author} {\bibinfo {author} {\bibfnamefont {E.}~\bibnamefont
  {Hennebicq}}, \bibinfo {author} {\bibfnamefont {G.}~\bibnamefont {Pourtois}},
  \bibinfo {author} {\bibfnamefont {G.~D.}\ \bibnamefont {Scholes}}, \bibinfo
  {author} {\bibfnamefont {L.~M.}\ \bibnamefont {Herz}}, \bibinfo {author}
  {\bibfnamefont {D.~M.}\ \bibnamefont {Russell}}, \bibinfo {author}
  {\bibfnamefont {C.}~\bibnamefont {Silva}}, \bibinfo {author} {\bibfnamefont
  {S.}~\bibnamefont {Setayesh}}, \bibinfo {author} {\bibfnamefont {A.~C.}\
  \bibnamefont {Grimsdale}}, \bibinfo {author} {\bibfnamefont {K.}~\bibnamefont
  {Mullen}}, \bibinfo {author} {\bibfnamefont {J.~L.}\ \bibnamefont {Bredas}},
  \ and\ \bibinfo {author} {\bibfnamefont {D.}~\bibnamefont {Beljonne}},\
  }\bibfield  {title} {\enquote {\bibinfo {title} {Exciton migration in
  rigid-rod conjugated polymers: An improved forster model},}\ }\href {\doibase
  10.1021/ja0488784} {\bibfield  {journal} {\bibinfo  {journal} {Journal of the
  American Chemical Society}\ }\textbf {\bibinfo {volume} {127}},\ \bibinfo
  {pages} {4744--4762} (\bibinfo {year} {2005})}\BibitemShut {NoStop}%
\bibitem [{\citenamefont {Singh}\ \emph {et~al.}(2009)\citenamefont {Singh},
  \citenamefont {Bittner}, \citenamefont {Beljonne},\ and\ \citenamefont
  {Scholes}}]{Singh09}%
  \BibitemOpen
  \bibfield  {author} {\bibinfo {author} {\bibfnamefont {J.}~\bibnamefont
  {Singh}}, \bibinfo {author} {\bibfnamefont {E.~R.}\ \bibnamefont {Bittner}},
  \bibinfo {author} {\bibfnamefont {D.}~\bibnamefont {Beljonne}}, \ and\
  \bibinfo {author} {\bibfnamefont {G.~D.}\ \bibnamefont {Scholes}},\
  }\bibfield  {title} {\enquote {\bibinfo {title} {Fluorescence depolarization
  in poly[2-methoxy-5-((2-ethylhexyl)oxy)-1,4-phenylenevinylene]: Sites versus
  eigenstates hopping},}\ }\href {<Go to ISI>://WOS:000272050200035} {\bibfield
   {journal} {\bibinfo  {journal} {Journal of Chemical Physics}\ }\textbf
  {\bibinfo {volume} {131}},\ \bibinfo {pages} {194905} (\bibinfo {year}
  {2009})}\BibitemShut {NoStop}%
\bibitem [{\citenamefont {Barford}, \citenamefont {Bittner},\ and\
  \citenamefont {Ward}(2012)}]{Barford12b}%
  \BibitemOpen
  \bibfield  {author} {\bibinfo {author} {\bibfnamefont {W.}~\bibnamefont
  {Barford}}, \bibinfo {author} {\bibfnamefont {E.~R.}\ \bibnamefont
  {Bittner}}, \ and\ \bibinfo {author} {\bibfnamefont {A.}~\bibnamefont
  {Ward}},\ }\bibfield  {title} {\enquote {\bibinfo {title} {Exciton dynamics
  in disordered poly(p-phenylenevinylene).\ 2.\ exciton diffusion},}\ }\href
  {\doibase 10.1021/jp310680u} {\bibfield  {journal} {\bibinfo  {journal}
  {Journal of Physical Chemistry A}\ }\textbf {\bibinfo {volume} {116}},\
  \bibinfo {pages} {10870--10870} (\bibinfo {year} {2012})}\BibitemShut
  {NoStop}%
\bibitem [{\citenamefont {Barford}(2007)}]{Barford07}%
  \BibitemOpen
  \bibfield  {author} {\bibinfo {author} {\bibfnamefont {W.}~\bibnamefont
  {Barford}},\ }\bibfield  {title} {\enquote {\bibinfo {title} {Exciton
  transfer integrals between polymer chains},}\ }\href {\doibase Artn
  13490510.1063/1.2714516} {\bibfield  {journal} {\bibinfo  {journal} {Journal
  of Chemical Physics}\ }\textbf {\bibinfo {volume} {126}},\ \bibinfo {pages}
  {134905} (\bibinfo {year} {2007})}\BibitemShut {NoStop}%
\bibitem [{\citenamefont {Wong}, \citenamefont {Bagchi},\ and\ \citenamefont
  {Rossky}(2004)}]{Wong04}%
  \BibitemOpen
  \bibfield  {author} {\bibinfo {author} {\bibfnamefont {K.~F.}\ \bibnamefont
  {Wong}}, \bibinfo {author} {\bibfnamefont {B.}~\bibnamefont {Bagchi}}, \ and\
  \bibinfo {author} {\bibfnamefont {P.~J.}\ \bibnamefont {Rossky}},\ }\bibfield
   {title} {\enquote {\bibinfo {title} {Distance and orientation dependence of
  excitation transfer rates in conjugated systems: Beyond the f\"orster
  theory},}\ }\href {\doibase 10.1021/jp037724s} {\bibfield  {journal}
  {\bibinfo  {journal} {Journal of Physical Chemistry A}\ }\textbf {\bibinfo
  {volume} {108}},\ \bibinfo {pages} {5752--5763} (\bibinfo {year}
  {2004})}\BibitemShut {NoStop}%
\bibitem [{\citenamefont {Das}\ and\ \citenamefont {Ramasesha}(2010)}]{Das10}%
  \BibitemOpen
  \bibfield  {author} {\bibinfo {author} {\bibfnamefont {M.}~\bibnamefont
  {Das}}\ and\ \bibinfo {author} {\bibfnamefont {S.}~\bibnamefont
  {Ramasesha}},\ }\bibfield  {title} {\enquote {\bibinfo {title} {Fluorescent
  resonant excitation energy transfer in linear polyenes},}\ }\href {<Go to
  ISI>://WOS:000276209700014} {\bibfield  {journal} {\bibinfo  {journal}
  {Journal of Chemical Physics}\ }\textbf {\bibinfo {volume} {132}},\ \bibinfo
  {pages} {124109} (\bibinfo {year} {2010})}\BibitemShut {NoStop}%
\bibitem [{\citenamefont {Barford}(2010)}]{Barford10c}%
  \BibitemOpen
  \bibfield  {author} {\bibinfo {author} {\bibfnamefont {W.}~\bibnamefont
  {Barford}},\ }\bibfield  {title} {\enquote {\bibinfo {title} {Beyond forster
  resonance energy transfer in linear nanoscale systems},}\ }\href {\doibase
  10.1021/jp107374r} {\bibfield  {journal} {\bibinfo  {journal} {Journal of
  Physical Chemistry A}\ }\textbf {\bibinfo {volume} {114}},\ \bibinfo {pages}
  {11842--11843} (\bibinfo {year} {2010})}\BibitemShut {NoStop}%
\bibitem [{\citenamefont {Hayes}, \citenamefont {Samuel},\ and\ \citenamefont
  {Phillips}(1995)}]{Hayes95}%
  \BibitemOpen
  \bibfield  {author} {\bibinfo {author} {\bibfnamefont {G.~R.}\ \bibnamefont
  {Hayes}}, \bibinfo {author} {\bibfnamefont {I.~D.~W.}\ \bibnamefont
  {Samuel}}, \ and\ \bibinfo {author} {\bibfnamefont {R.~T.}\ \bibnamefont
  {Phillips}},\ }\bibfield  {title} {\enquote {\bibinfo {title} {Exciton
  dynamics iu electroluminescent polymers studied by femtosecond time-resolved
  photoluminescence spectroscopy},}\ }\href {\doibase DOI
  10.1103/PhysRevB.52.R11569} {\bibfield  {journal} {\bibinfo  {journal}
  {Physical Review B}\ }\textbf {\bibinfo {volume} {52}},\ \bibinfo {pages}
  {11569--11572} (\bibinfo {year} {1995})}\BibitemShut {NoStop}%
\bibitem [{\citenamefont {Markov}\ \emph {et~al.}(2005)\citenamefont {Markov},
  \citenamefont {Amsterdam}, \citenamefont {Blom}, \citenamefont {Sieval},\
  and\ \citenamefont {Hummelen}}]{Markov05}%
  \BibitemOpen
  \bibfield  {author} {\bibinfo {author} {\bibfnamefont {D.~E.}\ \bibnamefont
  {Markov}}, \bibinfo {author} {\bibfnamefont {E.}~\bibnamefont {Amsterdam}},
  \bibinfo {author} {\bibfnamefont {P.~W.~M.}\ \bibnamefont {Blom}}, \bibinfo
  {author} {\bibfnamefont {A.~B.}\ \bibnamefont {Sieval}}, \ and\ \bibinfo
  {author} {\bibfnamefont {J.~C.}\ \bibnamefont {Hummelen}},\ }\bibfield
  {title} {\enquote {\bibinfo {title} {Accurate measurement of the exciton
  diffusion length in a conjugated polymer using a heterostructure with a
  side-chain cross-linked fullerene layer},}\ }\href {\doibase
  10.1021/jp0509663} {\bibfield  {journal} {\bibinfo  {journal} {Journal of
  Physical Chemistry A}\ }\textbf {\bibinfo {volume} {109}},\ \bibinfo {pages}
  {5266--5274} (\bibinfo {year} {2005})}\BibitemShut {NoStop}%
\bibitem [{\citenamefont {Lewis}\ \emph {et~al.}(2006)\citenamefont {Lewis},
  \citenamefont {Ruseckas}, \citenamefont {Gaudin}, \citenamefont {Webster},
  \citenamefont {Burn},\ and\ \citenamefont {Samuel}}]{Lewis06}%
  \BibitemOpen
  \bibfield  {author} {\bibinfo {author} {\bibfnamefont {A.~J.}\ \bibnamefont
  {Lewis}}, \bibinfo {author} {\bibfnamefont {A.}~\bibnamefont {Ruseckas}},
  \bibinfo {author} {\bibfnamefont {O.~P.~M.}\ \bibnamefont {Gaudin}}, \bibinfo
  {author} {\bibfnamefont {G.~R.}\ \bibnamefont {Webster}}, \bibinfo {author}
  {\bibfnamefont {P.~L.}\ \bibnamefont {Burn}}, \ and\ \bibinfo {author}
  {\bibfnamefont {I.~D.~W.}\ \bibnamefont {Samuel}},\ }\bibfield  {title}
  {\enquote {\bibinfo {title} {Singlet exciton diffusion in meh-ppv films
  studied by exciton-exciton annihilation},}\ }\href {\doibase
  10.1016/j.orgel.2006.05.009} {\bibfield  {journal} {\bibinfo  {journal}
  {Organic Electronics}\ }\textbf {\bibinfo {volume} {7}},\ \bibinfo {pages}
  {452--456} (\bibinfo {year} {2006})}\BibitemShut {NoStop}%
\bibitem [{\citenamefont {Scully}\ and\ \citenamefont
  {McGehee}(2006)}]{Scully06}%
  \BibitemOpen
  \bibfield  {author} {\bibinfo {author} {\bibfnamefont {S.~R.}\ \bibnamefont
  {Scully}}\ and\ \bibinfo {author} {\bibfnamefont {M.~D.}\ \bibnamefont
  {McGehee}},\ }\bibfield  {title} {\enquote {\bibinfo {title} {Effects of
  optical interference and energy transfer on exciton diffusion length
  measurements in organic semiconductors},}\ }\href {<Go to
  ISI>://WOS:000239764100119} {\bibfield  {journal} {\bibinfo  {journal}
  {Journal of Applied Physics}\ }\textbf {\bibinfo {volume} {100}},\ \bibinfo
  {pages} {034907} (\bibinfo {year} {2006})}\BibitemShut {NoStop}%
\bibitem [{\citenamefont {K\"ohler}\ and\ \citenamefont
  {B\"assler}(2015)}]{Kohler15}%
  \BibitemOpen
  \bibfield  {author} {\bibinfo {author} {\bibfnamefont {A.}~\bibnamefont
  {K\"ohler}}\ and\ \bibinfo {author} {\bibfnamefont {H.}~\bibnamefont
  {B\"assler}},\ }\href@noop {} {\emph {\bibinfo {title} {Electronic Processes
  in Organic Semiconductors: An Introduction}}},\ \bibinfo {edition} {1st}\
  ed.\ (\bibinfo  {publisher} {Wiley-VCH},\ \bibinfo {address} {Weinheim},\
  \bibinfo {year} {2015})\BibitemShut {NoStop}%
\bibitem [{\citenamefont {Spano}\ and\ \citenamefont
  {Yamagata}(2011)}]{Spano11}%
  \BibitemOpen
  \bibfield  {author} {\bibinfo {author} {\bibfnamefont {F.~C.}\ \bibnamefont
  {Spano}}\ and\ \bibinfo {author} {\bibfnamefont {H.}~\bibnamefont
  {Yamagata}},\ }\bibfield  {title} {\enquote {\bibinfo {title} {Vibronic
  coupling in j-aggregates and beyond: A direct means of determining the
  exciton coherence length from the photoluminescence spectrum},}\ }\href
  {\doibase 10.1021/jp104752k} {\bibfield  {journal} {\bibinfo  {journal}
  {Journal of Physical Chemistry B}\ }\textbf {\bibinfo {volume} {115}},\
  \bibinfo {pages} {5133--5143} (\bibinfo {year} {2011})}\BibitemShut {NoStop}%
\bibitem [{\citenamefont {Yamagata}\ and\ \citenamefont
  {Spano}(2014)}]{Spano14}%
  \BibitemOpen
  \bibfield  {author} {\bibinfo {author} {\bibfnamefont {H.}~\bibnamefont
  {Yamagata}}\ and\ \bibinfo {author} {\bibfnamefont {F.~C.}\ \bibnamefont
  {Spano}},\ }\bibfield  {title} {\enquote {\bibinfo {title} {Strong
  photophysical similarities between conjugated polymers and j-aggregates},}\
  }\href {\doibase 10.1021/jz402450m} {\bibfield  {journal} {\bibinfo
  {journal} {Journal of Physical Chemistry Letters}\ }\textbf {\bibinfo
  {volume} {5}},\ \bibinfo {pages} {622--632} (\bibinfo {year}
  {2014})}\BibitemShut {NoStop}%
\bibitem [{\citenamefont {Rebentrost}\ \emph {et~al.}(2009)\citenamefont
  {Rebentrost}, \citenamefont {Mohseni}, \citenamefont {Kassal}, \citenamefont
  {Lloyd},\ and\ \citenamefont {Aspuru-Guzik}}]{Rebentrost09}%
  \BibitemOpen
  \bibfield  {author} {\bibinfo {author} {\bibfnamefont {P.}~\bibnamefont
  {Rebentrost}}, \bibinfo {author} {\bibfnamefont {M.}~\bibnamefont {Mohseni}},
  \bibinfo {author} {\bibfnamefont {I.}~\bibnamefont {Kassal}}, \bibinfo
  {author} {\bibfnamefont {S.}~\bibnamefont {Lloyd}}, \ and\ \bibinfo {author}
  {\bibfnamefont {A.}~\bibnamefont {Aspuru-Guzik}},\ }\bibfield  {title}
  {\enquote {\bibinfo {title} {Environment-assisted quantum transport},}\
  }\href {<Go to ISI>://WOS:000263744500003} {\bibfield  {journal} {\bibinfo
  {journal} {New Journal of Physics}\ }\textbf {\bibinfo {volume} {11}},\
  \bibinfo {pages} {033003} (\bibinfo {year} {2009})}\BibitemShut {NoStop}%
\bibitem [{\citenamefont {Tozer}\ and\ \citenamefont
  {Barford}(2012)}]{Tozer12}%
  \BibitemOpen
  \bibfield  {author} {\bibinfo {author} {\bibfnamefont {O.~R.}\ \bibnamefont
  {Tozer}}\ and\ \bibinfo {author} {\bibfnamefont {W.}~\bibnamefont
  {Barford}},\ }\bibfield  {title} {\enquote {\bibinfo {title} {Exciton
  dynamics in disordered poly(p-phenylenevinylene).\ 1.\ ultrafast
  interconversion and dynamical localization},}\ }\href {\doibase
  10.1021/jp307040d} {\bibfield  {journal} {\bibinfo  {journal} {Journal of
  Physical Chemistry A}\ }\textbf {\bibinfo {volume} {116}},\ \bibinfo {pages}
  {10310--10318} (\bibinfo {year} {2012})}\BibitemShut {NoStop}%
\bibitem [{\citenamefont {Lee}\ and\ \citenamefont {Willard}(2019)}]{Lee19}%
  \BibitemOpen
  \bibfield  {author} {\bibinfo {author} {\bibfnamefont {E.~M.~Y.}\
  \bibnamefont {Lee}}\ and\ \bibinfo {author} {\bibfnamefont {A.~P.}\
  \bibnamefont {Willard}},\ }\bibfield  {title} {\enquote {\bibinfo {title}
  {Solving the trivial crossing problem while preserving the nodal symmetry of
  the wave function},}\ }\href {\doibase 10.1021/acs.jctc.9b00302} {\bibfield
  {journal} {\bibinfo  {journal} {Journal of Chemical Theory and Computation}\
  }\textbf {\bibinfo {volume} {15}},\ \bibinfo {pages} {4332--4343} (\bibinfo
  {year} {2019})}\BibitemShut {NoStop}%
\end{thebibliography}%

\end{document}